\documentclass{aa}  
\usepackage{graphicx}
\catcode`\@=11
\def\gtsima{\ifmmode{\mathrel{\mathpalette\@versim>}}
    \else{$\mathrel{\mathpalette\@versim>}$}\fi}
\def\ltsima{\ifmmode{\mathrel{\mathpalette\@versim<}}
    \else{$\mathrel{\mathpalette\@versim<}$}\fi}
\def\@versim#1#2{\lower 2.9truept \vbox{\baselineskip 0pt \lineskip 
    0.5truept \ialign{$\m@th#1\hfil##\hfil$\crcr#2\crcr\sim\crcr}}}
\catcode`\@=12
\def\lae{Ly$\alpha$ }
\usepackage{txfonts}
\begin{document}
   \title{He~II emitters in the VIMOS VLT Deep Survey: Population III
     star formation or peculiar stellar populations in galaxies at
     $2<z<4.6$?}  \titlerunning{He~II emitters in VVDS}

   \author{P. Cassata \inst{1}\fnmsep\thanks{Based on data obtained
with the European Southern Observatory Very Large Telescope, Paranal,
Chile, under Large Programs 070.A-9007 and 177.A-0837.  Based on
observations obtained with MegaPrime/MegaCam, a joint project of CFHT
and CEA/DAPNIA, at the Canada-France-Hawaii Telescope (CFHT) which is
operated by the National Research Council (NRC) of Canada, the
Institut National des Sciences de l'Univers of the Centre National de
la Recherche Scientifique (CNRS) of France, and the University of
Hawaii. This work is based in part on data products produced at
TERAPIX and the Canadian Astronomy Data Centre as part of the
Canada-France-Hawaii Telescope Legacy Survey, a collaborative project
of NRC and CNRS.}, 
O. Le F\`evre \inst{1}, 
S. Charlot \inst{2},
T. Contini \inst{3},
O. Cucciati \inst{4},
B. Garilli \inst{5},
G. Zamorani \inst{4},
C. Adami \inst{1},
S. Bardelli \inst{4},
V. Le Brun \inst{1},
B. Lemaux \inst{1},
D. Maccagni \inst{5},
A. Pollo \inst{6,7},
L. Pozzetti \inst{4},
L. Tresse \inst{1},
D. Vergani \inst{8},
A. Zanichelli \inst{4}
and E. Zucca \inst{4},
}
\authorrunning{P. Cassata et al.}  \offprints{P. Cassata}

   \institute{Aix Marseille Universit\'e, CNRS, LAM (Laboratoire
     d'Astrophysique de Marseille) UMR 7326, 13388, Marseille, France
     \email{paolo.cassata@oamp.fr} \and UPMC-CNRS, UMR7095, Institut
     d'Astrophysique de Paris, F-75014, Paris, France \and Institut de
     Recherche en Astrophysique et Plan\'etologie, CNRS, Universit\'e de
     Toulouse, 14 avenue E. Belin 31400 Toulouse, France \and
     INAF-Osservatorio Astronomico di Bologna - via Ranzani 1,
     I-40127, Bologna, Italy \and IASF-INAF - via Bassini 15, I-20133,
     Milano, Italy \and Astronomical Observatory of the Jagiellonian
     University, ul. Orla 171, 30-244 Krak\'ow, Poland\and National
     Centre for Nuclear Research, ul. Ho\.za 69, 00-681 Warsaw, Poland
     \and INAF – IASFBO, via P. Gobetti 101, 40129 Bologna, Italy
}

   \date{Received .....; accepted .....}

  \abstract 
{} 
  {The aim of this work is to identify He~II emitters at $2<z<4.6$ and
    to constrain the source of the hard ionizing continuum that powers
    the He~II emission.}
  {We assembled a sample of 277 galaxies with a highly reliable
    spectroscopic redshift at $2<z<4.6$ from the VIMOS-VLT Deep Survey
    (VVDS) Deep and Ultra-Deep data, and we identified 39
    He~II~$\lambda$1640 emitters. We studied their spectral
    properties, measuring the fluxes, equivalent widths (EW), and Full
    Width at Half Maximum (FWHM) for most relevant lines, including
    He~II~$\lambda$1640, Ly$\alpha$ line, Si~II~$\lambda$1527, and
    C~IV~$\lambda$1549.}
  {About 10\% of galaxies at $z\sim3$ and $i_{AB} \leq 24.75$ show
    He~II in emission, with rest frame equivalent widths EW$_0\sim$1
    -- 7\AA, equally distributed between galaxies with Ly$\alpha$ in
    emission or in absorption.  We find 11 (3.9\% of the global
    population) reliable He~II emitters with unresolved He~II lines
    (FWHM$_0<1200 km/s$), 13 (4.6\% of the global population) reliable
    emitters with broad He~II emission (FWHM$_0>1200 km/s$), 3 Active
    Galactic Nuclei (AGN), and an additional 12 possible He~II
    emitters. The properties of the individual broad emitters are in
    agreement with expectations from a Wolf-Rayet (W-R)
    model. Instead, the properties of the narrow emitters are not
    compatible with this model, nor with predictions of gravitational
    cooling radiation produced by gas accretion, unless this is
    severely underestimated by current models by more than two orders
    of magnitude. Rather, we find that the EW of the narrow He~II line
    emitters are in agreement with expectations for a Population III
    (PopIII) star formation, if the episode of star formation is
    continuous, and we calculate that a PopIII Star Formation Rate
    (SFR) of 0.1 -- 10 $M_{\odot}yr^{-1}$ alone is enough to sustain
    the observed He~II flux. }
  {We conclude that narrow He~II emitters are powered either by the
    ionizing flux from a stellar population rare at $z\sim0$ but much
    more common at $z\sim3$, or by PopIII star formation. As proposed
    by Tornatore and collaborators, incomplete Inter Stellar Medium
    (ISM) mixing may leave some small pockets of pristine gas at the
    periphery of galaxies from which PopIII may form, even down to
    $z\sim2$ or lower. If this interpretation is correct, we measure
    at $z\sim3$ a Star Formation Rate Density (SFRD) in PopIII stars
    of $10^{-6} M_{\odot} yr^{-1} Mpc^{-3}$, higher than, but
    qualitatively comparable to the value predicted by Tornatore and
    collaborators.}


   \maketitle
%

\section{Introduction}
Understanding the early phases of star formation in the Universe is a
major topic of recent astrophysical investigation. In particular,
despite the growing size of galaxy samples at $z>5$ -- 8, selected
either through the Lyman-break technique (e.g. Bouwens~et~al.~2007;
Bouwens~et~al.~2008; Bouwens~et~al.~2010; McLure~et~al.~2011) or via
strong Ly$\alpha$ emission detected in narrow band filters
(Hu~et~al.~2004; Tapken~et~al.~2006; Murayama~et~al.~2007;
Hibon~et~al.~2010; Hibon~et~al.~2012), some of which have
spectroscopic confirmation (Vanzella~et~al.~2009; Capak~et~al.~2011;
Curtis-Lake~et~al.~2012; Schenker~et~al.~2012), the population
responsible for the reionization of the Universe at this epoch has not
been identified yet. In this respect, zero metallicity stars and/or
proto-galaxies, which should be the first structures formed in the
life of the Universe and whose role in ionizing the Universe is
predicted to be very important, have not been identified yet. The
first population of stars formed from the pristine gas during the
reionization is the so-called Population III (hereafter Pop III). This
population is predicted to be of extremely low metallicity and to
produce strong UV ionizing continuum (Tumlinson~\&~Shull~2000;
Schaerer~2002). No evidence for Pop-III stars has been found so
far. Moreover, despite the high efficiency of the cold mode gas
accretion in cosmological simulation (Keres~et~al.~2005,
Dekel~et~al.~2009), which may be enough to sustain the observed rapid
increase of the Star Formation Rate Density (SFRD) of the Universe
between $z\sim8$ and $z\sim2$, scarce evidence for gas infalling onto
an overdensity has been reported so far. Steidel~et~al.~(2010) found
only partial evidence of inflow around galaxies at $z\sim2$;
Cresci~et~al.~(2010) claimed that the inverted metallicity gradient
observed in z$\sim3$ galaxies could be due to cold flows;
Giavalisco~et~al.~(2011) detected cold gas around z$\sim1.6$ galaxies
that is possibly infalling into an overdensity.

Coincidently, the footprint of both PopIII star formation and gas
cooling during gravitational accretion is the presence of
Ly$\alpha\lambda~1216$ and He~II~$\lambda~1640$ emission lines in the
spectra of the sources. Dual Ly$\alpha\lambda~1216$ and
He~II~$\lambda~1640$ emitters have been proposed by various authors as
candidates hosting PopIII star formation (Tumlinson~et~al.~2001;
Schaerer~2003; Raiter,~Schaerer~\&~Fosbury~2010). In Pop III star
formation regions, where an extreme top-heavy initial mass function
(IMF) is expected, very massive stars with high effective temperatures
are formed. These stars, unlike normal PopII and PopI stars, produce
photons shorter by $\lambda=228\AA~$ that can ionize $He^+$. In a
region with primeval composition and in the absence of other metals,
the H and He lines become the dominant line coolant for the gas, and
thus the gas emits strong Ly$\alpha+He~II$ emission. On the other
hand, Fardal~et~al.~(2001) showed that the pristine gas recently
accreted onto an overdensity cools down and emits hydrogen or helium
line radiation; Yang~et~al.~(2006) claimed that dual
Ly$\alpha\lambda~1216$ and He~II~$\lambda~1640$ can be used to trace
the infall of pristine gas onto an overdensity. If the accretion mode
is cold, as predicted by many theoretical studies (Fardal~et~al.~2001;
Keres~et~al.~2005), the gas temperature is below the halo virial
temperature, reaching $T\sim10^5K$; if the gas is pristine (zero
metallicity), H and He lines are the most efficient gas coolants, and
thus Ly$\alpha\lambda~1216$ and He~II~$\lambda~1640$ emission is
produced.

Strong Ly$\alpha+$He~II emission lines are commonly found in other
astrophysical objects, such as W-R stars, AGN, and supernovae driven
winds. However, several diagnostics can be used to distinguish cooling
radiation and PopIII star formation from these other mechanisms: AGN,
W-R stars and emitters powered by supernovae driven winds typically
show other emission lines in their spectra, such as C~III and C~IV
(Reuland~et~al.~2007; Leitherer~et~al.~1996;
Allen~et~al.~2008). Moreover, W-R stars are associated with strong
winds, and thus produce broad emission lines (a few 1000 km/s,
Schaerer~2003). Brinchmann,~Pettini~\&~Charlot~(2008) showed that the
broad He~II emission detected in the composite spectrum of $z\sim3$
galaxies by Shapley~et~al.~(2003) can indeed be reproduced by W-R
models.  In W-R galaxies, or young star clusters with heavy star
formation, He~II emission may be observed in either a W-R stellar mode
and broadened by the W-R winds, or in a nebular narrow-line mode as
the strong UV emission from these stars photoionizes the surronding
medium (Kudritzki~2002). Nebular He~II emission also appears in some
star forming regions in which the source of ionisation is not clearly
identified as W-R or O stars, but this is a rare event in the local
universe (Kehrig~et~al.~2011).

Unfortunately, no direct diagnostics can be used to distinguish
between some extreme PopI populations, PopIII, or cooling radiation,
to explain He~II~$\lambda~1640$ emission; rather, only indirect
inference can be used to discriminate these possibilities.
Schaerer~et~al.~(2003) predicted that a PopIII region forming stars at
a rate of 1 $M_{\odot}yr^{-1}$, dependent on the IMF (top-heavy or
less extreme) and the metallicity ($0<Z<10^{-5}$), will produce He~II
luminosities between $\sim1.5\times10^{39}$ and $\sim5\times10^{41}$
erg/s. On the other hand, Yang~et~al.~(2006) predicted that pristine
gas infalling into overdensities at $z\sim$2 -- 3 would produce
similar He~II luminosities, between $\sim1\times10^{40}$ and
$\sim1\times10^{41}$ erg/s. In principle, the two mechanisms predict
different $L_{Ly\alpha}/L_{He~II}$ ratios: around $\sim100$ for the
PopIII regions and around $\sim10$ for the cooling radiation. However,
since Ly$\alpha$, unlike He~II, is a resonant line with a large cross
section, the observed Ly$\alpha$ emission depends on various aspects
such as the geometry of the system, its dynamical state, or the
presence of dust. The combination of these effects results in an
average escape fraction $f_{esc}$ of the Ly$\alpha$ photons around
10\% at $2<z<4$ (Hayes~et~al.~2011). However, individual galaxies at
those redshift show $f_{esc}$ values that fluctuate significantly
around the average value.  As a result, the observed ratio
$L_{Ly\alpha}/L_{He~II}$ can be very different from the intrinsic one
and thus can not be used as a diagnostics.

The quest for Population III stars or gravitational cooling emission
at high-$z$ has so far been quite unproductive. Searches of dual
Ly$\alpha$+He~II emitters beyond $z\sim5$ are difficult, as the
He~II~$\lambda~1640$ line is redshifted in the near infrared:
Nagao~et~al.~(2005) found no He~II in a deep near-infrared
spectroscopic observation of a strong Ly$\alpha$ emitter at z=6.6. New
multi-object high-performance near-infrared spectrographs coming
online in the next months will give new insights into this aspect.

Several studies at $z\sim$~4 -- 5 have identified objects with
extremely large Ly$\alpha$ equivalent widths (EW$_0>$250\AA~), that
are expected in the case of a top-heavy IMF, a very young age ($<10^7$ 
yrs) and/or a very low metallicity (Malhotra\&Rhoads~2002;
Shimasaku~et~al.~2006). However, there was no evidence for He~II in
emission in these galaxies, either in individual or in stacked
spectra. A dedicated survey for dual Ly$\alpha$+He~II emitters at
$z\sim4-5$ (Nagao~et~al.~2008) found no convincing candidates.

Searches of zero metallicity objects at lower redshift where
observations should be easier in principle have to face the question
whether significant PopIII star formation is expected to take place at
these low redshifts. Tornatore,~Ferrara,~and~Schneider~(2007) showed
that Population III star formation continues down to $z\sim2.5$
because of inefficient heavy element transport by outflows, that
leaves pockets of pristine gas untouched in the periphery of collapsed
structures.  Population III stars can form in these pokets, even if at
a rate 10$^4$ times slower than PopII.
Fumagalli,~O'Meara,~\&~Prochaska~(2011) detected two gas clouds with
primordial composition (Z$<10^{-4}$) at redshift z$\sim3$.


Motivated by these predictions, some authors succeeded in identifying
$1.5<z<3$ Ly$\alpha+$He~II emitters, but none could unambiguously
conclude which powerful source of ionization produced these
emission lines.  Prescott,~Dey~\&~Jannuzi~(2009) discovered a
Ly$\alpha$ nebula at $z\sim1.67$, and they concluded that W-R stars or
shocks could not be the sources of ionization. Scarlata~et~al.~(2009)
studied in great detail an extended Ly$\alpha$ blob at $z\sim2.38$,
but could not determine whether the source of Ly$\alpha$+He~II emission
is an AGN or instead cooling radiation.


The aim of this paper is to look for He~II emitters in the VIMOS VLT
Deep Survey (VVDS) Deep and Ultra-Deep data at $2<z<4.6$, measuring how
frequent they are. The VVDS provides a large unbiased i-band
magnitude-selected spectroscopic sample of galaxies with measured
redshifts (Le F\`evre et al., 2005, and Le F\`evre et al., 2013, in
prep.), offering an excellent basis to perform a census of He~II
emitters. On the galaxies with He~II detection, we measure their
properties such as line fluxes, equivalent width and line width in an
attempt to constrain the mechanisms powering the emission.

Throughout the paper, we use a standard Cosmology with $\Omega_M=0.3$,
$\Omega_{\Lambda}=0.7$, and $h=0.7$. In Sect.~2 we present the
observations; in Sect. 3 we introduce the sample of He~II emitters;
in Sect. 4 we analyze the main mechanisms that can produce He~II
emission and we compare them with the properties of our sample; in
Sect. 5 we discuss our results; and in Sect.~6 we draw our
conclusions.

\begin{figure*}[!ht]
  \centering
\includegraphics[width=.95\textwidth]{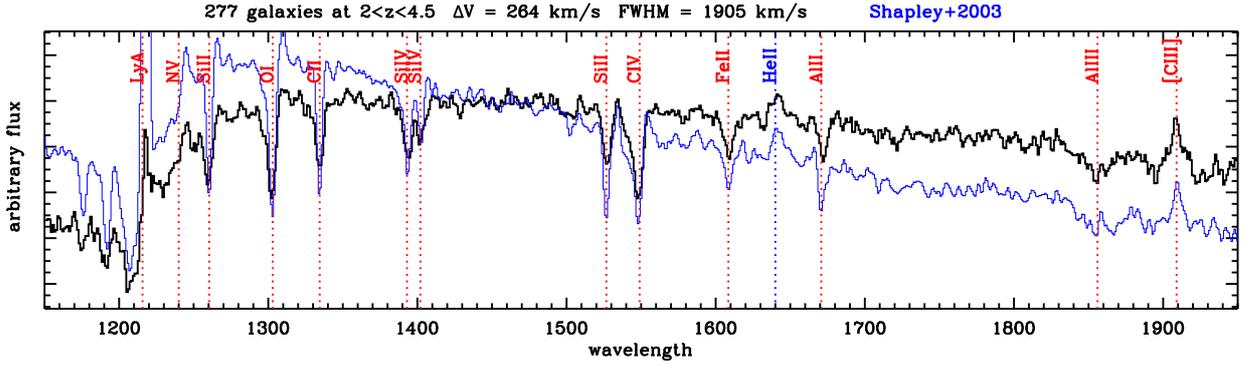}
\caption{Stack of all 277 spectra at $2<z<4.6$ for which He~II falls
  in the spectral region, in comparison with the composite spectrum by
  Shapley~et~al.~(2003), in blue.}
\label{spec_all}
\end{figure*}

\begin{figure*}[!h]
  \centering
\includegraphics[width=1\textwidth]{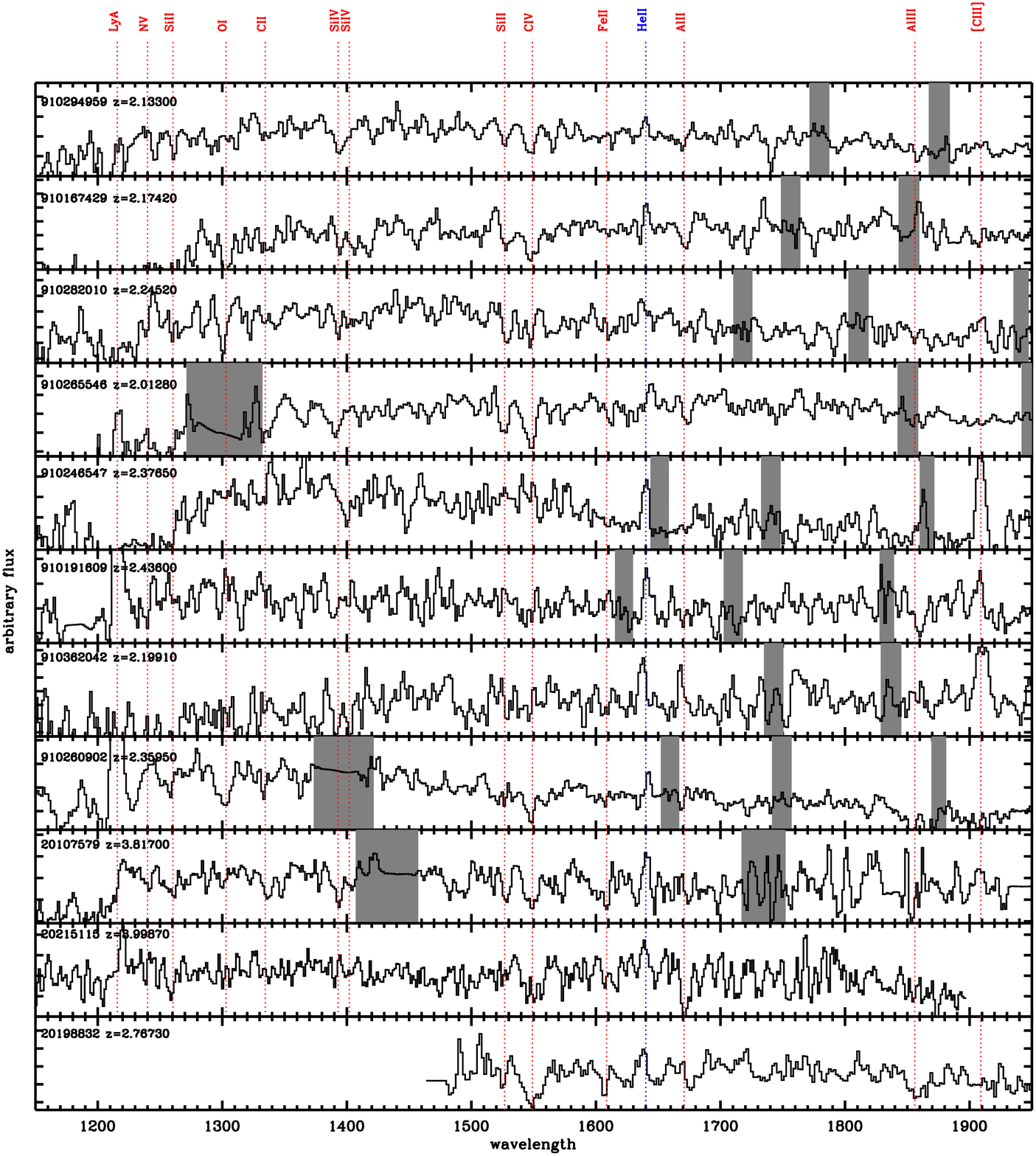}
\caption{Spectra of the reliable He~II emitters with narrow He~II
  emission (FWHM$<1000km/s$). The grey bands show regions of the
  spectra that are contaminated by strong skylines or zero orders. We
  also report for each galaxy the id, the velocity difference between
  the He~II and the systemic redshift, the width (FWHM) of the He~II
  line, and the systemic redshift.}
\label{spec_narrow}
\end{figure*}

\begin{figure*}[!ht]
  \centering
\includegraphics[width=.95\textwidth]{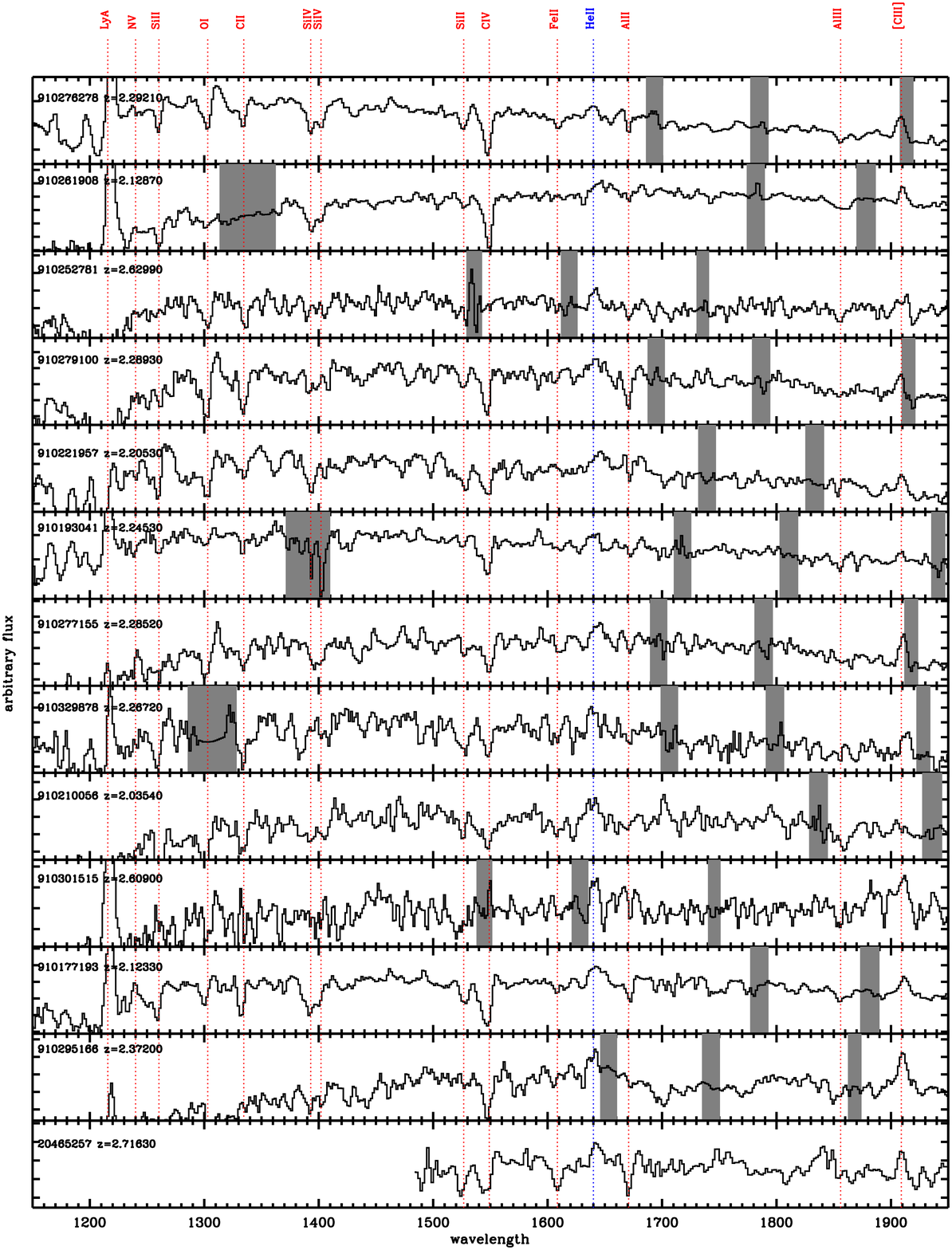}
\caption{Spectra of the reliable He~II emitters with broad He~II
  emission (FWHM$>1800km/s$).  The grey bands show regions of the
  spectra that are contaminated by strong skylines or zero orders. We
  also report for each galaxy the id, the velocity difference between
  the He~II and the systemic redshift, the width (FWHM) of the He~II
  line, and the systemic redshift. }
\label{spec_broad}
\end{figure*}

\begin{figure*}[!ht]
  \centering
\includegraphics[width=1\textwidth]{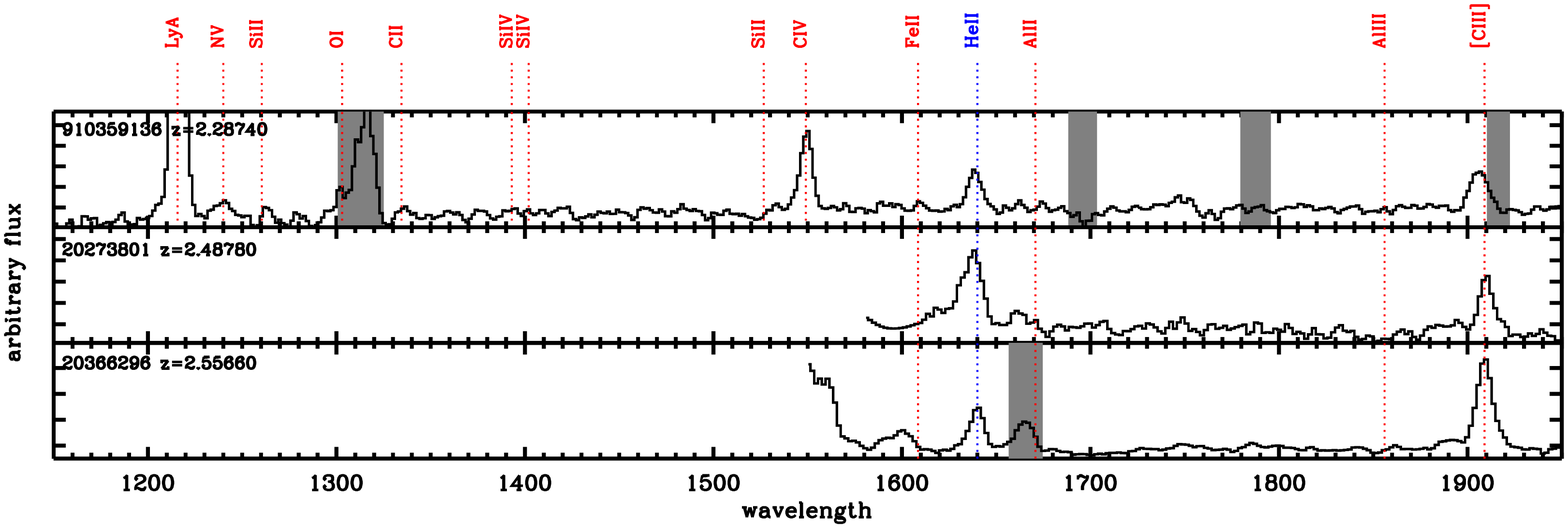}
\caption{Spectra of the AGN with He~II emission. The grey bands show
  regions of the spectra that are contaminated by strong skylines or
  zero orders. We also report for each galaxy the id, the velocity
  difference between the He~II and the systemic redshift, the width
  (FWHM) of the He~II line, and the systemic redshift.}
\label{spec_agn}
\end{figure*}

\begin{figure*}[!ht]
  \centering
\includegraphics[width=1\textwidth]{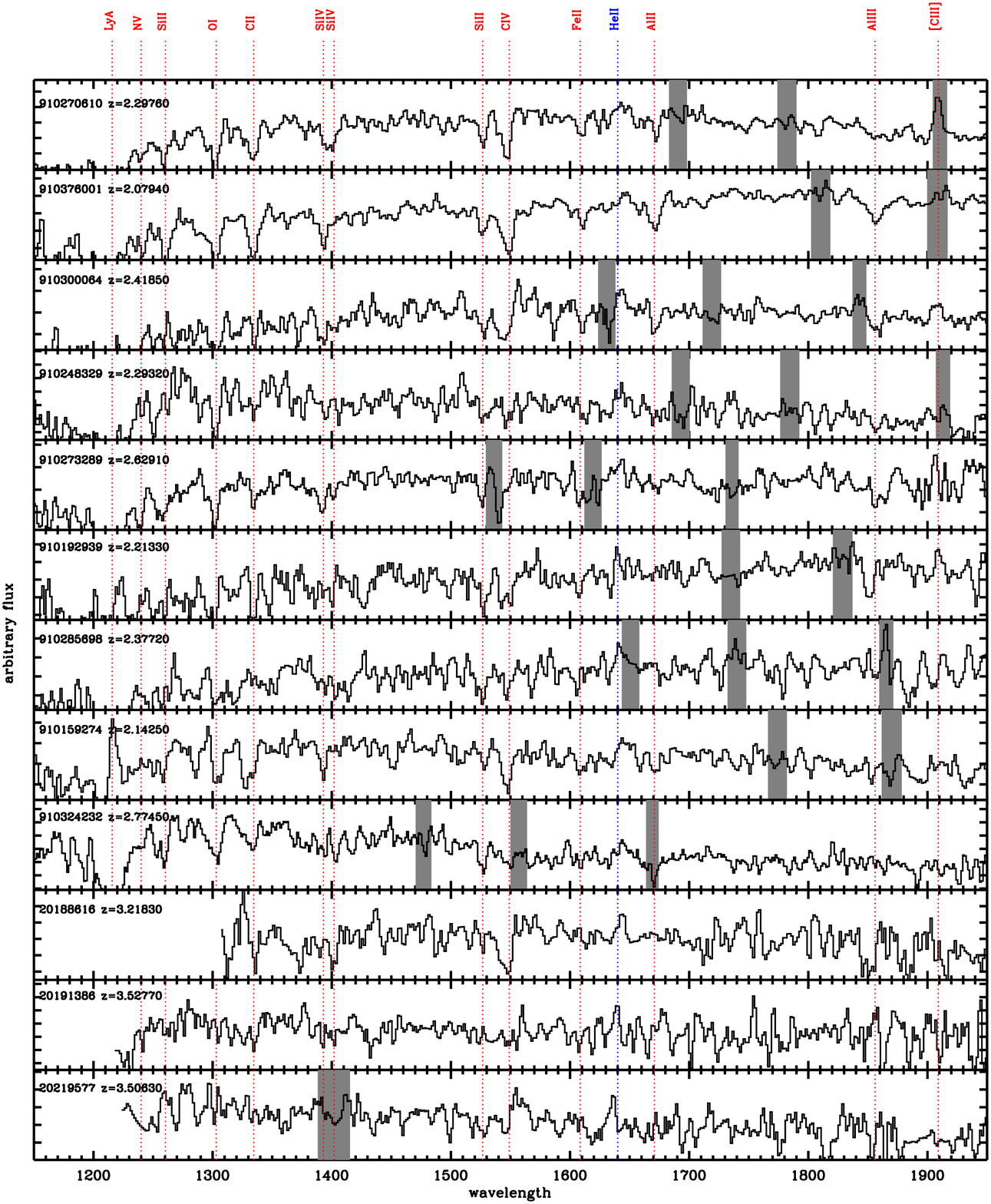}
\caption{Spectra of the possible He~II emitters.  The grey bands show
  regions of the spectra that are contaminated by strong skylines or
  zero orders. We also report for each galaxy the id, the velocity
  difference between the He~II and the systemic redshift, the width
  (FWHM) of the He~II line, and the systemic redshift.}
\label{spec_uncertain}
\end{figure*}

\begin{figure*}[!ht]
  \centering
\includegraphics[width=.8\columnwidth]{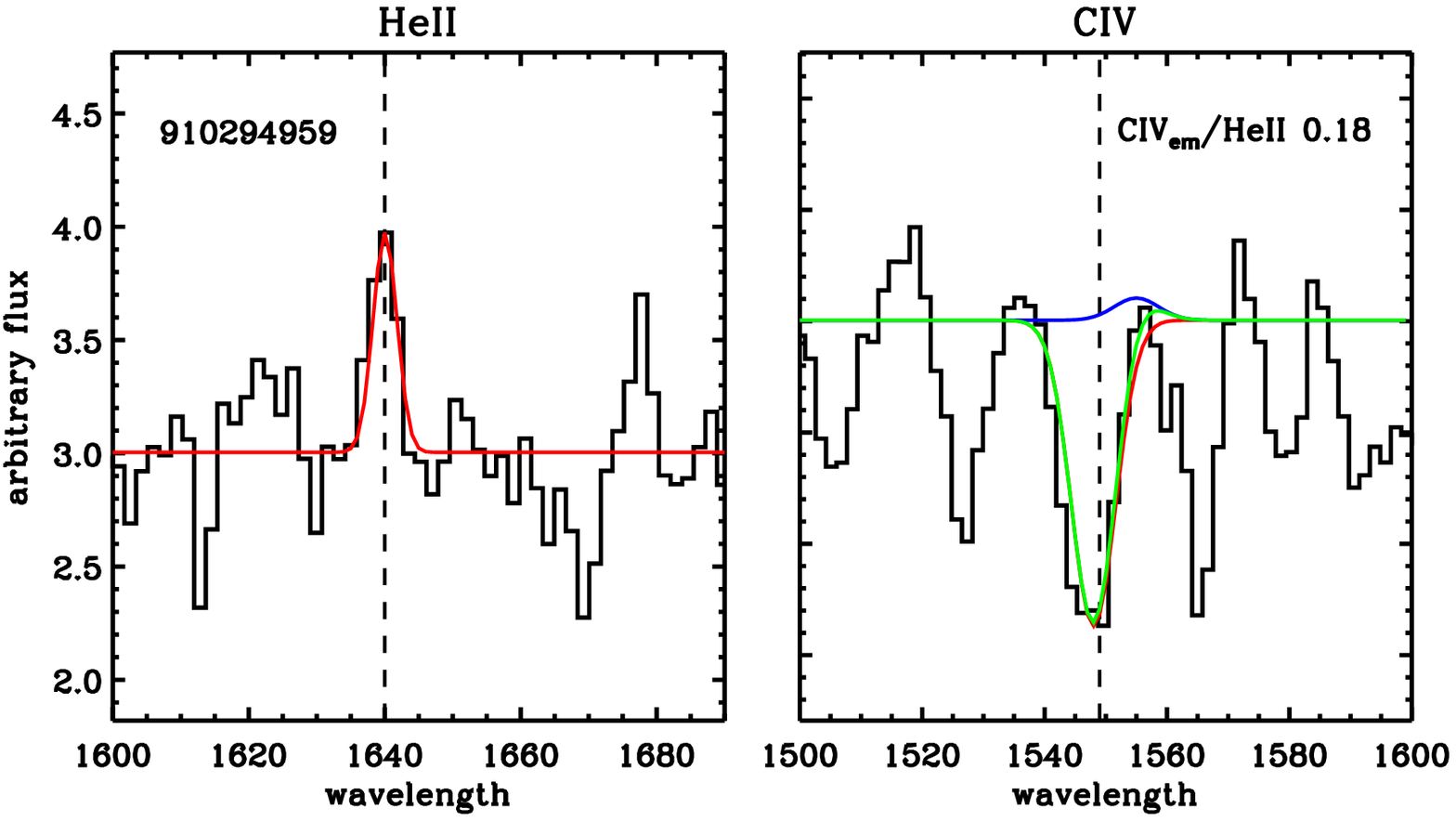}
\includegraphics[width=.8\columnwidth]{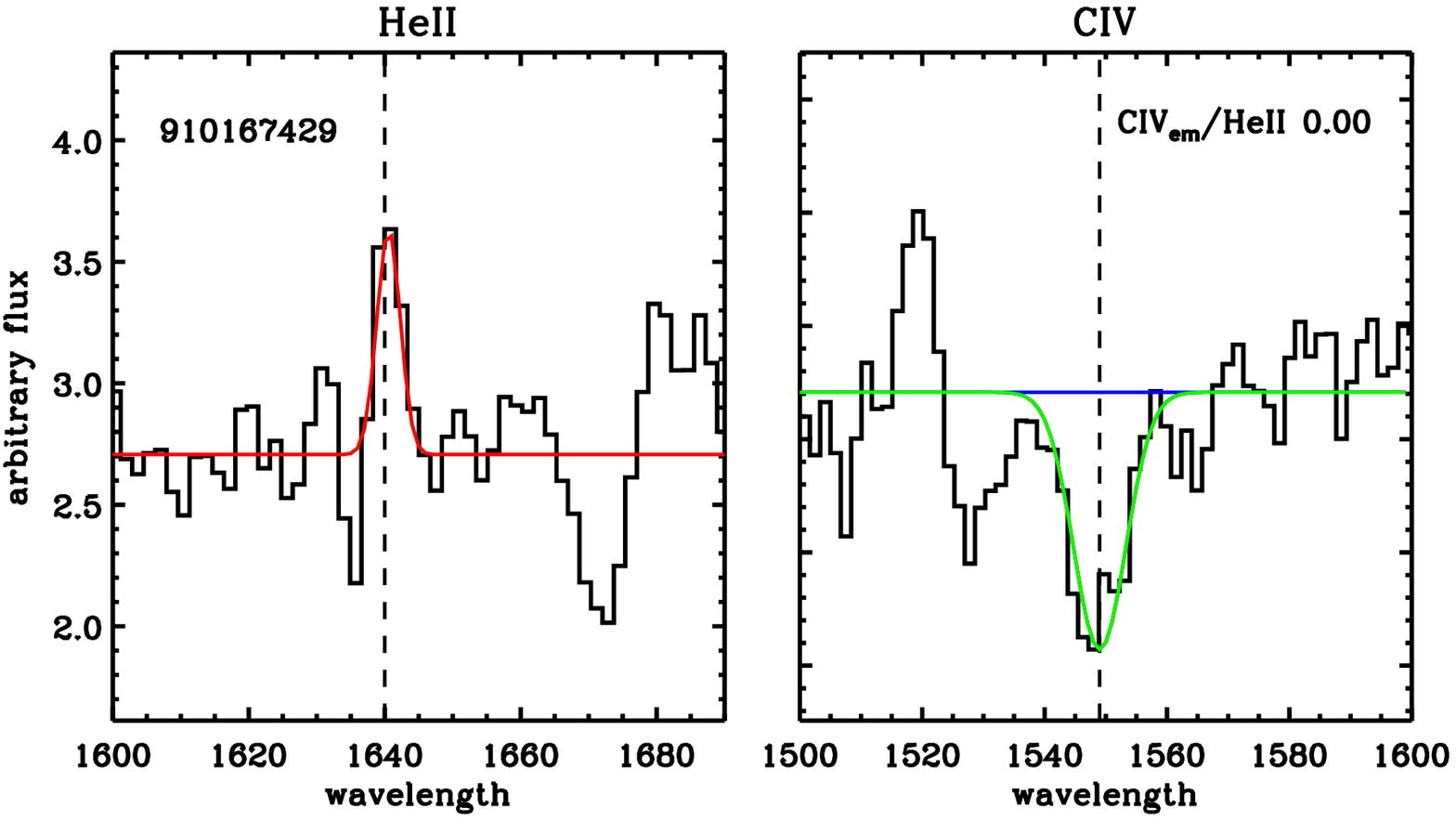}
\includegraphics[width=.8\columnwidth]{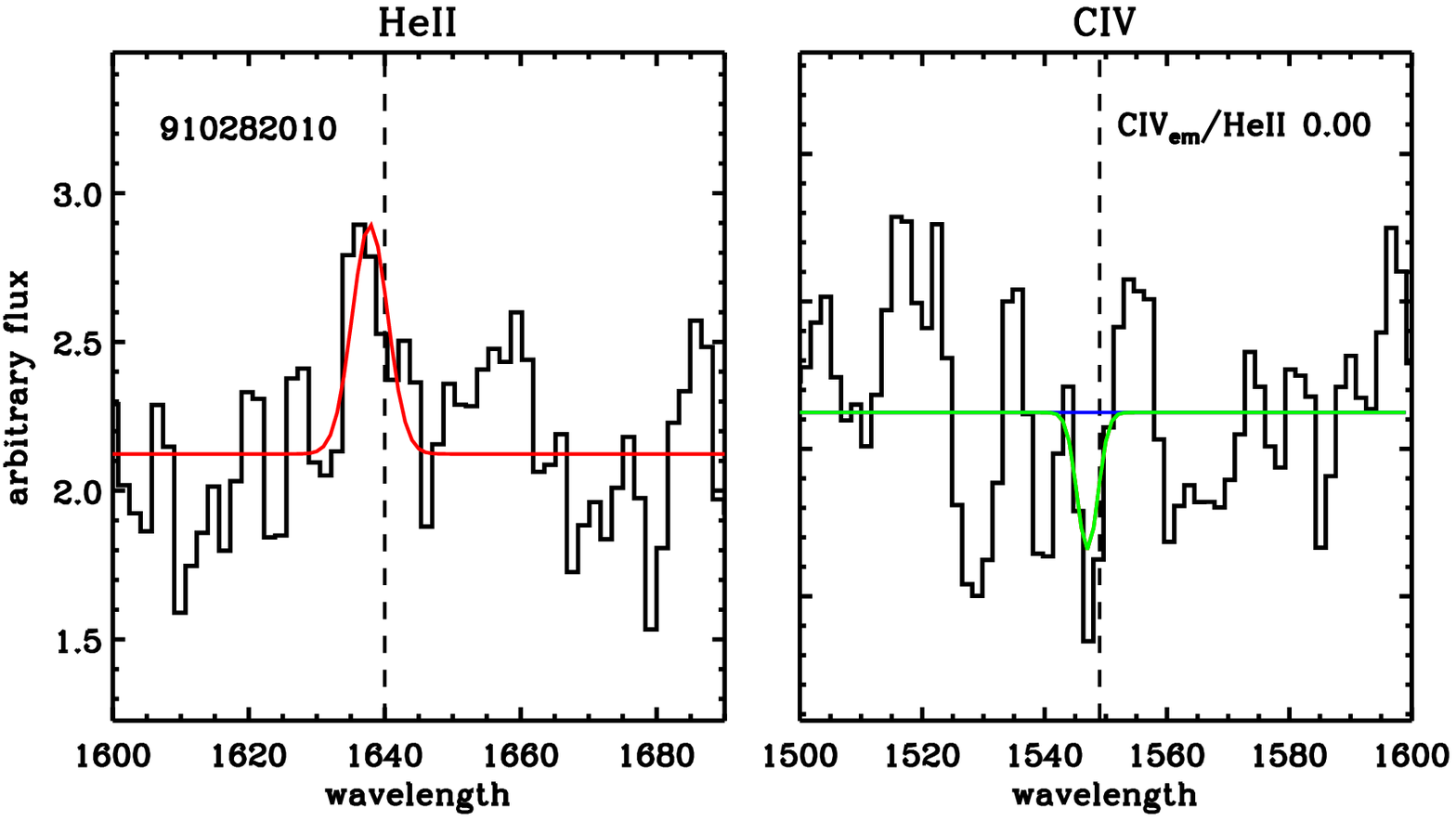}
\includegraphics[width=.8\columnwidth]{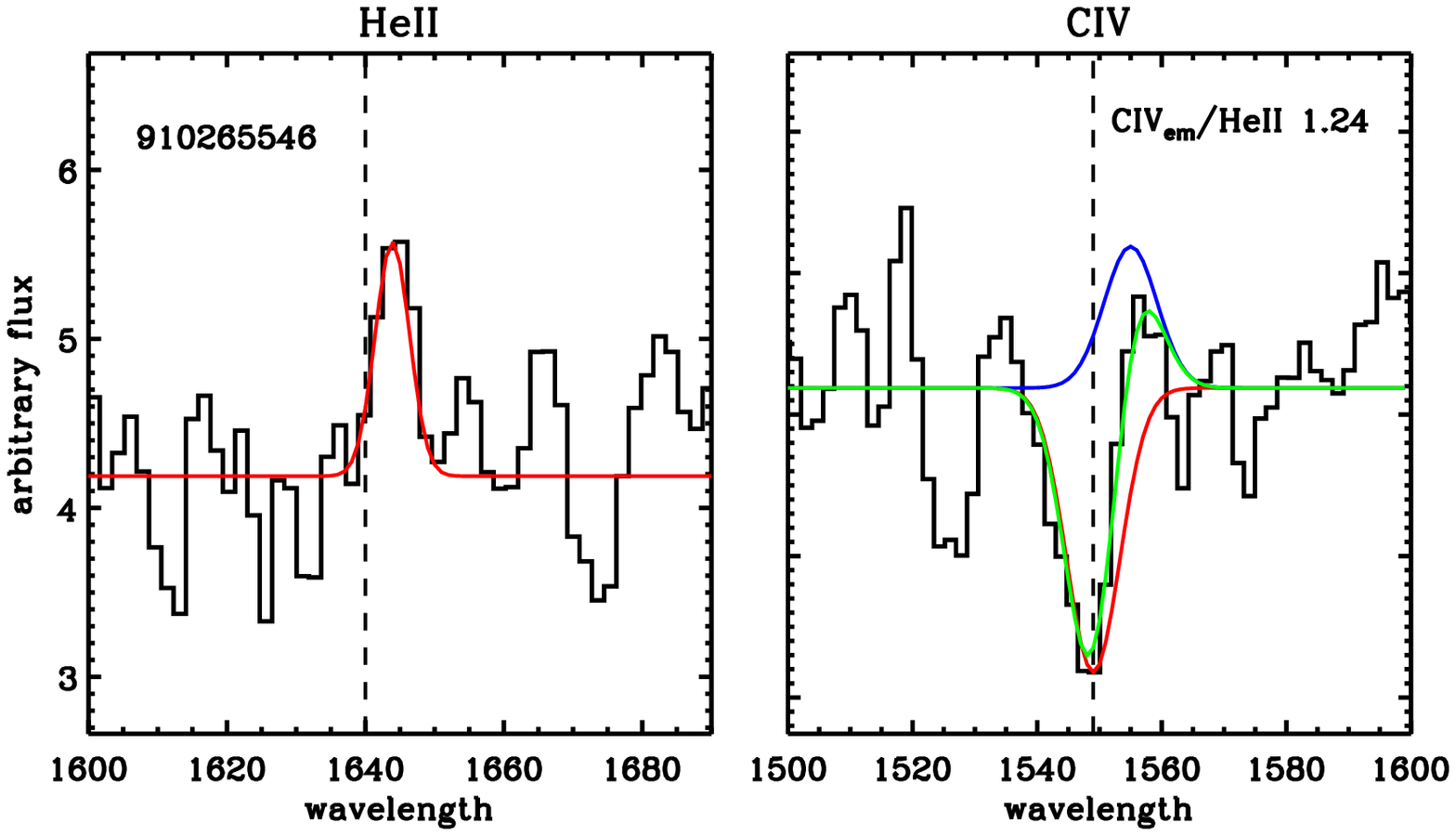}
\includegraphics[width=.8\columnwidth]{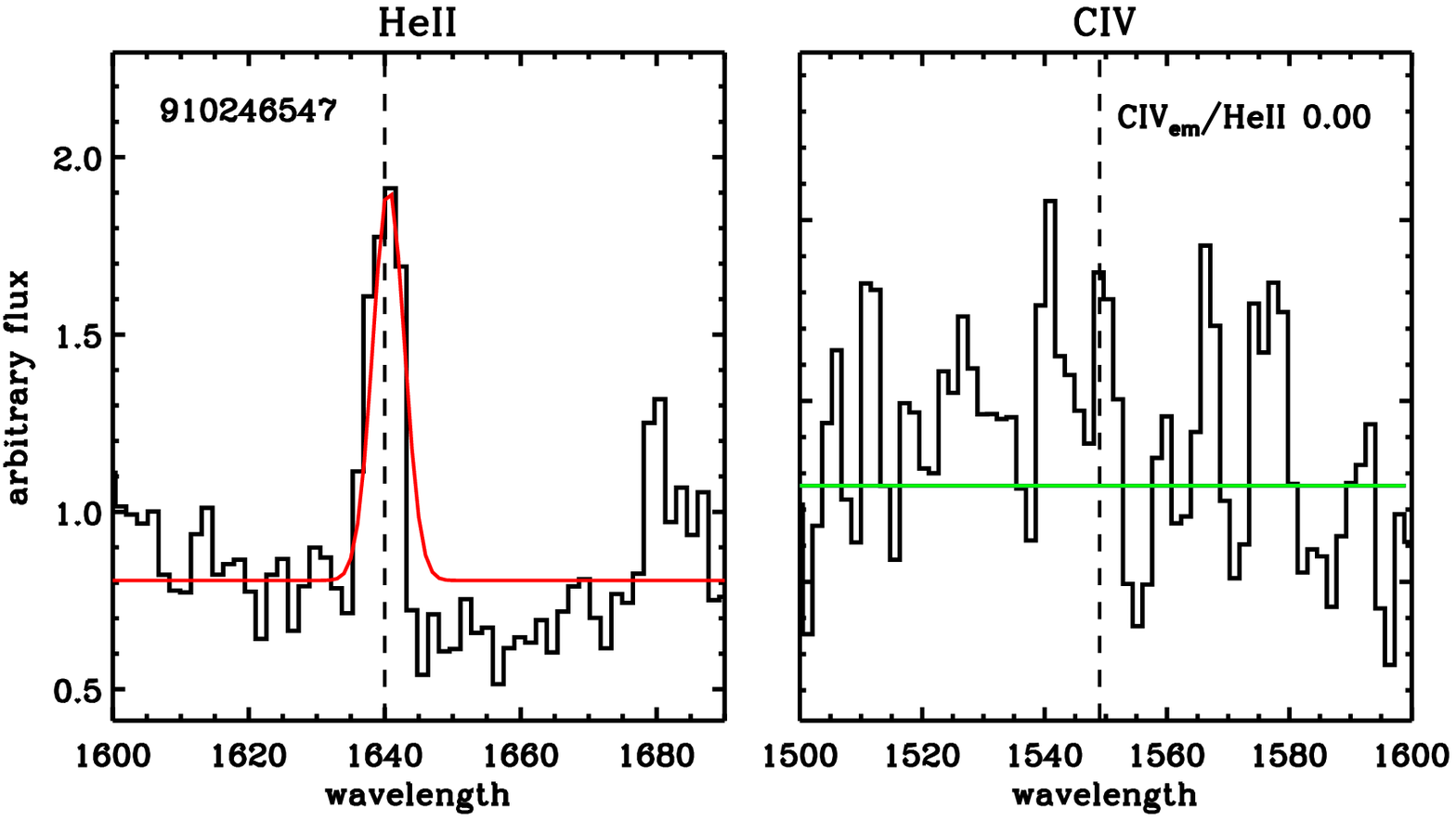}
\includegraphics[width=.8\columnwidth]{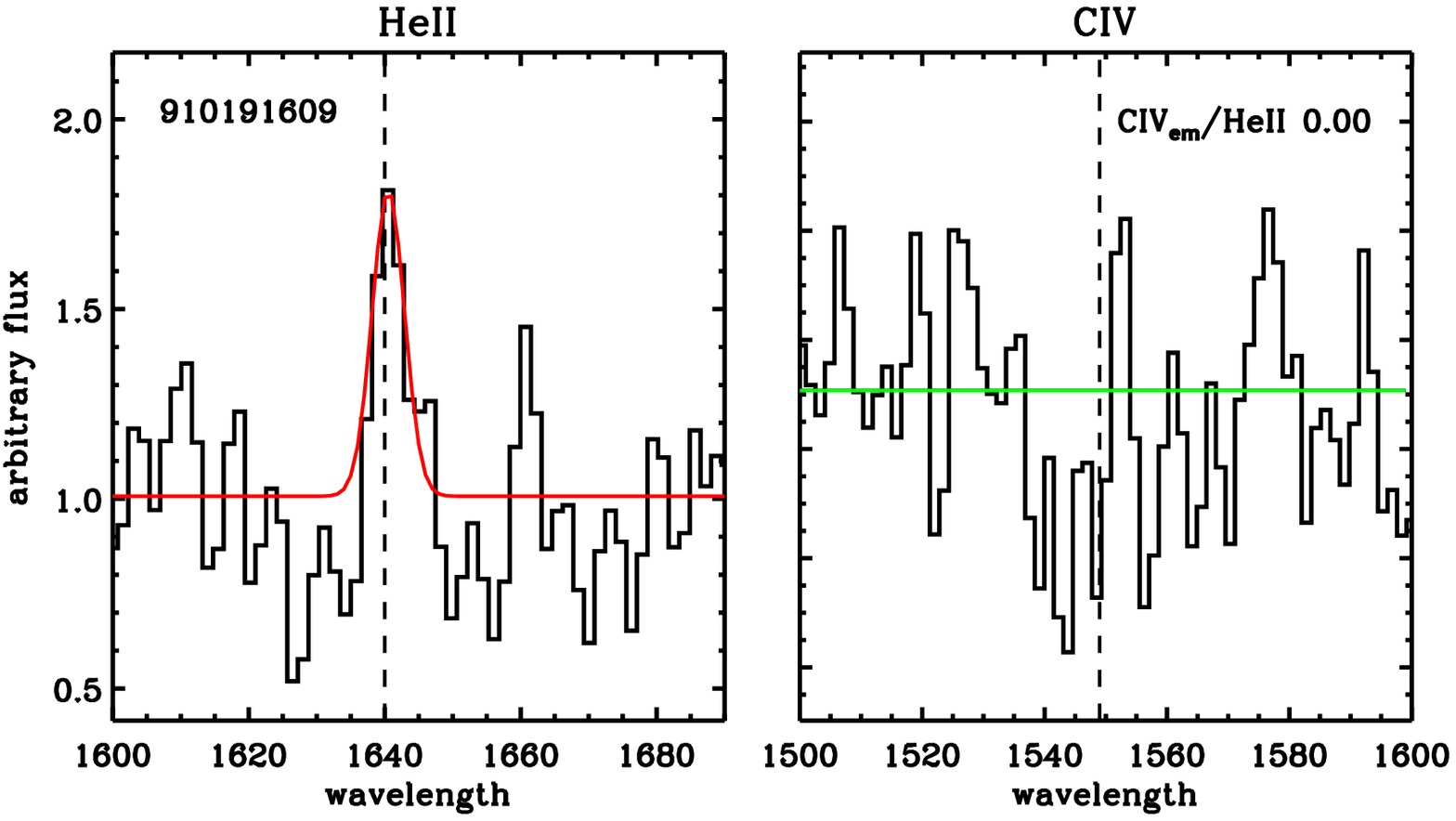}
\includegraphics[width=.8\columnwidth]{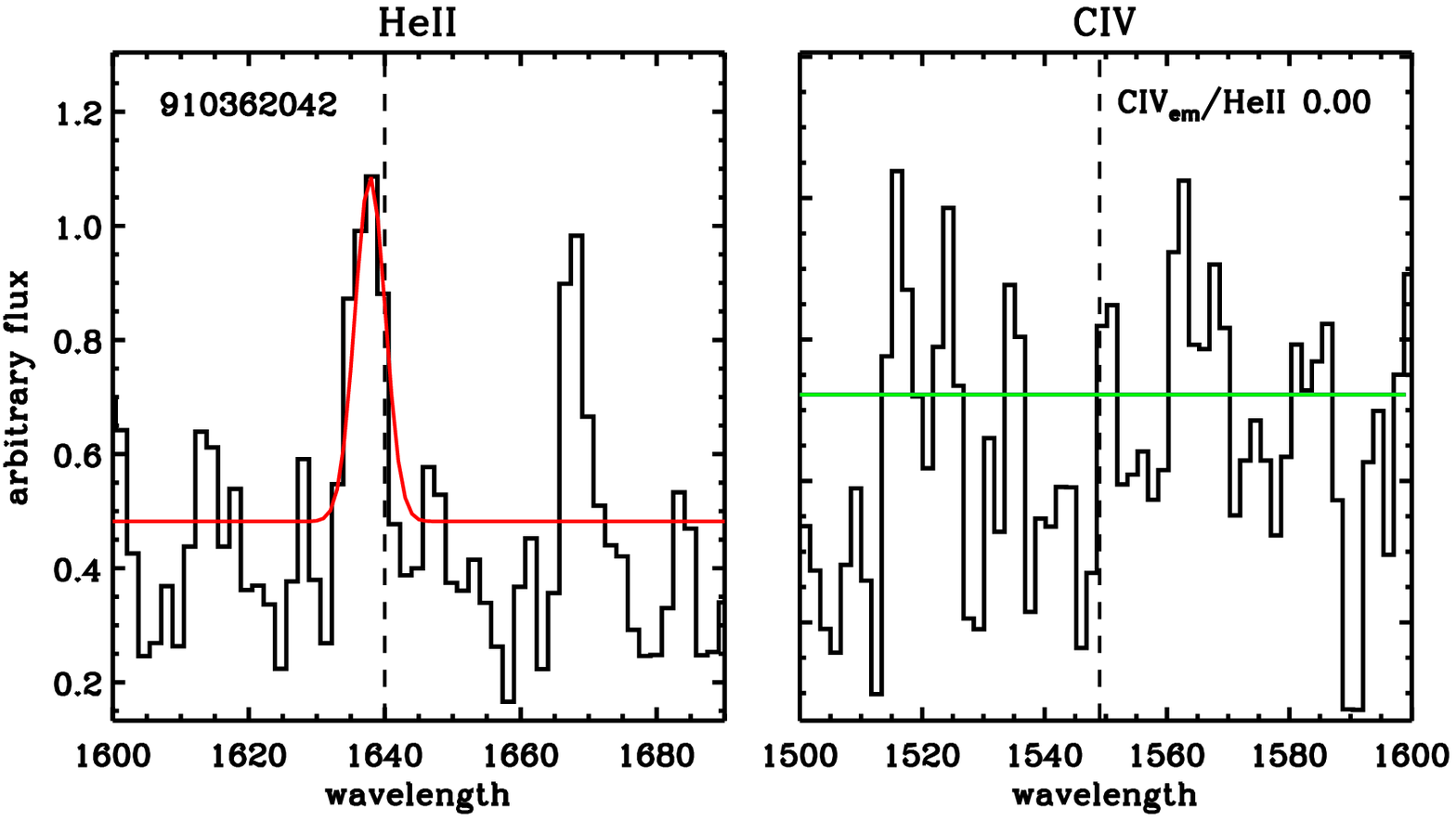}
\includegraphics[width=.8\columnwidth]{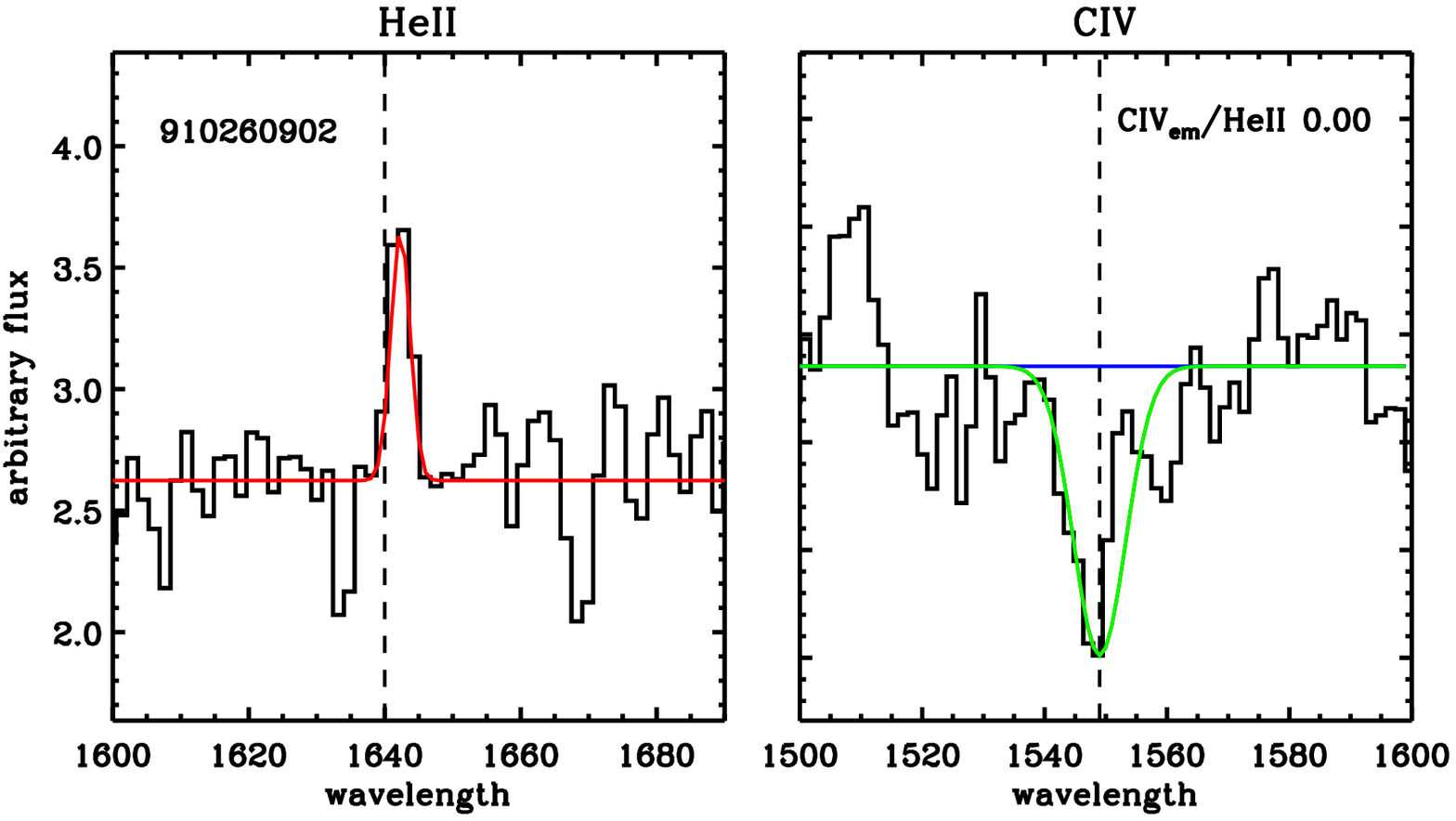}
\includegraphics[width=.8\columnwidth]{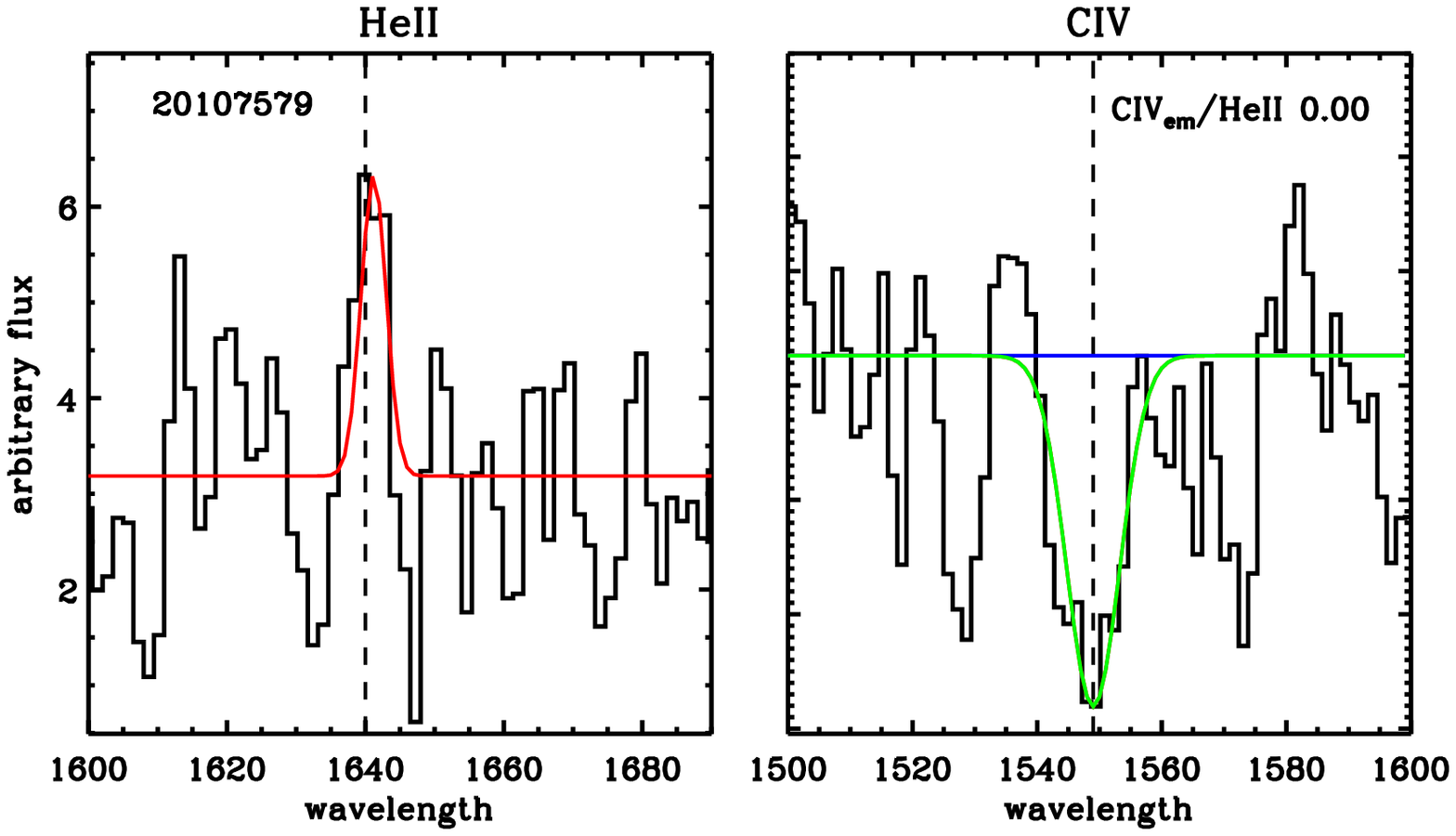}
\includegraphics[width=.8\columnwidth]{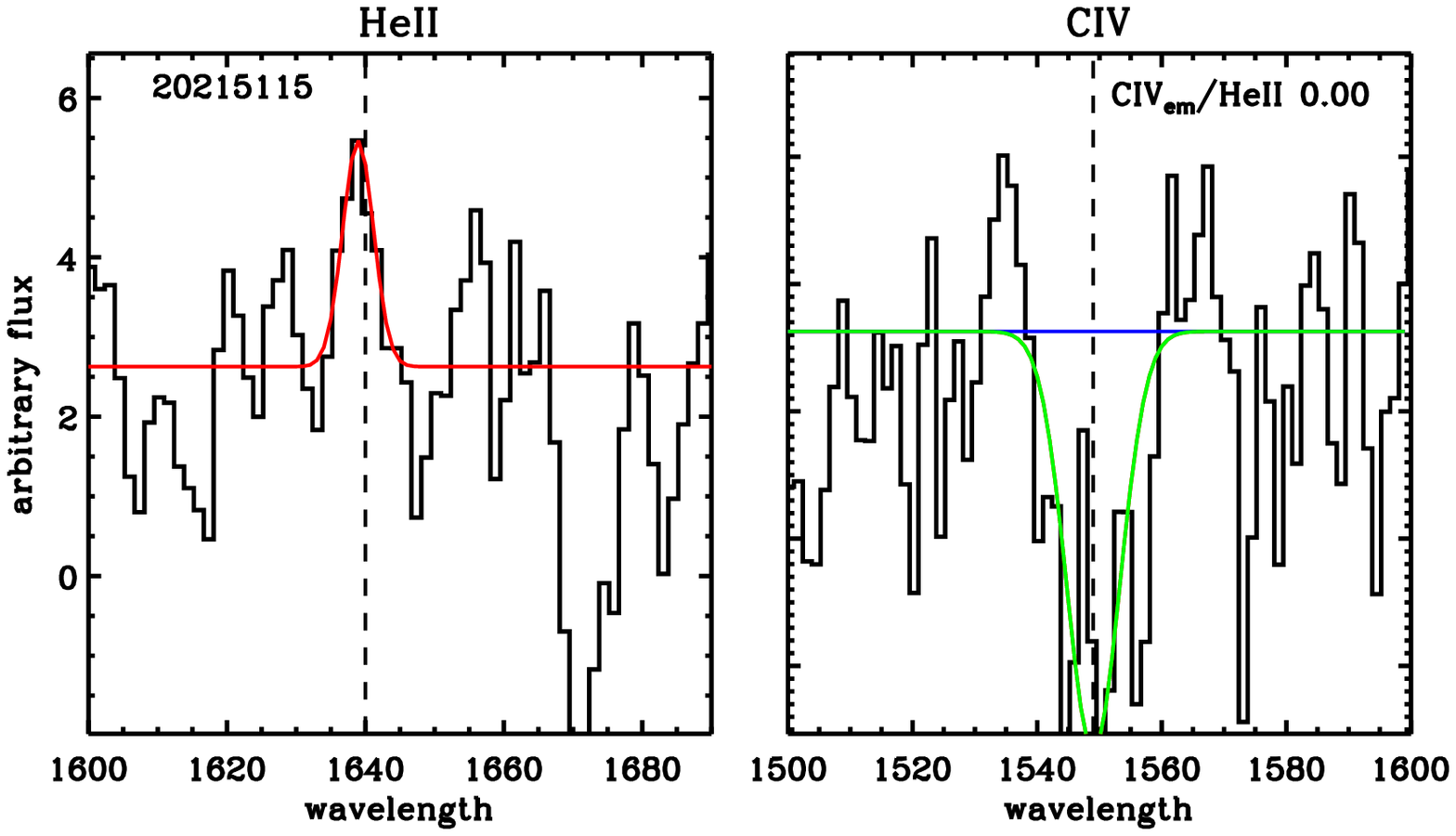}
\includegraphics[width=.8\columnwidth]{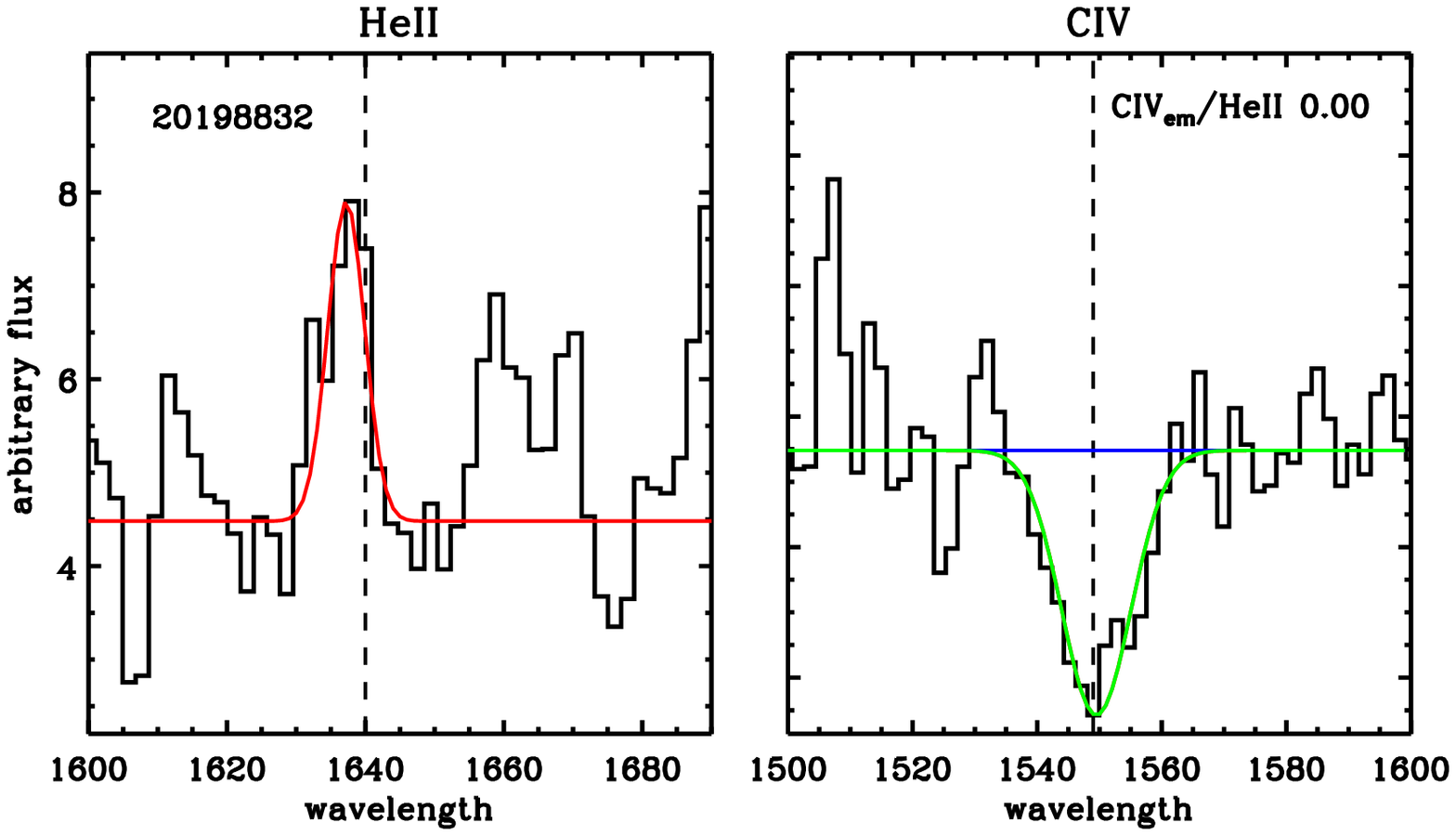}
\caption{Close--up of the region around He~II (left side of each
  sub--figure) and around CIV (right side of each sub--figure) for
  each of the narrow He~II emitters. In the pairs of panels, the red
  line shows a Gaussian fit to either He~II or CIV. When a single
  Gaussian is not enough to reproduce the CIV absorption, we add a
  second Gaussian in emission (shown in blue); the resulting profile,
  mimicking a P-Cygni profile, is shown in green. The number in the
  right--side panels is the ratio between the flux of the C~IV
  component in emission and the flux of He~II.}
\label{zoom_narrow}
\end{figure*}

\begin{figure*}[!ht]
  \centering
\includegraphics[width=.7\columnwidth]{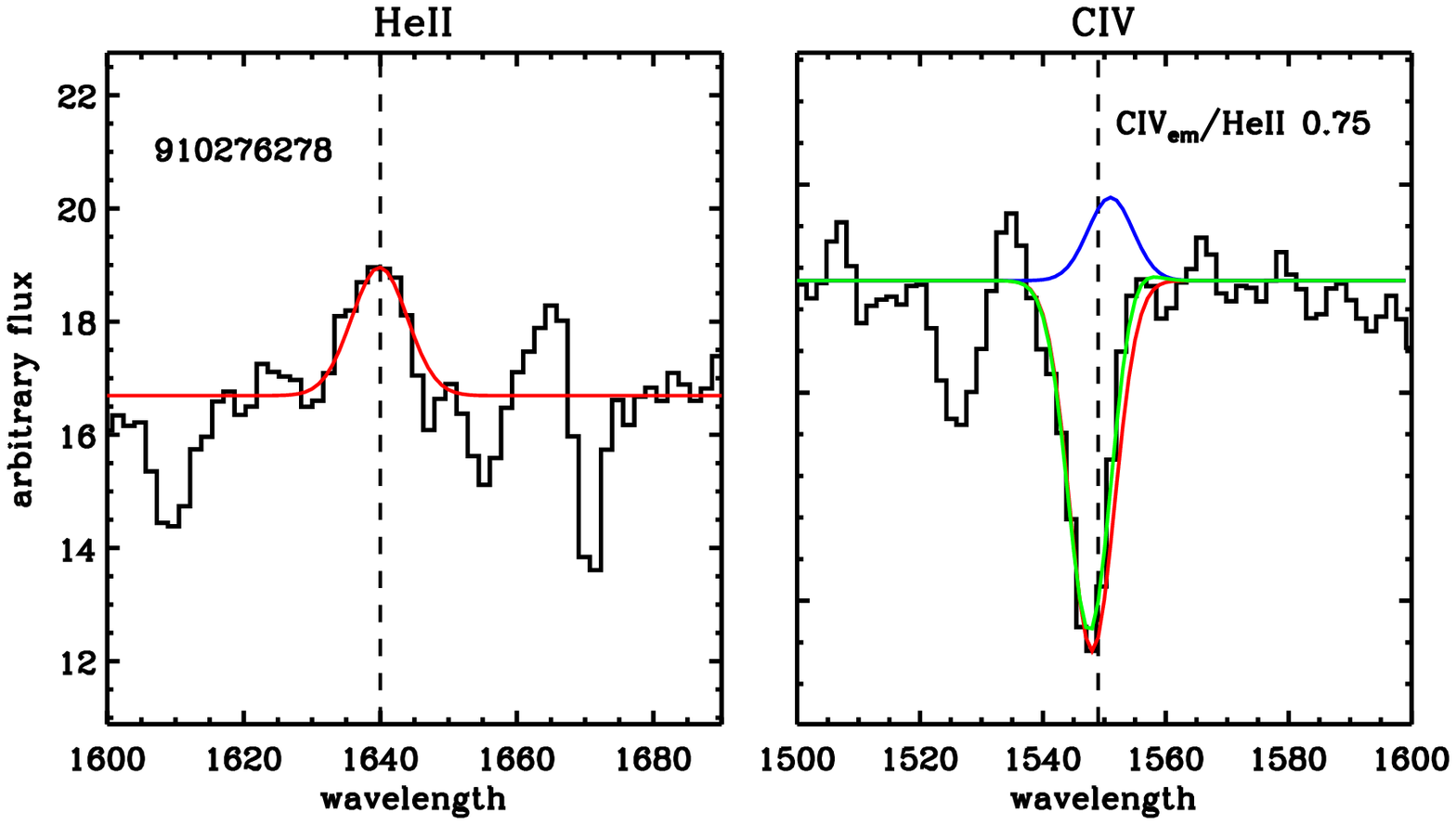}
\includegraphics[width=.7\columnwidth]{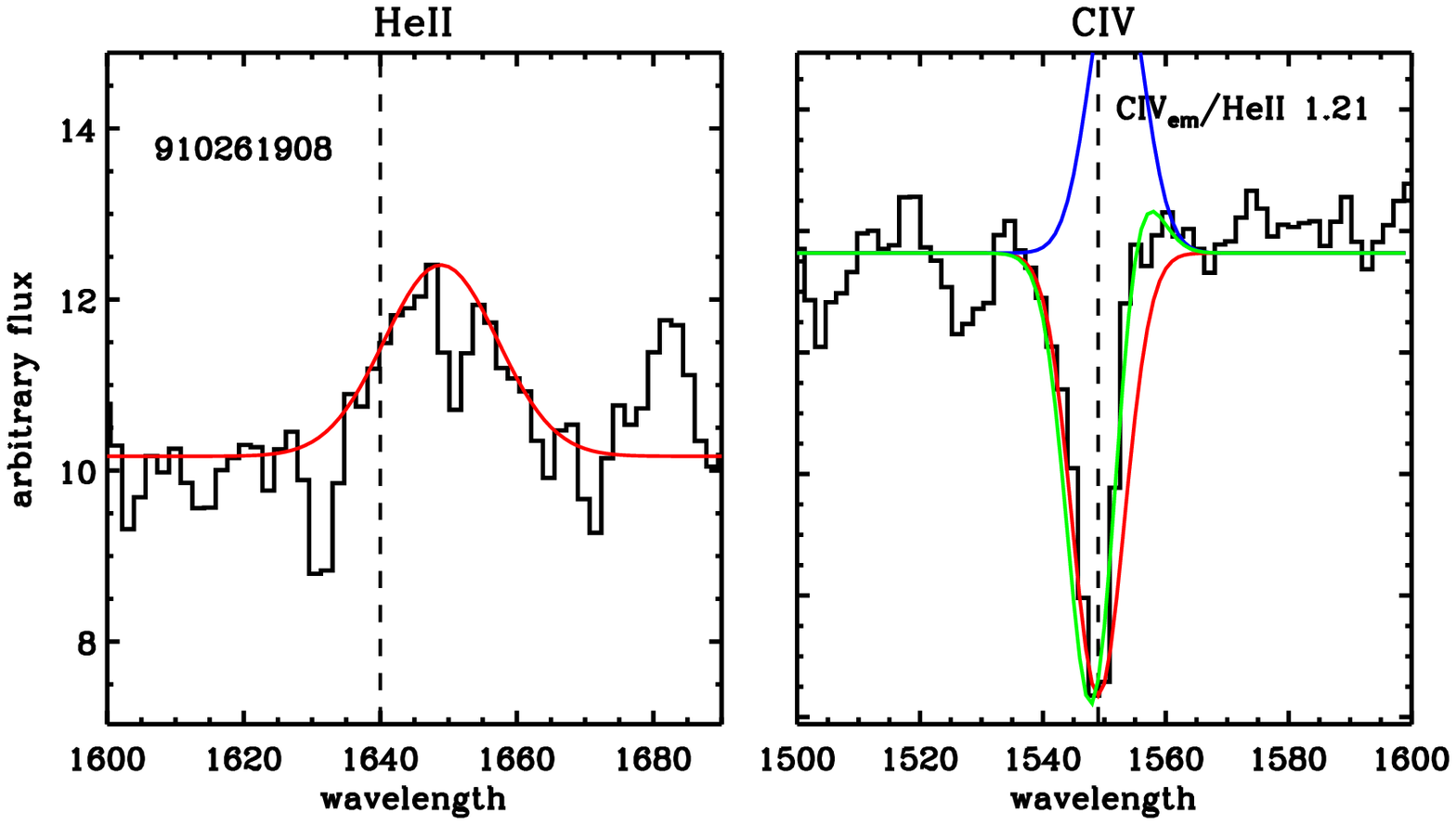}
\includegraphics[width=.7\columnwidth]{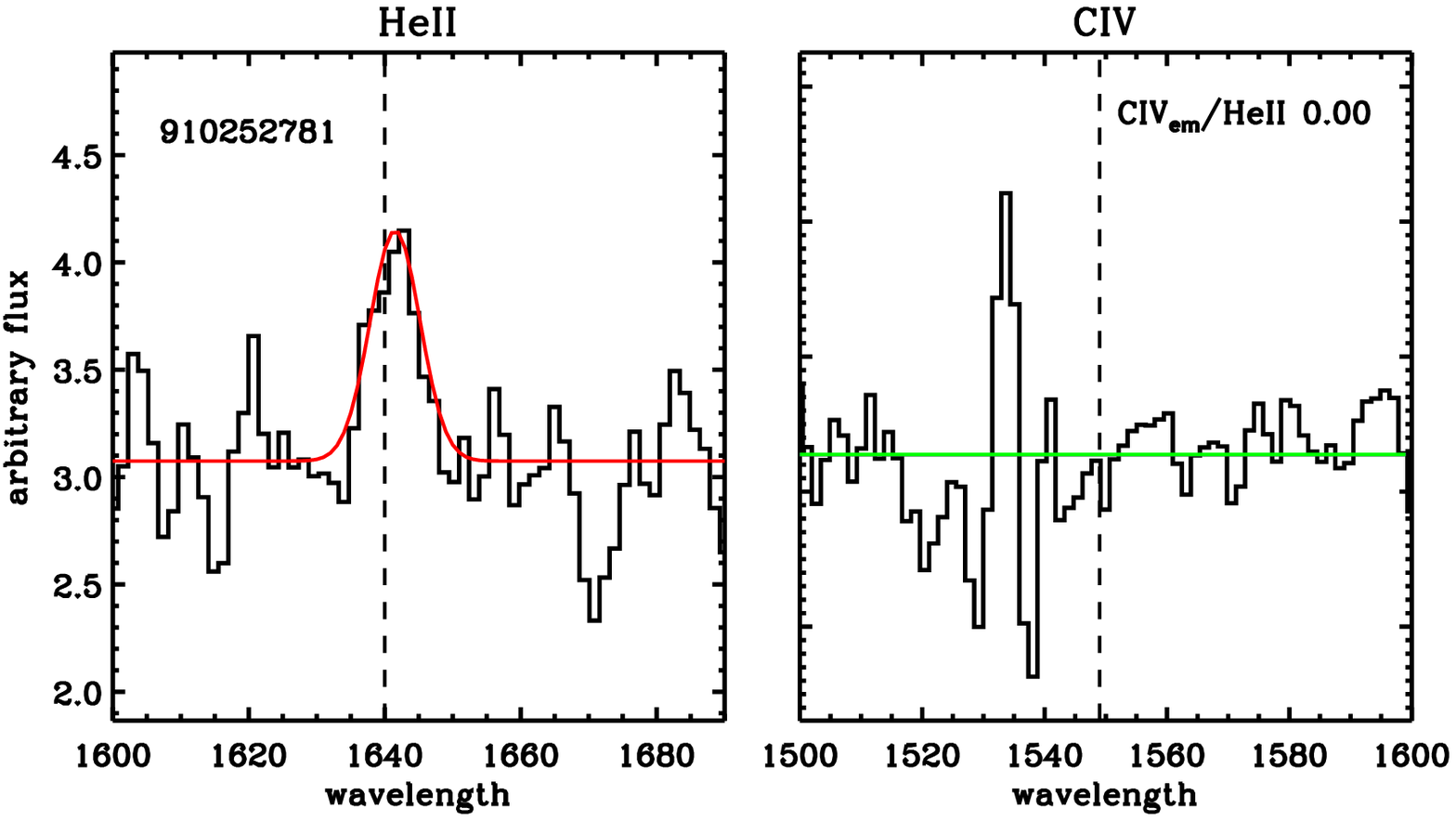}
\includegraphics[width=.7\columnwidth]{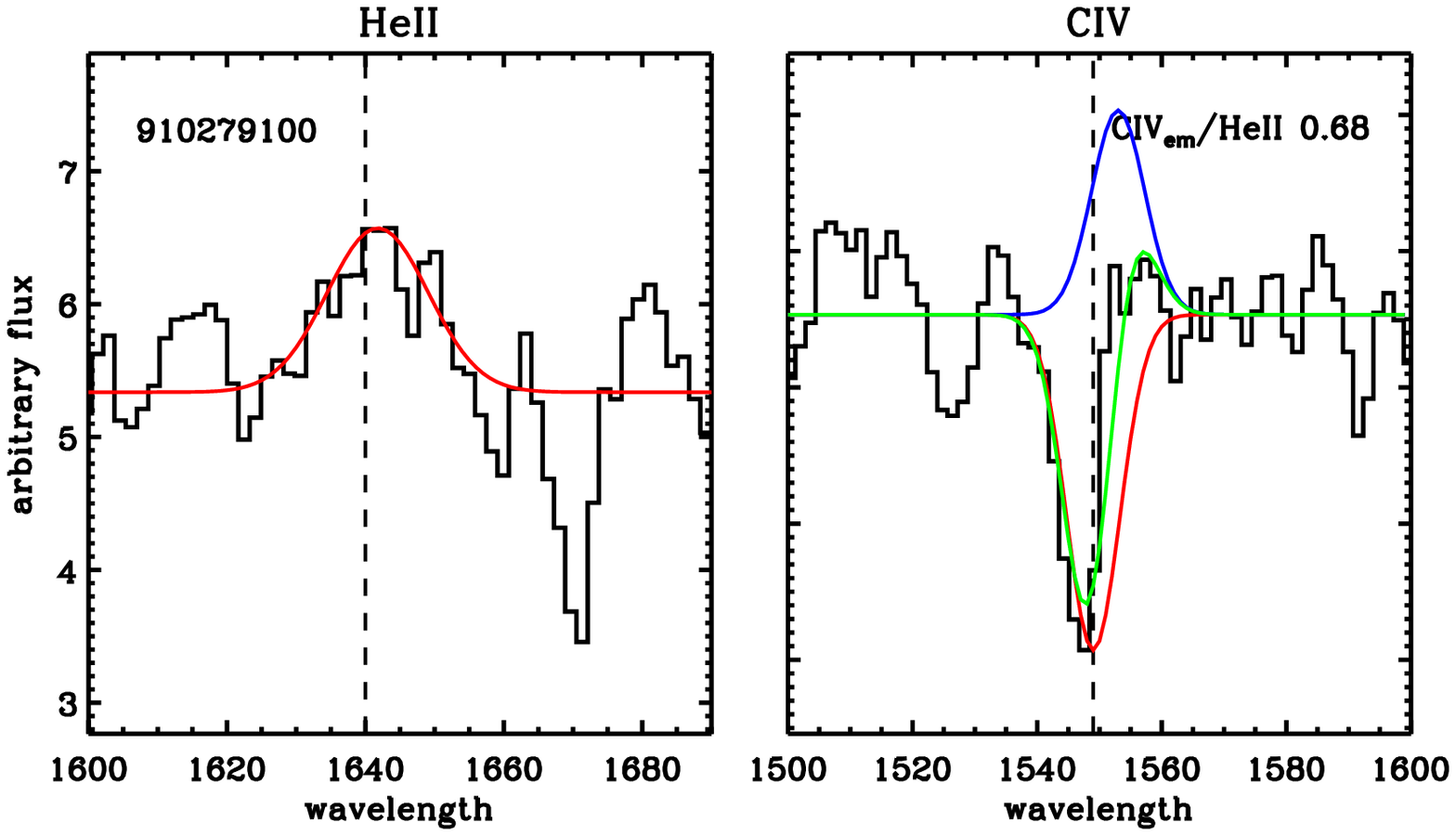}
\includegraphics[width=.7\columnwidth]{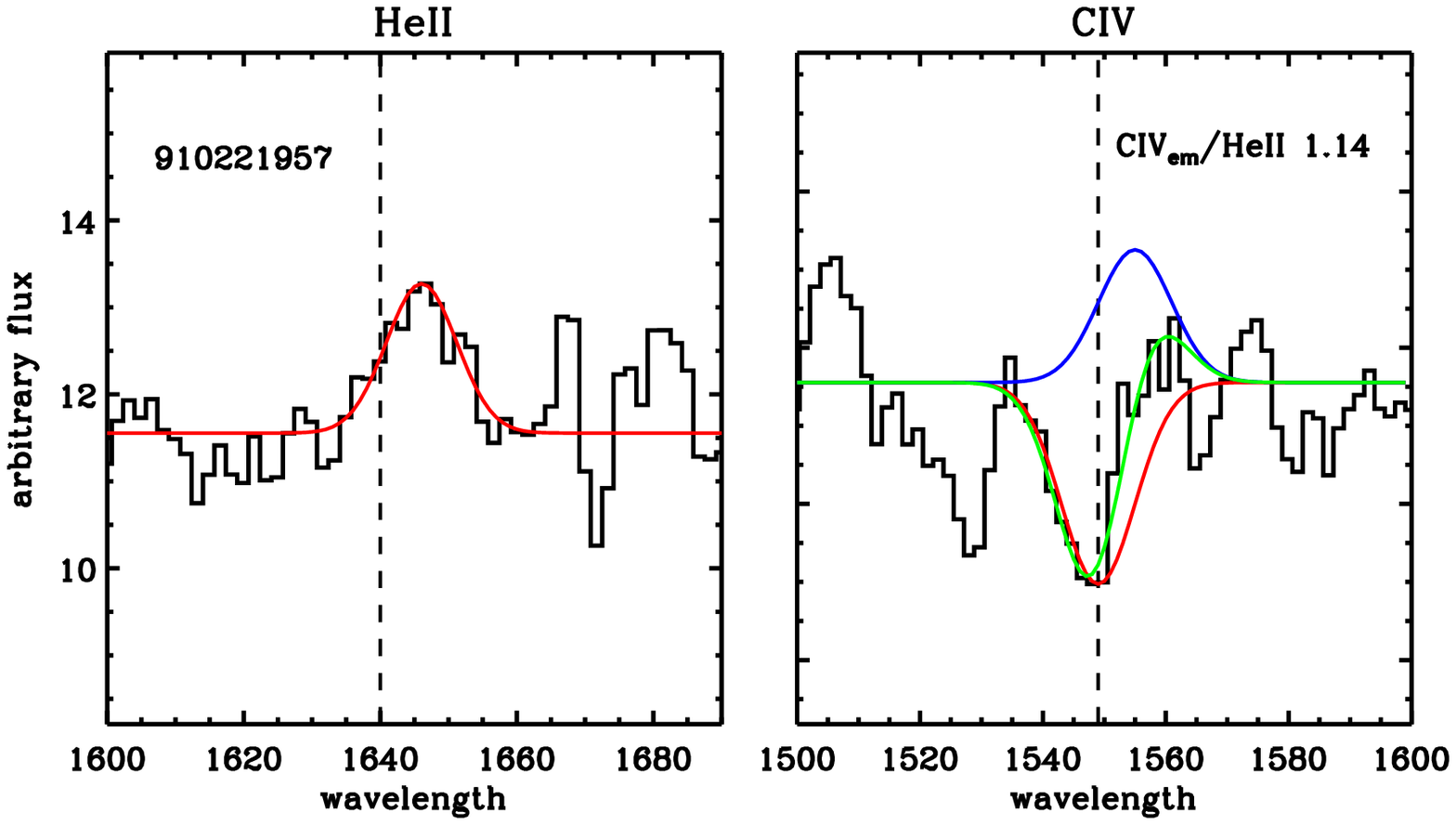}
\includegraphics[width=.7\columnwidth]{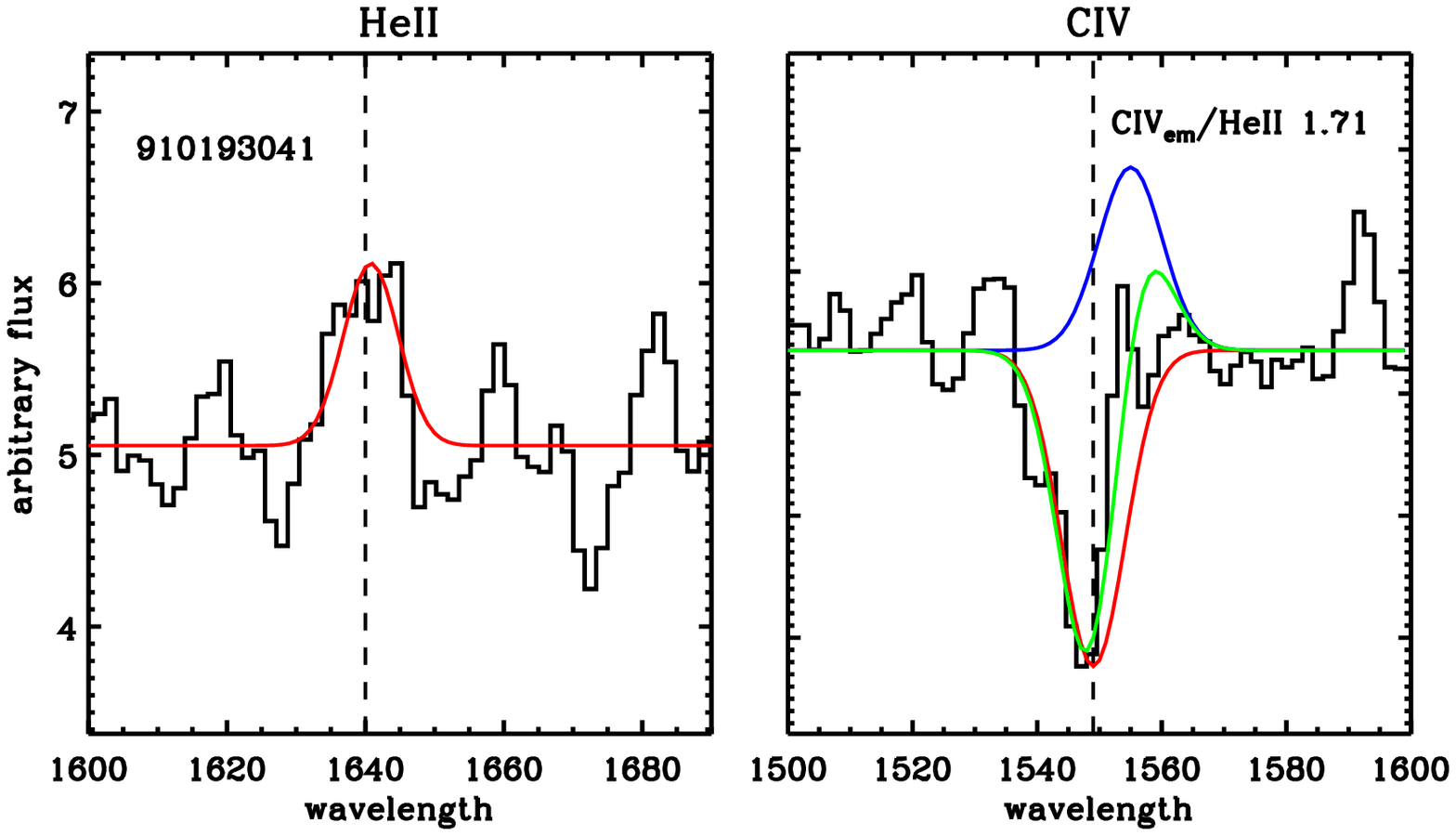}
\includegraphics[width=.7\columnwidth]{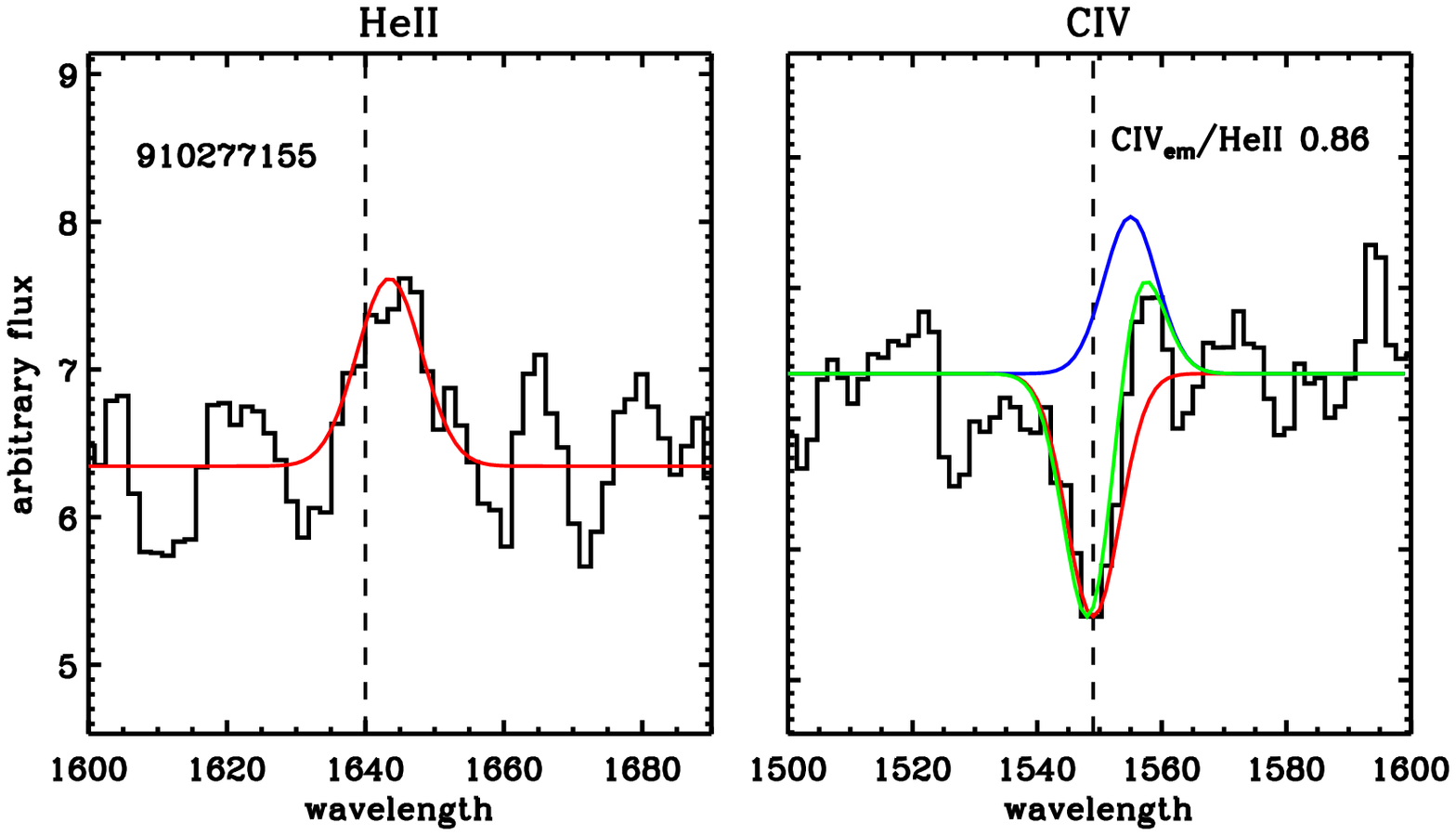}
\includegraphics[width=.7\columnwidth]{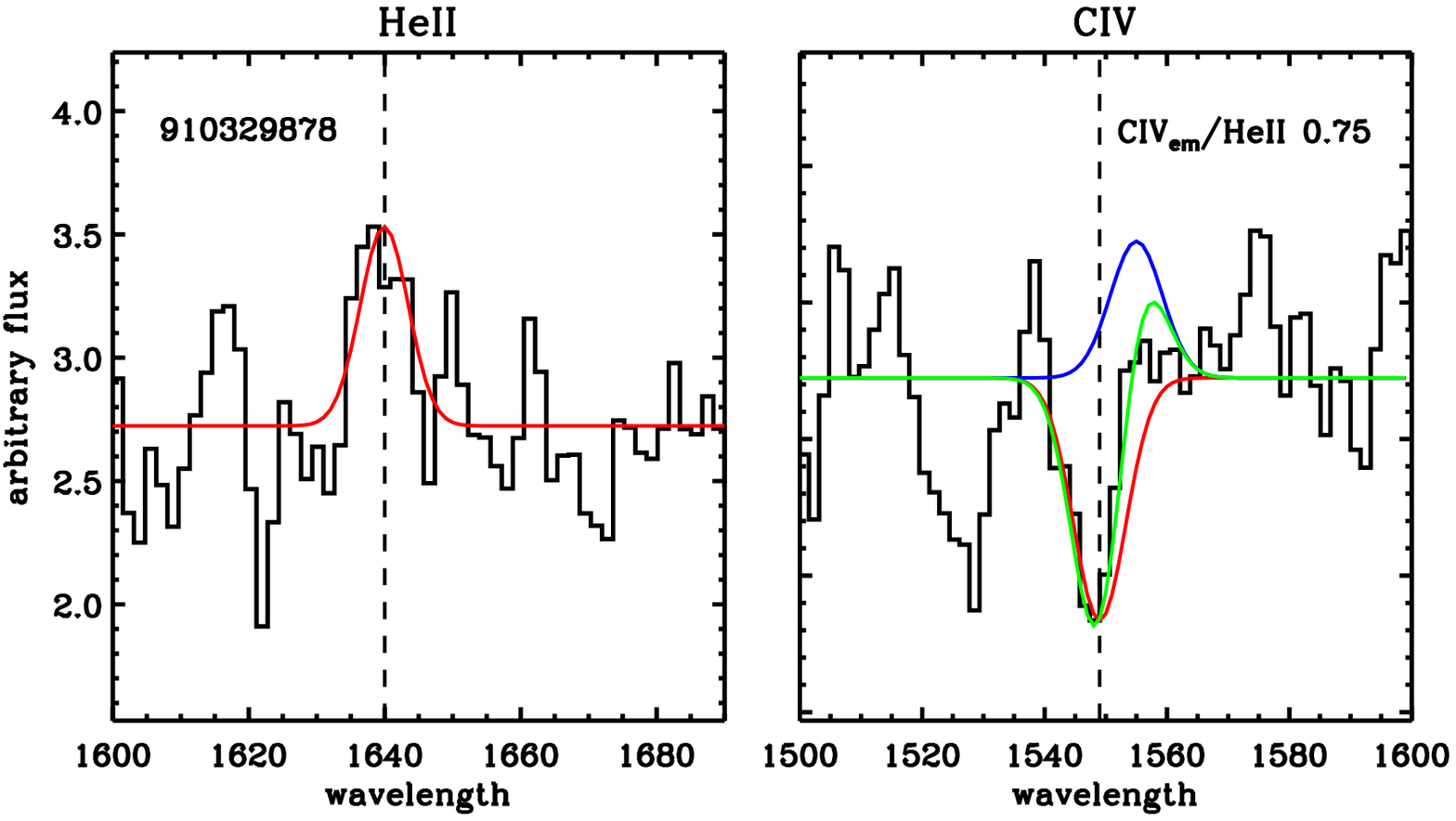}
\includegraphics[width=.7\columnwidth]{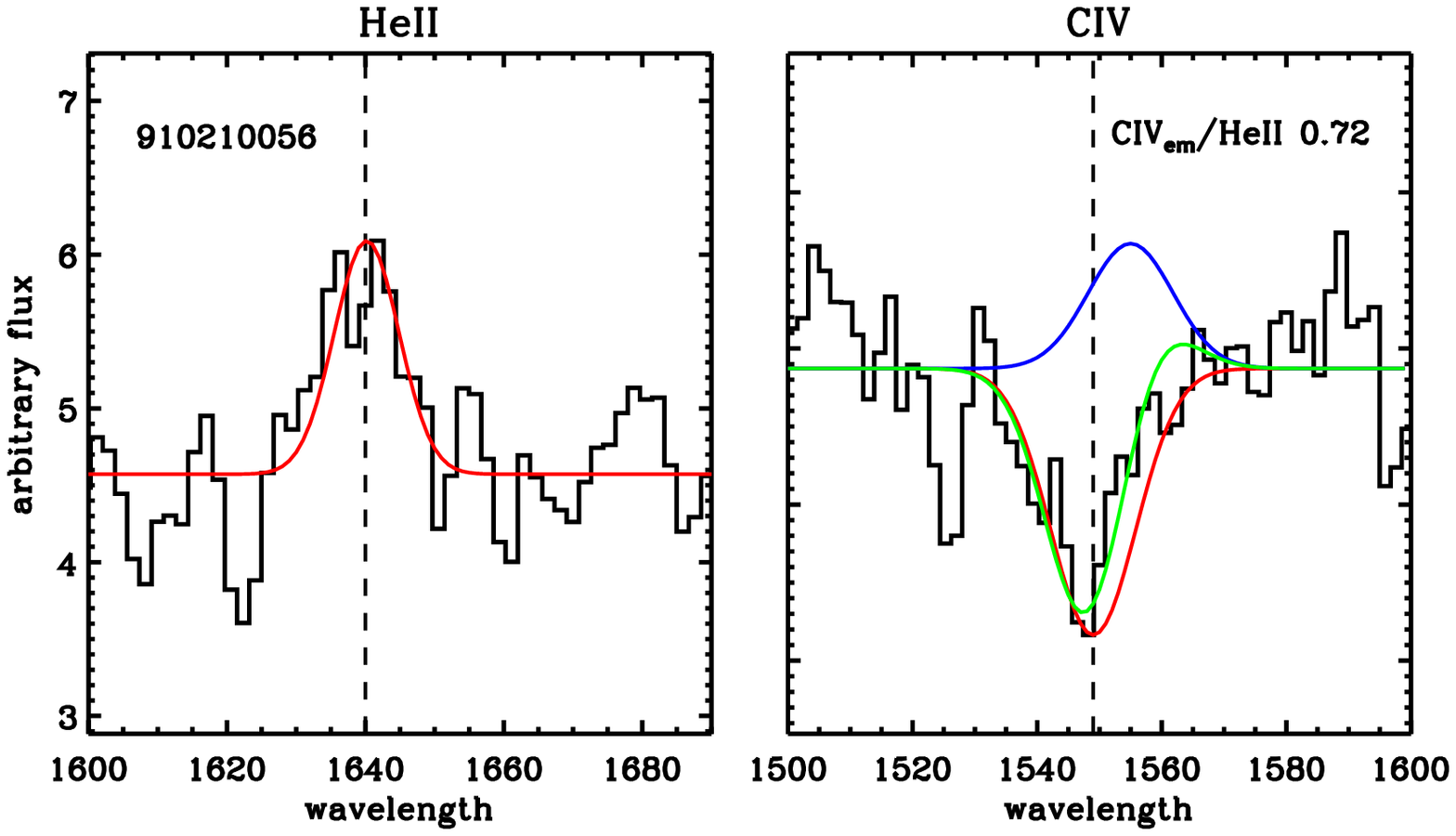}
\includegraphics[width=.7\columnwidth]{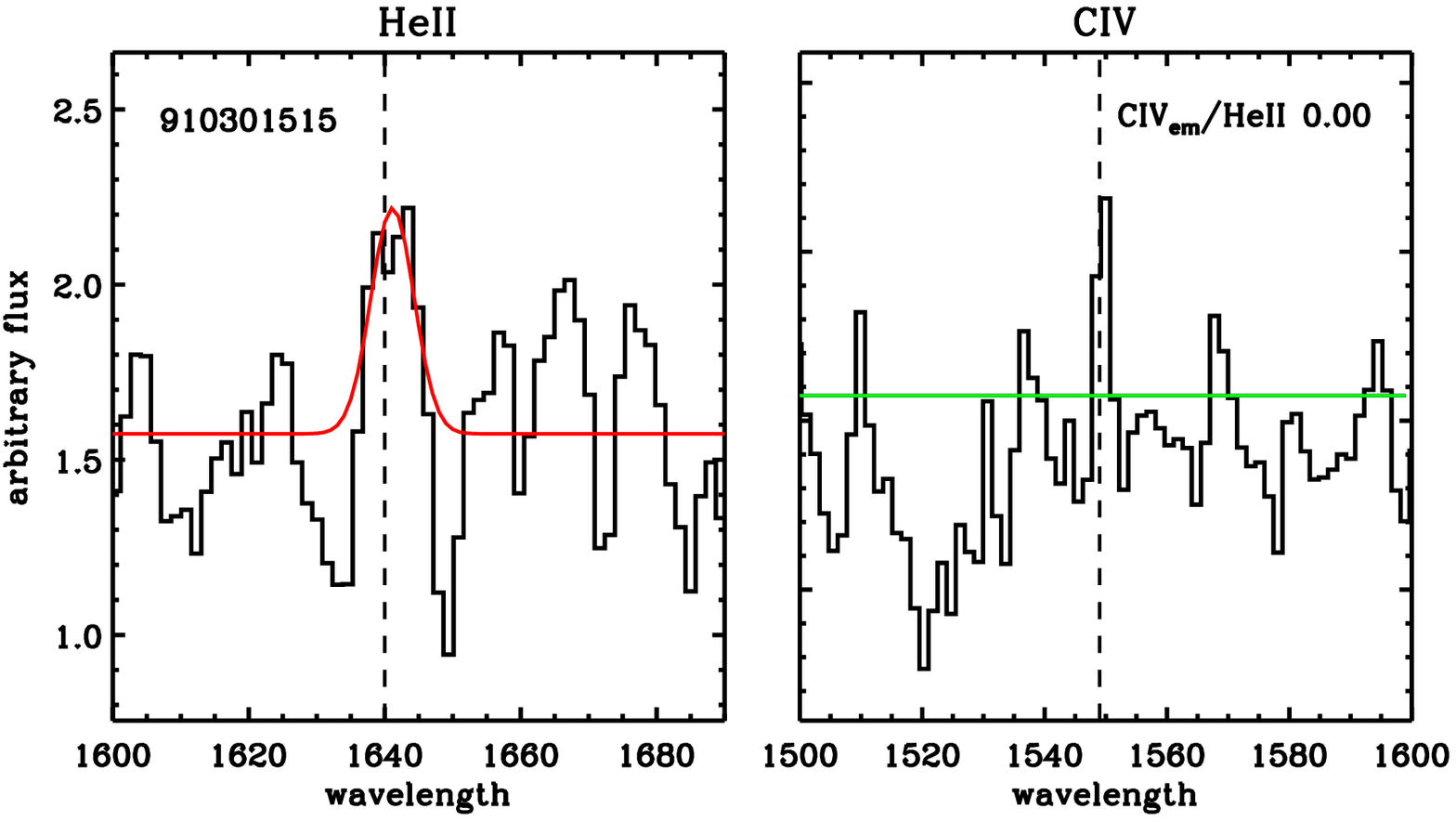}
\includegraphics[width=.7\columnwidth]{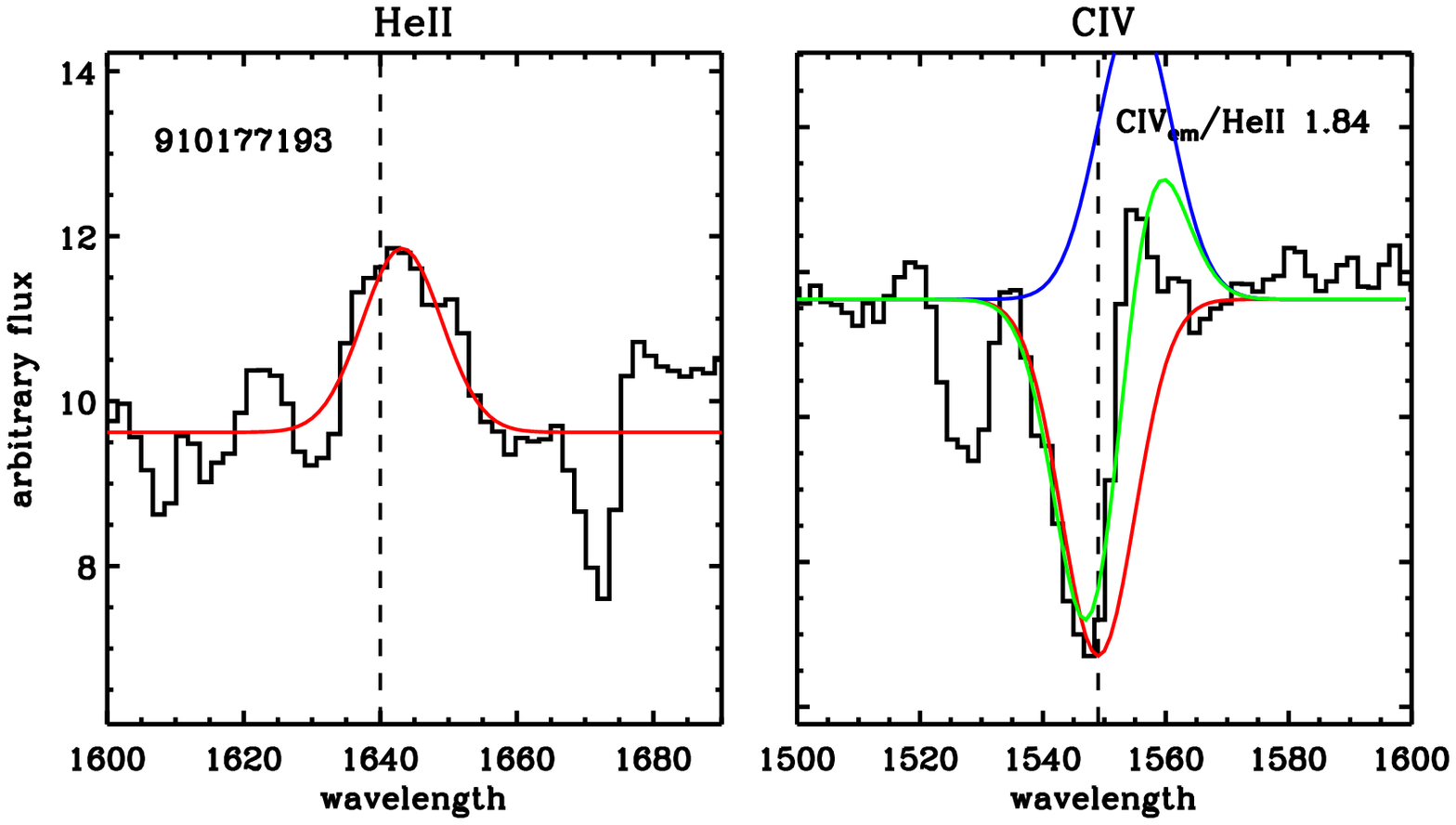}
\includegraphics[width=.7\columnwidth]{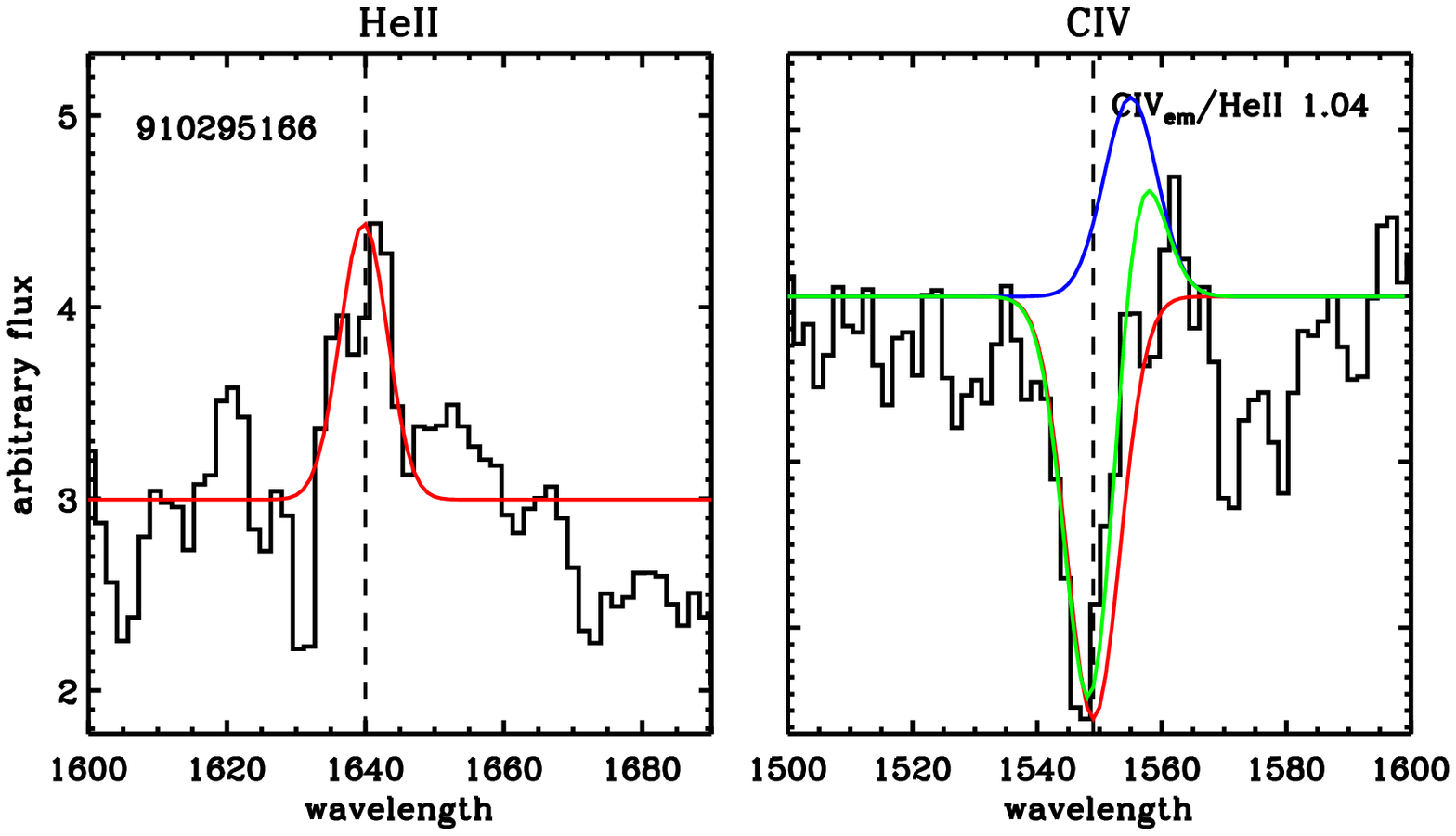}
\includegraphics[width=.7\columnwidth]{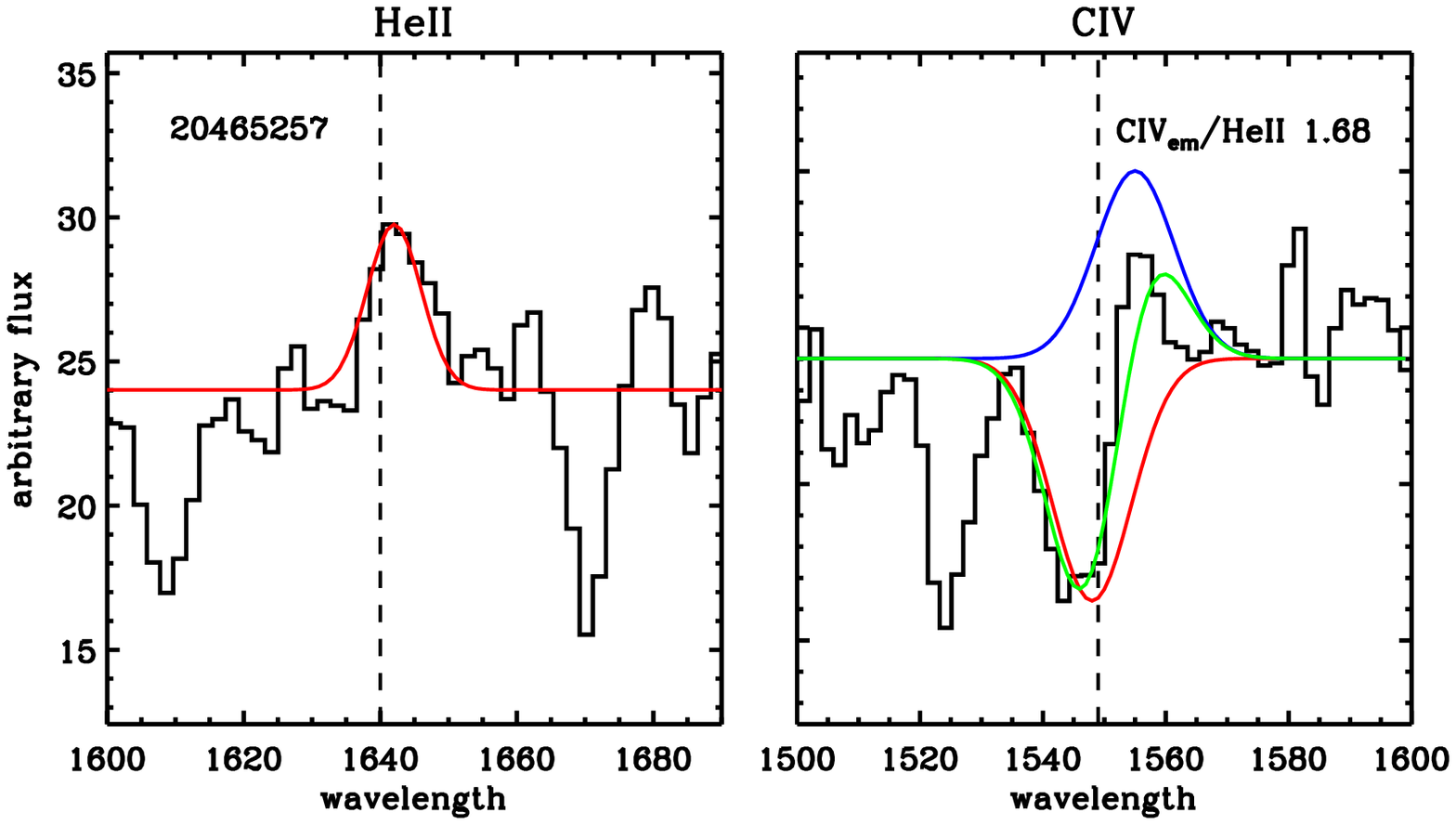}
\caption{Close--up of the region around He~II (left side of each
  sub--figure) and around CIV (right side of each sub--figure) for
  each of the broad He~II emitters. In the pairs of panels, the red line
  shows a Gaussian fit to either He~II or CIV. When a single Gaussian
  is not enough to reproduce the CIV absorption, we add a second
  Gaussian in emission (shown in blue); the resulting profile,
  mimicking a P-Cygni profile, is shown in green. The number in the
  right--side panels is the ratio between the flux of the C~IV
  component in emission and the flux of He~II.}
\label{zoom_broad}
\end{figure*}

\begin{figure*}[!ht]
  \centering
\includegraphics[width=.8\columnwidth]{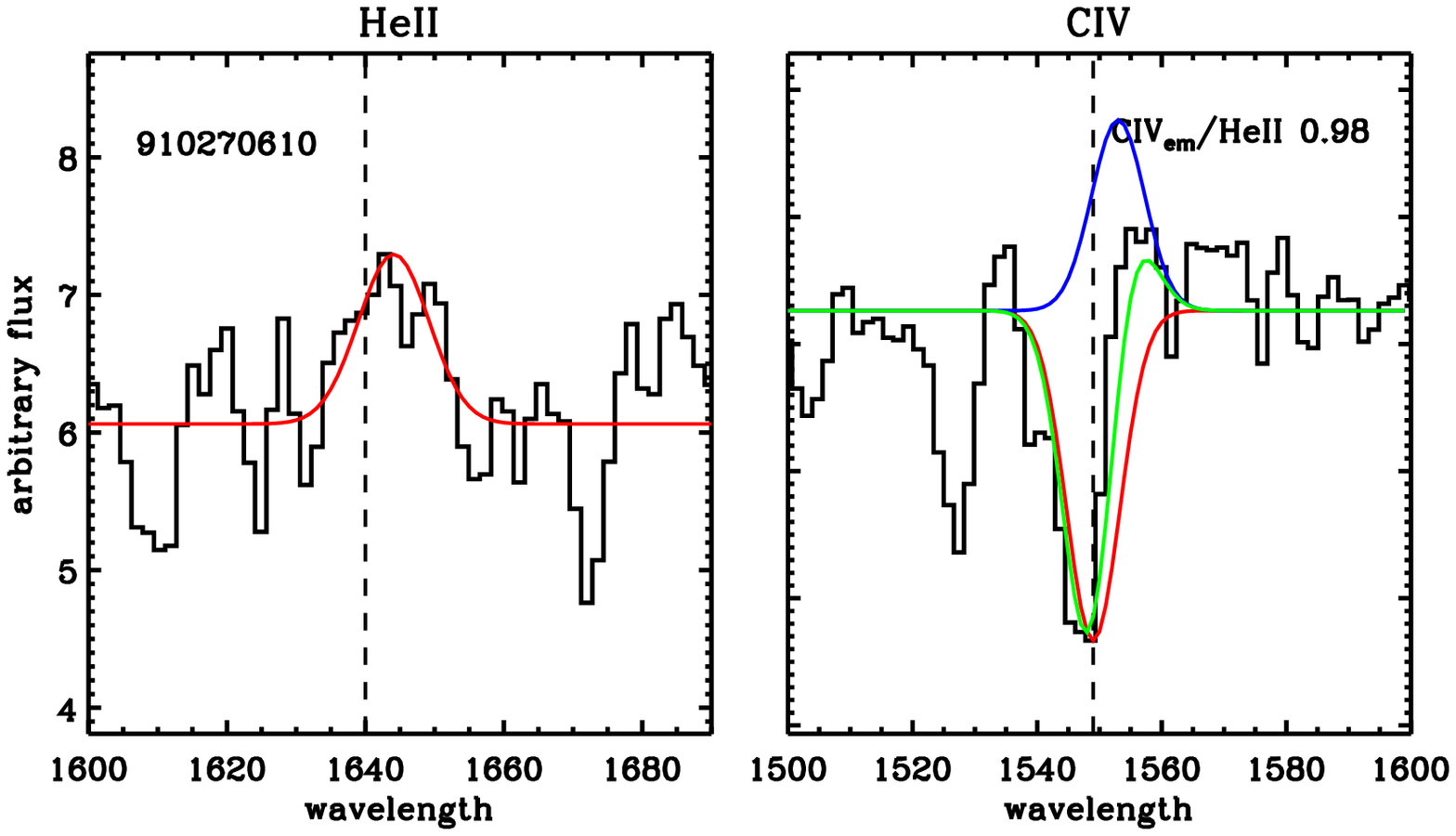}
\includegraphics[width=.8\columnwidth]{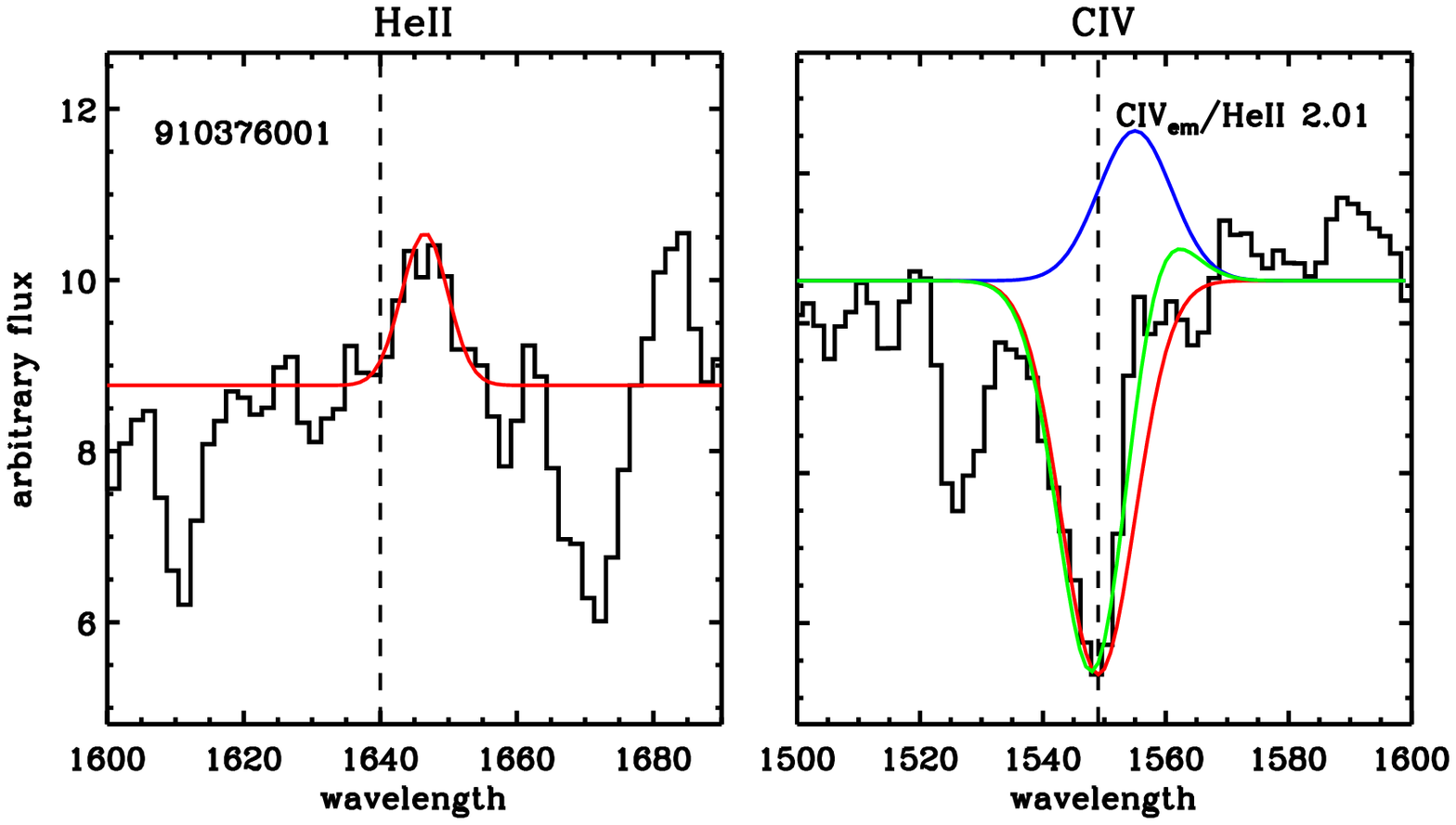}
\includegraphics[width=.8\columnwidth]{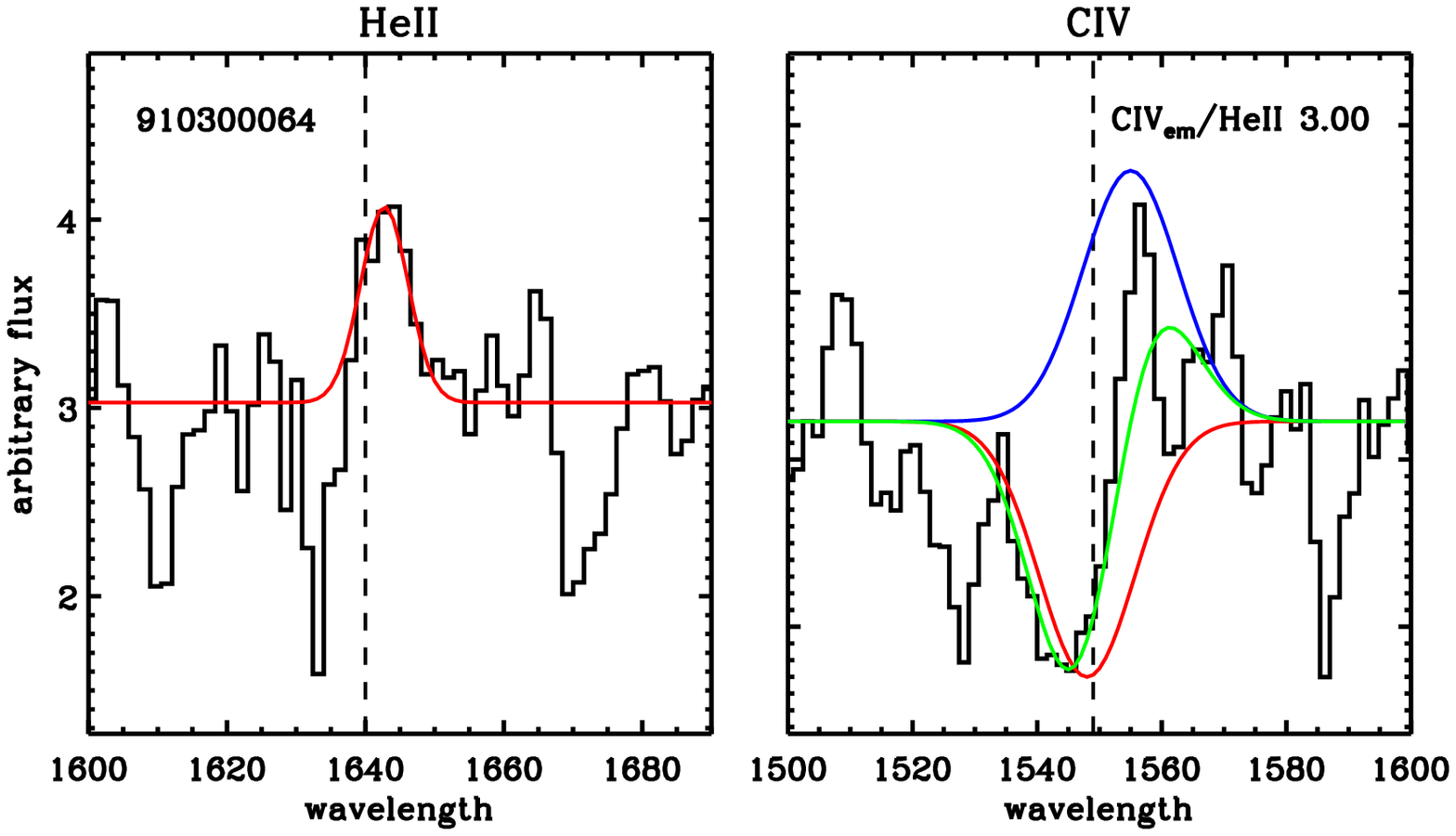}
\includegraphics[width=.8\columnwidth]{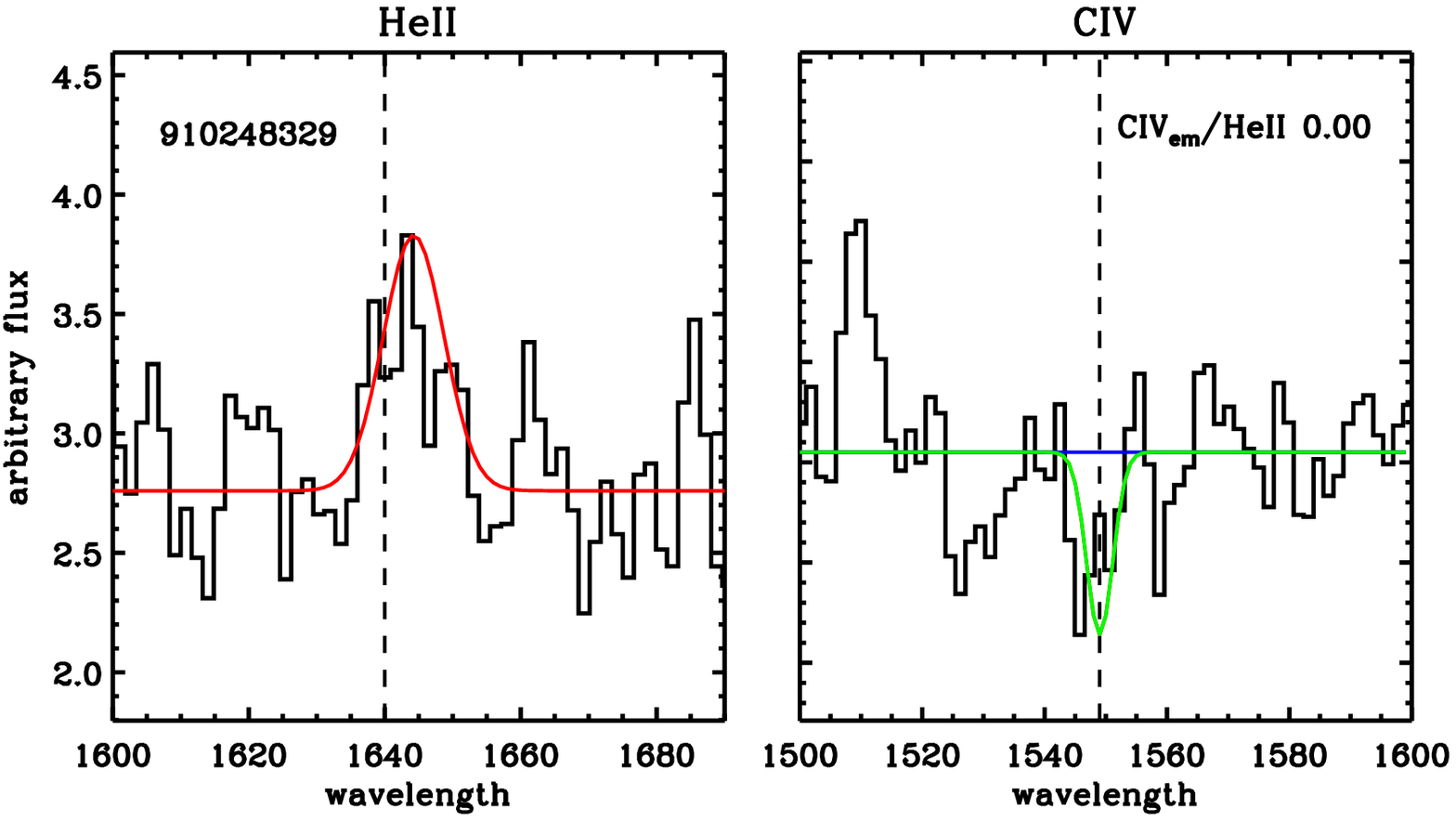}
\includegraphics[width=.8\columnwidth]{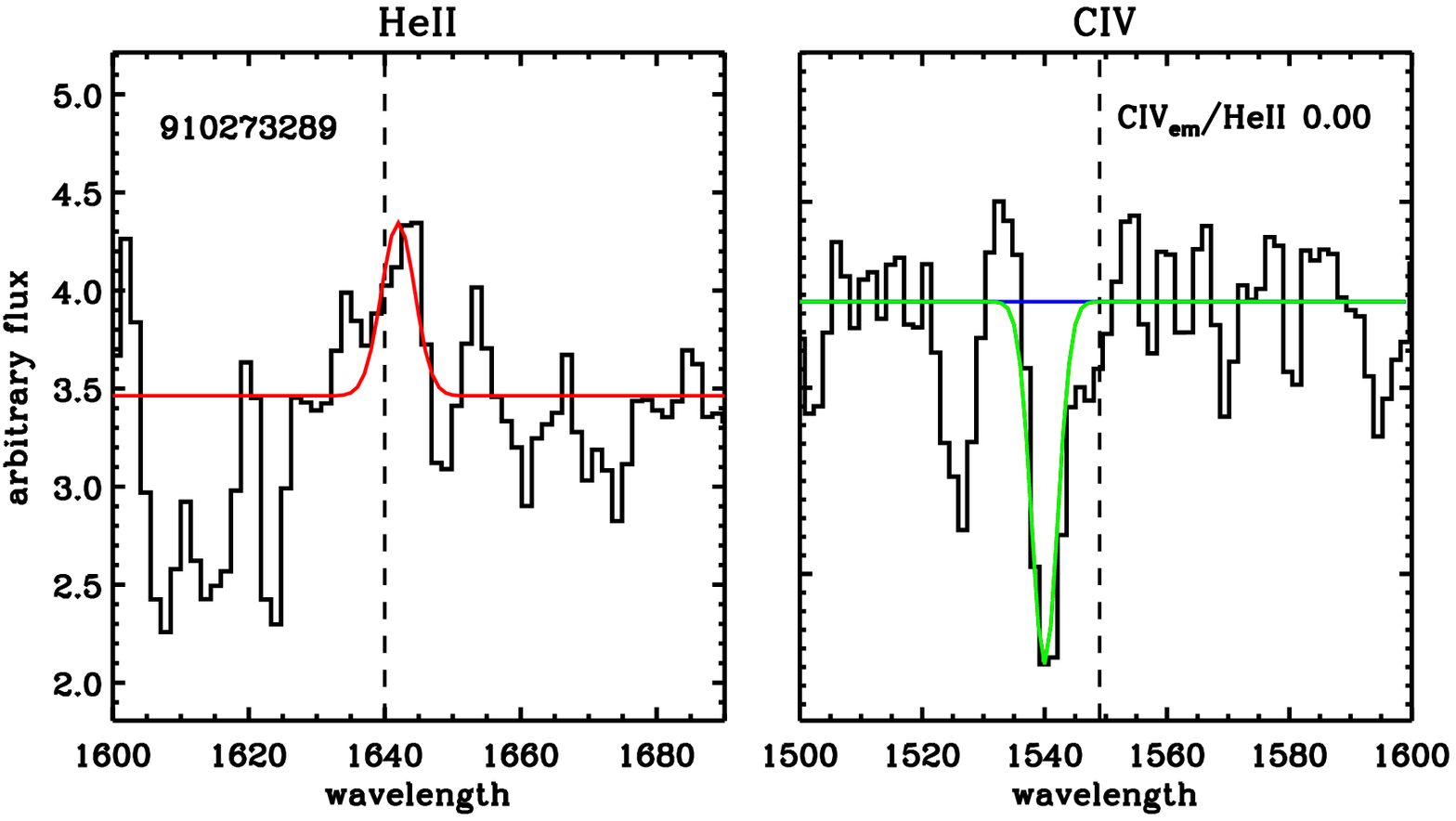}
\includegraphics[width=.8\columnwidth]{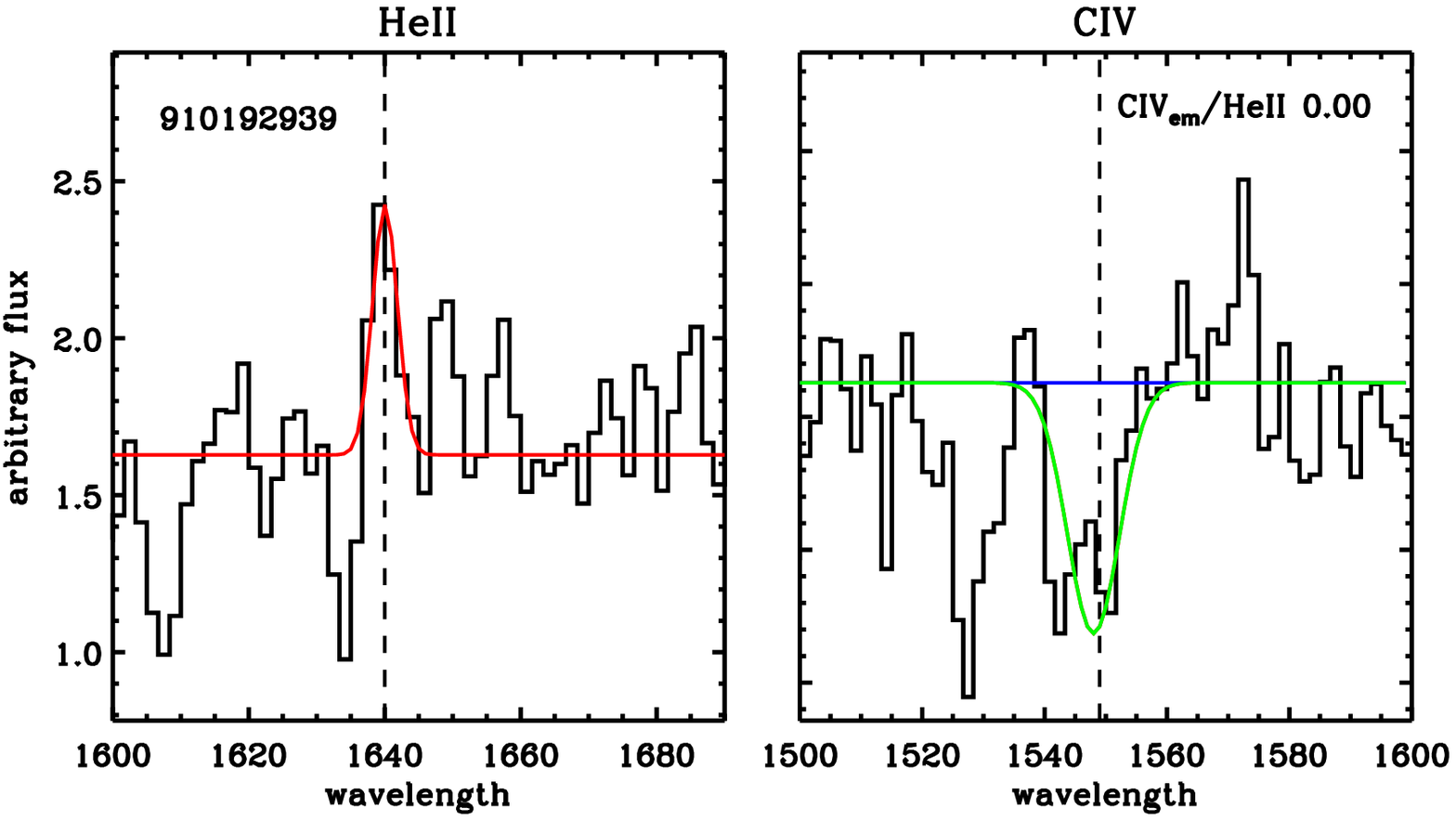}
\includegraphics[width=.8\columnwidth]{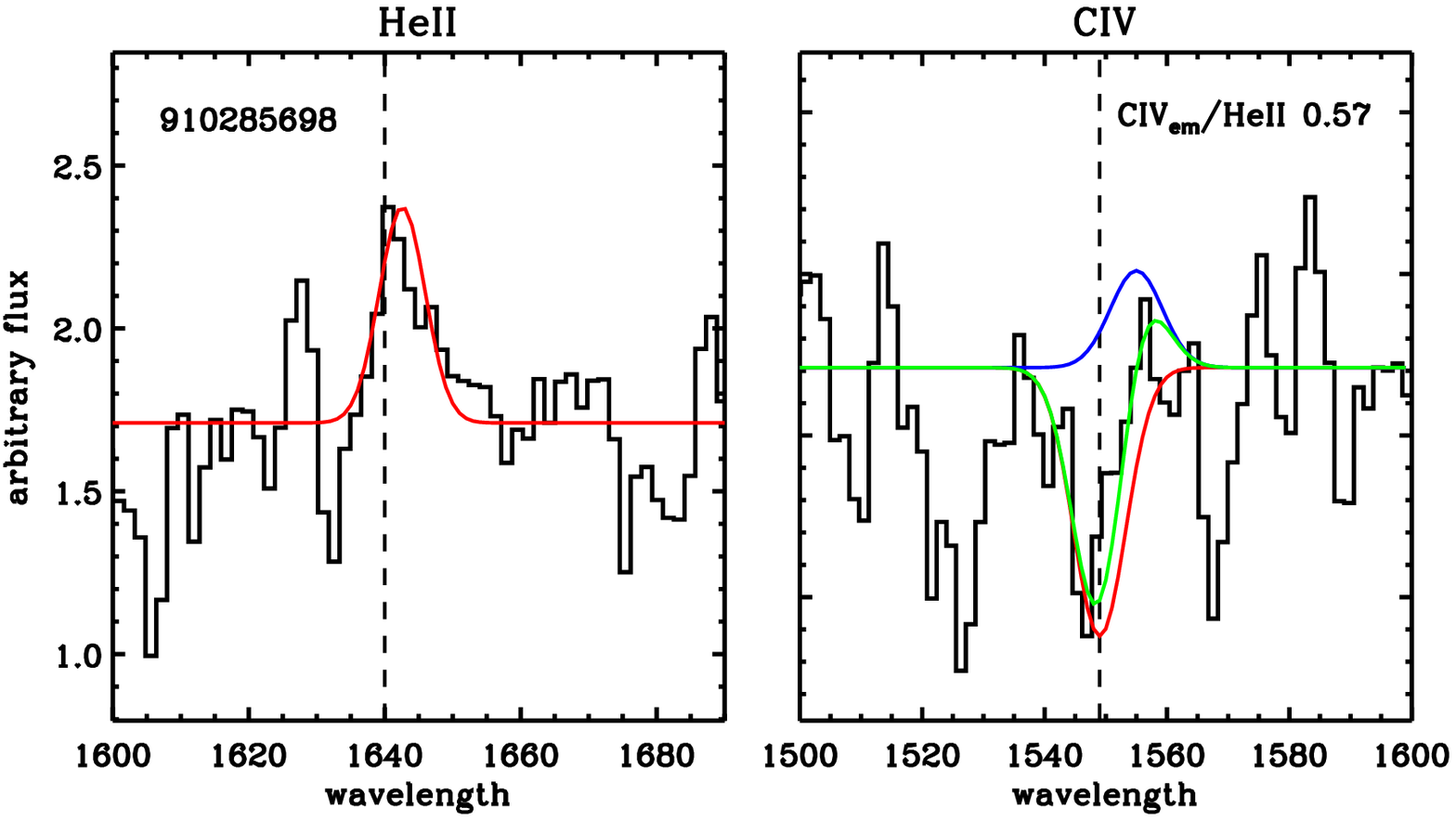}
\includegraphics[width=.8\columnwidth]{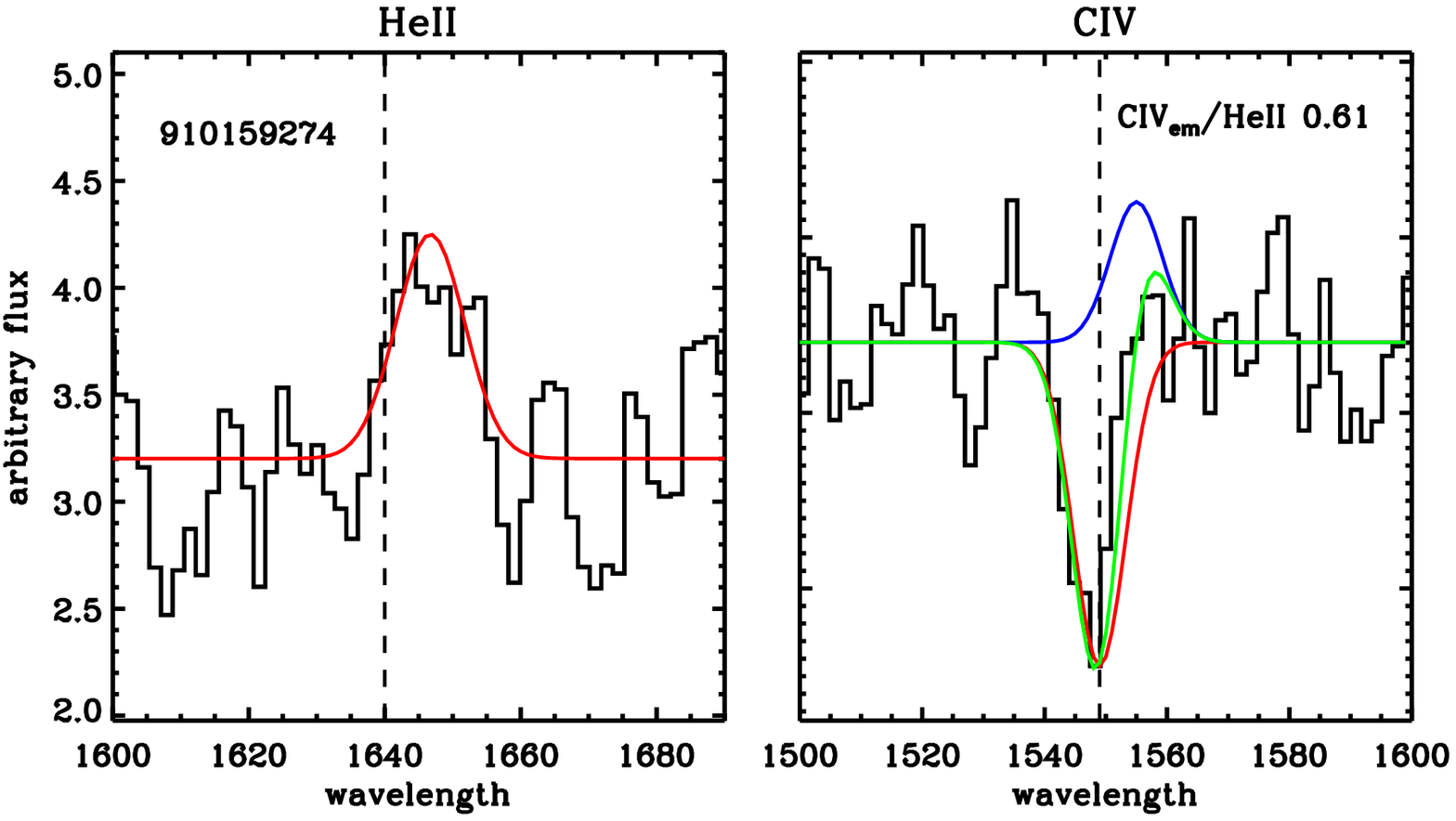}
\includegraphics[width=.8\columnwidth]{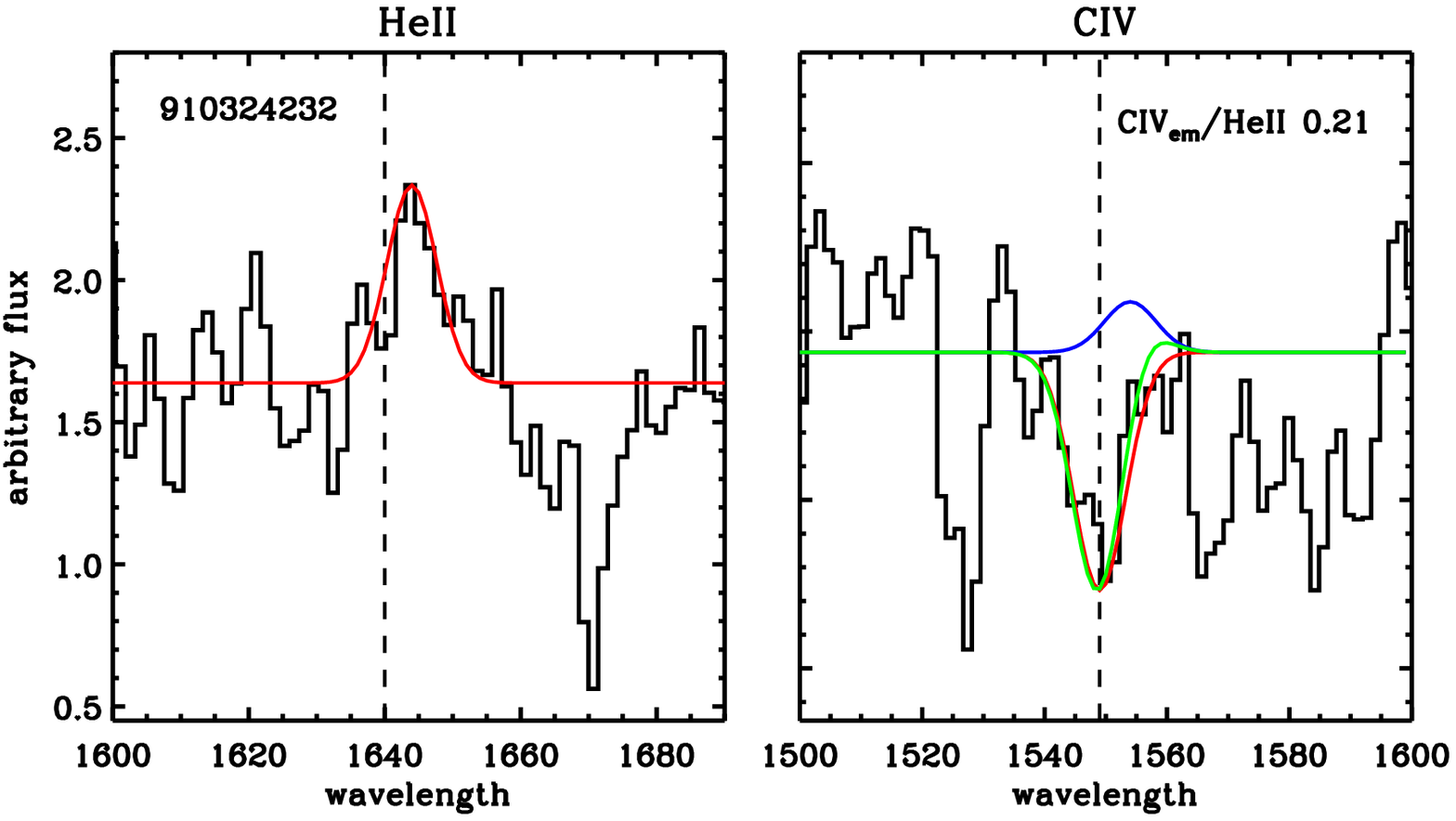}
\includegraphics[width=.8\columnwidth]{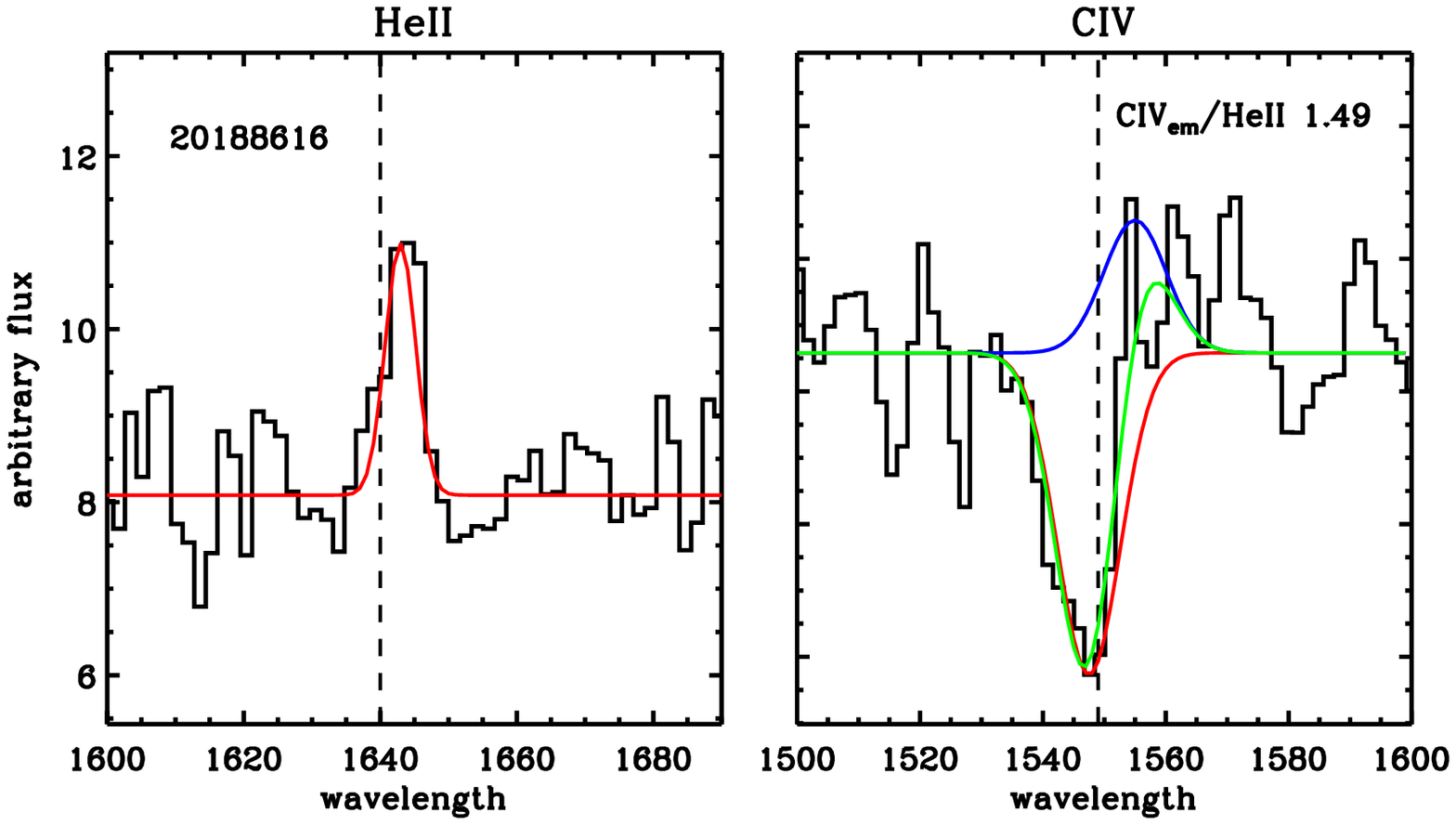}
\includegraphics[width=.8\columnwidth]{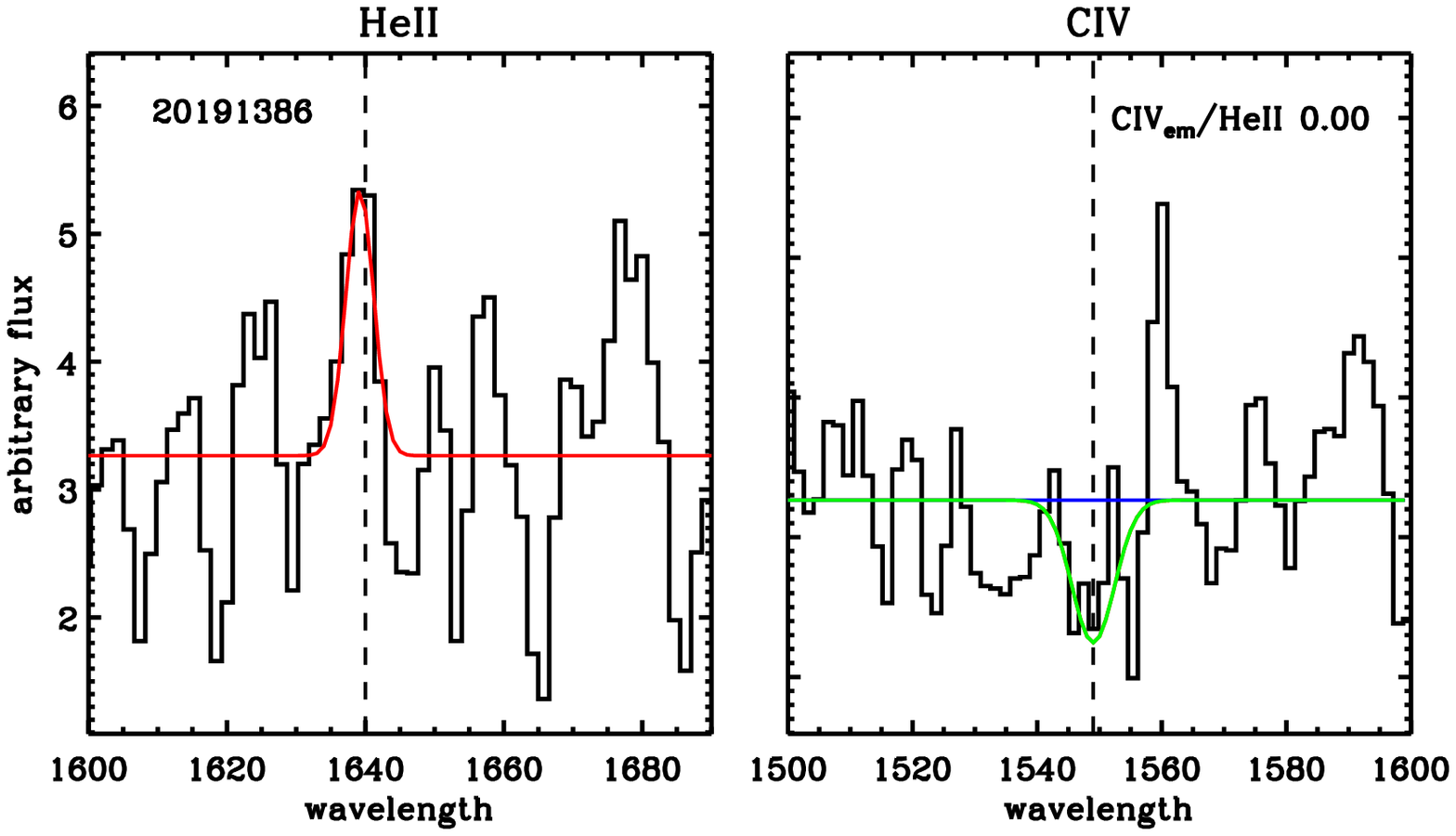}
\includegraphics[width=.8\columnwidth]{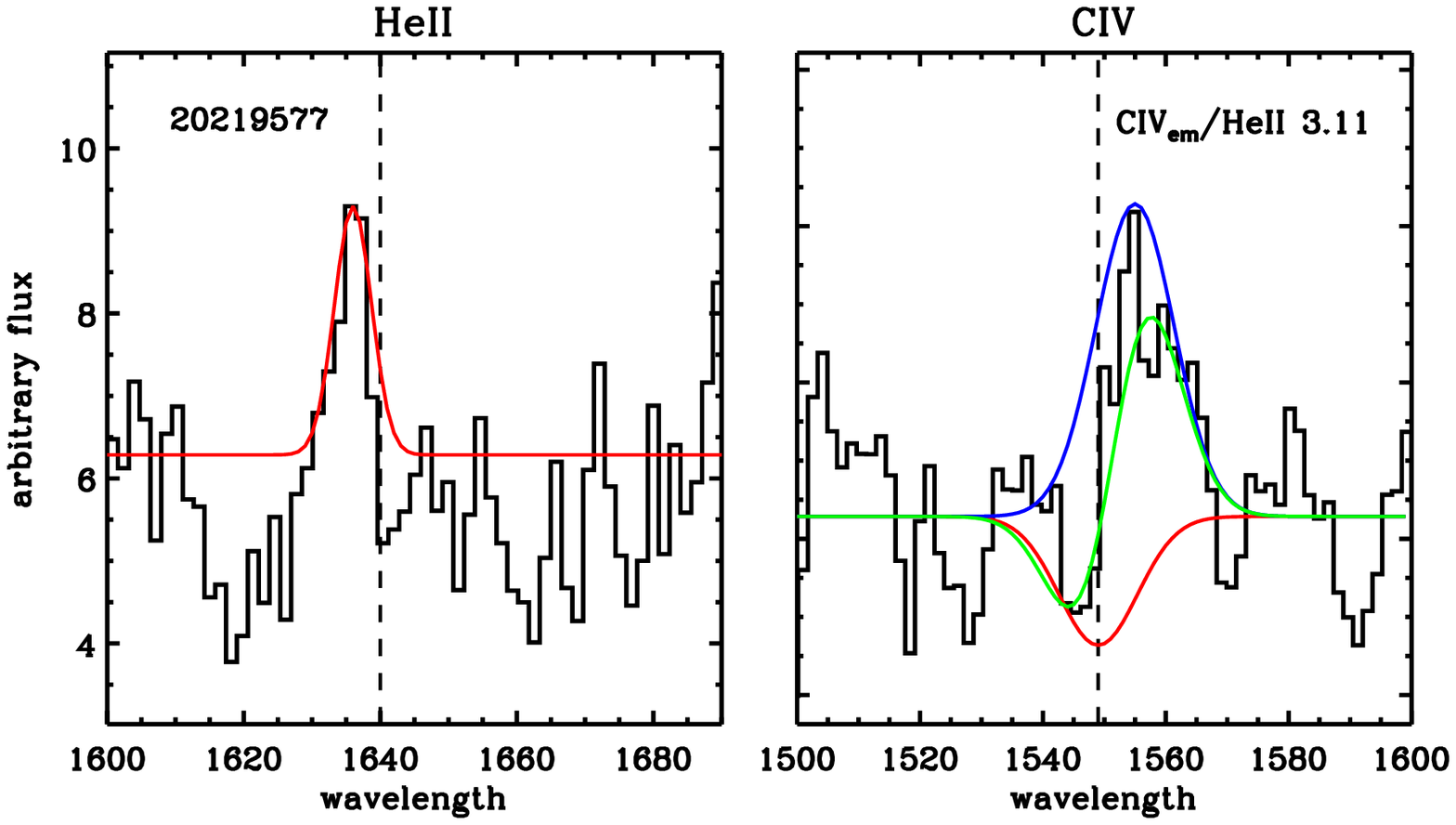}
\caption{Close--up of the region around He~II (left side of each
  sub--figure) and around CIV (right side of each sub--figure) for
  each of the probable He~II emitters. In the pairs of panels, the red
  line shows a Gaussian fit to either He~II or CIV. When a single
  Gaussian is not enough to reproduce the CIV absorption, we add a
  second Gaussian in emission (shown in blue); the resulting profile,
  mimicking a P-Cygni profile, is shown in green.  The number in the
  right--side panels is the ratio between the flux of the C~IV
  component in emission and the flux of He~II.}
\label{zoom_prob}
\end{figure*}



\section{Observations}\label{sect_obs}

\subsection{Spectroscopy}

The VVDS has exploited the high multiplex capabilities of the VIMOS
instrument on the ESO-VLT (Le F\`evre et al., 2003) to collect more
than 35000 spectra of galaxies between $z\sim$0 and $z\sim5$ (Le
F\`evre et al., 2005; Garilli~et~al.~2008, Le~F\`evre~et~al.~2013, in
prep.). In the VVDS-Deep 0216-04 field, more than $\sim$10000 spectra
have been collected for galaxies with $I_{AB} \leq24$, observed with
the LR-Red grism across the wavelength range $5500<\lambda<9350$\AA,
with integration times of 16000 seconds, over an area of 0.62
deg$^2$. In addition, the VVDS Ultra-Deep (Le~F\`evre~et~al.~2013, in
prep.) has collected $\sim$1000 spectra for galaxies with $i_{AB}
\leq24.75$, obtained with LR-blue and LR-red grisms, with integration
times of 65000 seconds for each grism, over an area of 0.16
deg$^2$. This produces spectra with a wavelength range
$3600<\lambda<9350$\AA~.  For both the Deep and Ultra-Deep surveys,
the slits have been designed to be 1'' in width, providing a good
sampling of the 1'' typical seeing of Paranal, and between $\sim$4''
and $\sim$15'' in length, which allows for good sky determination on
both sides of the main target. The resulting resolution power is
$R\simeq230-250$ for both the LR-blue and LR-red grisms, providing a
theoretical spectral resolution of $\sim22 \AA~$ at 5000\AA~. Using
the calibration lamp emission line spectra, we measured a true
resolution (FWHM) of $\sim22$\AA~, corresponding to a velocity
resolution of $\sim$1150 km/s at $z\sim2.5$. However, as the
atmospheric seeing during the observations was often better than the
slit width, the galaxies in our sample never completely filled the
slit: we measured on the 2D spectra the FWHM in the spatial direction
for the 39 He~II emitters, and we found values between 0.5'' and
1.2'', with a typical value of 0.8''. These values are very similar to
the FWHM of the seeing, meaning that the spectra are barely resolved
in the spatial direction.  Assuming that objects are circular, we can
assume that their spatial extent across the slit is also 0.8''. As a
result, the effective resolution in the spectral direction can be
smaller than the nominal one. We have therefore taken $\sim$18\AA~as
the average spectra resolution, corresponding to $\sim1000$ km/s at
$\lambda=1640$\AA~at $z\sim2.5$.

We deconvolved the observed line FWHM into intrinsic FWHM$_0$ applying
the simple equation

\begin{equation}
FWHM_0=\sqrt{FWHM_{obs}^2-FWHM_{instr}^2},
\end{equation}

\noindent where FWHM$_{instr}$ is the instrumental
resolution. Assuming that FWHM$_{instr}$=1000 km/s (as derived from
the mean spectral resolution of 18\AA~ reported above), an emitter
with observed FWHM=1200 km/s will have an intrinsic FWHM=663 km/s, and
an emitter with observed FWHM=2000 km/s will have an intrinsic
FWHM=1732 km/s. Of course, with our assumption, the conversion is
meaningless for those galaxies with observed FWHM$<1000$ km/s.

\subsection{Photometry and SED fitting}\label{sect:photometry}

The VVDS 0216-04 field benefits from extensive deep photometry.  The
field was first observed with the CFH12K camera in the $BVRI$ bands
(Le F\`evre~et~al., 2004; McCracken~et~al., 2003), in the $U$ band
(Radovich~et~al.~2004) and in the $K_s$ band
(Iovino~et~al.~2005). More recent and significantly deeper
observations were obtained as part of the CFHT Legacy Survey in $ugri$
and $z'$ (Goranova~et~al.~2009; Coupon~et~al.~2009), and as part of
the WIRDS survey in $J$, $H$, and $K_s$ bands (Bielby~et~al.,
2012). We use the T0005 data release of the CFHTLS observations that
reach 5$\sigma$ point source limiting magnitudes of 26.8, 27.4, 27.1,
26.1 and 25.7 in the $u^*$, $g'$, $r'$, $i'$, and $z'$ bands,
respectively. The WIRDS observations reach a $5\sigma$ point source
limit of 24.2, 24.1, ad 24 in the $J$, $H$, and $K_s$ bands,
respctively.

Additional NIR imaging is available as a part of the $Spitzer$
Wide-Area InfraRed Extragalactic survey (SWIRE; Lonsdale~et~al.~2003),
providing imaging in the 3.6, 4.5, 5.8, 8.0 and 24 $\mu$m channels
reaching 5$\sigma$ limits of 22.8, 22.1, 20.2, 20.1 and 18.5,
respectively. Moreover, Herschel data are available as a part of the
$Herschel$ Multi-tiered Extragalactic Survey (HerMES;
Oliver~et~al.~2012): the coverage is quite shallow, giving 5$\sigma$
limits of about 12 mJy in the 250, 350, and 500 $\mu$m channels.

X-ray observations of the VVDS 0216-04 field are also available as a
part of the XMM Medium Deep Survey (XMDS; Chiappetti~et~al.~2005),
reaching 3$\sigma$ limits of 6$\times10^{-15}$ and 8$\times10^{-15}$
erg s$^{-1}$ cm$^{-2}$ in the soft and hard bands, respectively. This
limit translates to about $L_{X}\sim10^{44}$ erg s$^{-1}$ cm$^{-2}$ at
$z=2$, corresponding to a moderately powerful Seyfert.

We performed the spectral energy distribution (SED) fitting using the
code ALF (Algorithm for Luminosity Function, Ilbert~et~al.~2005) that
includes the routines of the code Le
Phare\footnotetext{http://www.cfht.hawaii.edu/~arnouts/LEPHARE/lephare.
  html}.  We used the template library from Bruzual~\&~Charlot (2003),
and we also included in the fit the dust attenuation in the form
defined by Calzetti~et~al.~(2000). The SED fitting was performed on
the observed $u^*Bg'Vr'i'Iz'JHK_s$ photometric broadbands. We refer
the reader to Cucciati~et~al.~(2012) for more
details. Walcher~et~al.~(2008) demonstrated that, for a sample of
galaxies with spectroscopic redshift $z<1.2$ drawn from the VVDS
sample, the SFR inferred from a similar SED fitting procedure (though
with slightly worse photometry than in this work) agrees well with the
one measured via O~II and/or H$\alpha$ lines.

\section{The He~II emitters}\label{dataset}
\begin{table*}
 \centering
\begin{tabular}{|c|c c|c c|c c|c c|}
\hline
\multicolumn{1}{|c}{} & \multicolumn{2}{c}{Stack} & \multicolumn{2}{c}{Stack} & \multicolumn{2}{c}{Stack} & \multicolumn{2}{c|}{Stack}\\ 
\multicolumn{1}{|c}{} & \multicolumn{2}{c}{narrow} & \multicolumn{2}{c}{broad} & \multicolumn{2}{c}{possible} & \multicolumn{2}{c|}{all}\\ 
\multicolumn{1}{|c}{} & \multicolumn{2}{c}{(11 galaxies)} & \multicolumn{2}{c}{(13 galaxies)} & \multicolumn{2}{c}{(12 galaxies)} & \multicolumn{2}{c|}{(277 galaxies)}\\ 
Line &EW$_0$ & FWHM  &EW$_0$ & FWHM  &EW$_0$ & FWHM  &EW$_0$ & FWHM \\
\hline
Ly$\alpha$$\lambda$1216 &  -8.4  &  2063  &  8.0 & 1021 & -9.6 & 2097 & 0.7   & 1000\\
Si~II$\lambda$1260      &  -1.6  &  --    & -1.7 & 984  & -1.2 & 565  & -1.6  & 1500\\ 
O~I$\lambda$1303        &   --   &  --    & -2.2 & 1130 & -3.8 & 1222 & -2.5  & 1542\\ 
C~II$\lambda$1334       &  -0.5  &  912   & -1.5 & 893  & -2.0 &  784 & -1.7  & 1457\\ 
Si~IV$\lambda$1393      &  -2.2  &  1239  & -2.2 & 1659 & -1.8 & 1076 & -1.9  & 1772\\ 
Si~IV$\lambda$1402      &  -1.9  &  1519  & -1.1 & 1093 & -1.5 & 1148 & -1.6  & 2166\\ 
Si~II$\lambda$1527      &  -1.6  &  1212  & -1.8 & 1495 & -2.4 & 1185 & -1.6  & 1259\\ 
C~IV$\lambda$1549       &  -5.9  &  2756  & -3.7 & 1586 & -4.5 & 1932 & -3.0  & 1724\\ 
He~II$\lambda$1640      &   4.1  &  993   &  2.7 & 1914 & 3.3  & 2220 &  1.0  & 1776\\ 
C~III$\lambda$1909      &   7.8  &  1631  &  3.8 & 1768 & --   & --   &  3.4  & 2064\\ 
\hline
\end{tabular}
\caption{Rest-frame equivalent widths and FWHM for absorption and
  emission lines in the composite spectra of Figs.~\ref{spec_all} and
  \ref{spec_stack}.}\label{tab:stacks}
\end{table*}

\subsection{Properties of the parent sample of star-forming galaxies
at $2<z<4.6$}  

In this study we concentrate on galaxies with secure redshifts (flags
2, 3, 4, 9, see Le F\`evre et al., 2005) in the range $2<z<4.6$, for
which the He~II line at $\lambda_{rest}=1640 \AA~$ is redshifted to
$\sim$5000 -- 9200\AA~. The Ultra-Deep spectra cover this wavelength
range completely, while the Deep spectra only cover this range from
$z=2.3$ to $z=4.6$. In total, these criteria select 277 galaxies, 66
of which are from the Deep survey. The galaxies in the sample show
different lines in their UV rest-frame domain, such as Ly$\alpha$
(1216\AA~), Si~II (1260\AA~), O~I doublet (1303\AA~), C~II (1334\AA~),
Si~IV doublet (1397\AA~), C~IV doublet (1549\AA~), Al~II (1671\AA~),
He~II (1640\AA~), and C~III (1909\AA~). For each galaxy, the redshift
was estimated by measuring the observed wavelength of the lines
present in the observed spectrum. Since it has been observationally
shown that Ly$\alpha$ is often redshifted with respect to interstellar
lines (Steidel~et~al.~2010), Ly$\alpha$ was not considered a good
tracer of the systemic redshift of galaxies. For each galaxy, we have
at least $\sim3$ high signal-to-noise interstellar lines to perform
the redshift determination.

In Fig.~\ref{spec_all} we report the average spectrum resulting from
the stack of all the 277 galaxies at $2<z<4.6$ for which the He~II
line falls in the region covered by the spectroscopy, and we compare
them with the stack of about 1000 Lyman-break galaxies by
Shapley~et~al.~(2003, hereafter S03). The stack was produced
running the {\tt scombine} routine in IRAF, which produced a variance
weighted average. The spectra normalized to the median flux density
estimated between $\lambda=1500$\AA~ and $\lambda=2000$\AA. A sigma
clipping algorithm (lsigma=3, hsigma=3) was applied during the stacking
process and the spectra were combined using mean stacking.

\begin{table*}
 \centering
\begin{tabular}{|c|c c c c c c c c |}
\hline
  id  & redshift & mag$_i$ & mag$_K$ & $\log$(M/M$_{\odot}$) & log(SFR [M$_{\odot}$ yr$^{-1}$] & FWHM$_{obs}$(He~II) [km/s] & $\beta$ slope & type \\
\hline
910294959  &  2.133  &  24.62  &  23.51  &  9.55  &  0.76  &   756 &   $-1.89\pm   0.09$  & N\\
910167429  &  2.174  &  24.09  &  22.37  &  10.2  &  0.18  &   732 &   $-0.72\pm   0.08$  & N\\
910282010  &  2.245  &  24.58  &  23.87  &  9.58  &  1.36  &  1117 &   $-0.63\pm   0.11$  & N\\
910265546  &  2.012  &  23.76  &  -99.9  &  9.82  &  2.84  &  1064 &   $-0.88\pm   0.05$  & N\\
910246547  &  2.376  &  24.50  &  23.35  &  9.35  &  2.35  &   993 &   $-0.05\pm   0.23$* & N\\
910191609  &  2.436  &  24.05  &  24.84  &  9.16  &  2.18  &  1017 &   $-0.38\pm   0.16$* & N\\
910362042  &  2.199  &  24.63  &  24.82  &  9.53  &  2.56  &   953 &   $-0.10\pm   0.24$* & N\\
910260902  &  2.359  &  23.48  &  22.22  &  10.5  &  1.47  &   601 &   $-1.82\pm   0.12$  & N\\
20107579   &  3.817  &  23.45  &  22.98  &  10.2  &  1.21  &   774 &   $-0.09\pm   0.43$* & N\\
20215115   &  3.998  &  24.05  &  24.68  &  8.49  &  1.51  &   957 &   $-2.19\pm   1.03$* & N\\
20198832   &  2.767  &  -99.9  &  -99.9  &  10.0  &  0.85  &   778 &   $-1.55\pm   0.18$  & N\\
\hline                                                                              
average N     &  2.592  &  24.12  &  23.62  &  9.67  &  1.57  &   885 & -1.28 & \\
\hline
910276278  &  2.292  &  22.92  &  22.19  &  10.2  &  1.27  &  1746  &    $-1.43\pm   0.04$  & B\\
910261908  &  2.128  &  23.25  &  22.60  &  9.82  &  1.51  &  2278  &    $-0.82\pm   0.04$  & B\\
910252781  &  2.629  &  24.19  &  23.47  &  9.28  &  2.29  &  1576  &    $-0.88\pm   0.11$  & B\\
910279100  &  2.289  &  23.08  &  21.87  &  10.4  &  1.38  &  3132  &    $-1.59\pm   0.08$  & B\\
910221957  &  2.205  &  22.89  &  22.03  &  10.2  &  1.66  &  2213  &    $-1.09\pm   0.04$  & B\\
910193041  &  2.245  &  23.78  &  23.61  &  9.34  &  1.36  &  1885  &    $-1.24\pm   0.06$  & B\\
910277155  &  2.285  &  23.52  &  23.02  &  9.90  &  1.44  &  1998  &    $-1.37\pm   0.05$  & B\\
910329878  &  2.267  &  24.47  &  23.61  &  9.56  &  0.61  &  1515  &    $-1.90\pm   0.10$  & B\\
910210056  &  2.035  &  23.61  &  22.47  &  9.75  &  2.02  &  2917  &    $-1.17\pm   0.05$  & B\\
910301515  &  2.609  &  24.41  &  23.06  &  9.94  &  1.56  &  1356  &    $-1.11\pm   0.14$  & B\\
910177193  &  2.123  &  22.88  &  21.66  &  10.3  &  1.97  &  2707  &    $-0.66\pm   0.03$  & B\\
910295166  &  2.372  &  22.01  &  20.96  &  10.9  &  1.92  &  1456  &    $-2.02\pm   0.11$  & B\\
20465257   &  2.716  &  22.56  &  22.31  &  9.55  &  2.55  &  2062  &    $-1.41\pm   0.08$  & B\\
\hline
average B     &  2.323  &  23.35  &  22.53  &  9.93  &  1.65  &  2064 & -1.28 & \\
\hline
910270610  &  2.297  &  23.44  &  -99.9  &  10.1  &  1.31  &  1564  &  $-1.00\pm   0.04$  & P\\
910376001  &  2.079  &  22.70  &  20.53  &  10.7  &  3.05  &  2072  &  $-0.26\pm   0.05$  & P\\
910300064  &  2.418  &  23.94  &  22.94  &  10.0  &  1.63  &  1480  &  $-0.78\pm   0.07$  & P\\
910248329  &  2.293  &  24.17  &  22.96  &  10.0  &  0.45  &  1571  &  $-1.76\pm   0.14$  & P\\
910273289  &  2.629  &  24.00  &  23.16  &  9.30  &  2.31  &  1048  &  $-1.42\pm   0.16$  & P\\
910192939  &  2.213  &  22.76  &  21.97  &  9.89  &  2.90  &   603  &  $ 0.30\pm   0.10$  & P\\
910285698  &  2.377  &  24.24  &  23.11  &  9.98  &  1.85  &  1449  &  $-0.20\pm   0.15$  & P\\
910159274  &  2.142  &  24.38  &  23.26  &  9.12  &  1.50  &  2709  &  $-1.38\pm   0.11$  & P\\
910324232  &  2.774  &  23.81  &  -99.9  &  10.3  &  0.81  &  1028  &  $-2.61\pm   0.15$  & P\\
20188616   &  3.218  &  23.02  &  22.22  &  10.7  &  1.44  &   878  &  $-4.92\pm   0.28$  & P\\
20191386   &  3.527  &  23.99  &  22.59  &  10.7  &  1.06  &  8624  &  $-0.85\pm   0.35$  & P\\
20219577   &  3.506  &  23.61  &  22.83  &  8.58  &  1.60  &  1165  &  $-2.00\pm   0.21$  & P\\
\hline
average P     &  2.623  &  23.67  &  22.56  &  9.95  &  1.66  &  2015 & -1.40 & \\
\hline
910359136  &  2.287  &  24.57  &  21.29  &  10.9  &  0.35  &  1454  &   $-0.51\pm   0.13$  & A\\
20273801   &  2.487  &  22.84  &  21.23  &  10.7  &  1.49  &  2209  &   $-2.01\pm   0.19$  & A\\
20366296   &  2.556  &  22.68  &  21.41  &  10.3  &  3.30  &  1649  &   $-1.36\pm   0.12$  & A\\  
\hline
average A     &  2.623  &  23.36  &  21.31  &  10.7  &  1.71  &  1770 & -1.29 & \\
\hline
\end{tabular}
\caption{Properties of the 39 galaxies with He~II in emission. We
  report redshift, $i'$ and $K_s$ observed magnitudes, stellar mass,
  star--formation rate (as measured from SED fitting), $\beta$ slope,
  and type (N: narrow He~II emitters; B: broad He~II emitters; P:
  probable He~II emitters; A: AGN). We also report the average values
  for each group. Objects for which the $\beta$ slope is marked with
  an asterisk have barely detected continua, and they have not been
  used to compute the average value of the narrow
  emitters.}\label{tab:emit}
\end{table*}

In Table~\ref{tab:stacks} we report equivalent widths and line widths
(FWHM) for the most relevant absorption and emission lines of the
stacked spectra. In the composite spectrum, a broad He~II emission at
$\lambda=$1640\AA~is evident: we measure a rest-frame equivalent width
of $\sim$1.0\AA, and FWHM $\sim$ 1800 km/s. The presence of the He~II
line in this composite implies that the He~II emission is a common
feature in the spectra of star forming galaxies at these
redshifts. Our composite spectrum is different to the S03
composite. The main difference is the slope of the continuum: our
spectrum is much redder than S03. This difference is probably due to
the different selection criteria. The Lyman-break galaxy technique
used in S03 tends to select bluer objects, while our flux-limited
sample is unbiased with respect to the color. Moreover, the S03
composite has stronger Ly$\alpha$ emission, stronger P-Cygni features
around N~V, O~I, and C~IV lines, and a slightly stronger He~II
emission (EW$_0\sim$1.5\AA).

In order to test the robustness of our stacking procedure, we
experimented with other stacking parameters, varying the type of
scaling used before stacking (e.g., using the mean or median, applying
a moderate or strong sigma clipping, changing the region in which the
continuum fluxes are evaluated, etc.). We find that the set of
parameters used for our stacked spectra gives the best signal-to-noise
(S/N) continuum; moreover, the equivalent widths of the relevant lines
are always within 20\% of the values shown in Table~1.


\subsection{Searching for individual galaxies with He~II emission}

We then visually inspected the 277 spectra to identify bona fide He~II
emitters responsible for the emission feature in the combined
spectra. We excluded all galaxies for which the He~II emission at
$1640\AA~$ falls in a spectral region dominated by bright OH airglow
emission lines, which dominate the background at $\lambda>7500\AA~$,
and so are not easy to subtract from the combined object+sky spectra
in low-resolution spectra. We identified a preliminary list of 114
potential emitters, and we cleaned it by removing the less convincing
cases (objects with lines at the location of weaker OH airglow
emission, or at spectral locations with known higher background noise)
which resulted in 39 bona fide He~II emitters. Among these, 9 are from
the DEEP survey and the remaining 30 from the Ultra-DEEP one. Three of
them were identified as AGN during the redshift measurement process.
In order to test the robustness of our sample and to estimate the
fraction of false positive lines that could be due to a stochastic
noise peak, we multiplied all of the 277 spectra by -1 and we
re-checked them to look for fake He~II lines. We found a list of 10
fake emitters, and after cleaning this sample as we did for the real
sample, we ended up with only 2 possible emitters. This experiment
shows that 5\% at most of our 39 emitters could be fake detections.

For each He~II emitter, we measured the equivalent width, the line
flux, the line observed FWHM, and the S/N of the continuum around the
He~II line using the {\it onedspec} task of IRAF. We then divided the
sample into four subsamples: the {\it reliable} narrow He~II emitters
(11 objects, with FWHM$_{He~II}<$1200 km/s); the {\it reliable} broad
He~II emitters (13 objects, with FWHM$_{He~II}>$1200 km/s); the {\it
  possible} He~II emitters (12 objects, for which the detection is not
certain, based on the line shape, position with respect to sky lines
and/or the line S/N ratio) and AGN (3 objects). We chose a discriminating
observed He~II FWHM of 1200 km/s because this corresponds to an
intrinsic (deconvolved) FWHM $\sim$663 km/s, a very conservative lower
limit for the line width expected in the case of Wolf-Rayet stars (see
Sects.~\ref{sect:analysis} and \ref{sect:WR}).

\subsection{Spectro-photometric properties of the He~II emitters}
We report in Table~\ref{tab:emit} the photometric properties of the 39
He~II emitters in the sample, together with the UV continuum slope
$\beta$ and width of the He~II line that allowed us to classify the
reliable emitters between narrow and broad.  The $\beta$ slopes were
computed fitting a power law ($F_{\lambda}\propto\lambda^{\beta}$) to
the spectra between $\lambda=1500~\AA$ and
$\lambda=2500~\AA$. However, for some objects (marked with an asterisk
in Table~2) the continuum is barely detected and so the slope is
poorly constrained. Most of the objects (27) are at $2<z<2.5$; the
highest redshift object is at redshift $z\sim4$; and 4 emitters have
$3.5<z<4.5$. The He~II emitters span a broad range of properties:
He~II is found equally in massive objects ($M^*/M_{\odot}>10^{10.5}$)
and in less massive ones ($M^*/M_{\odot}\sim10^{9}$), both on highly
star--forming objects ($SFR>1000 M_{\odot} yr^{-1}$ and in normal
star--forming galaxies ($SFR\sim 1-10 M_{\odot} yr^{-1}$). The He~II
emitters have $\beta$ slopes between -0.5 and -2.5, but we do not find
any correlation between the slope and the strength of \lae or He~II
lines. Moreover, we do not find any particular correlation between
these properties and the width of the He~II line: the only clear
difference is that the narrow emitters are typically fainter (both in
$i'$ and $K_s$ bands) than the others. The 1 arcsec typical seeing of
the CFHTLS images, corresponding to about 8 kpc at $z\sim2.5$, did not
allow us to perform a detailed morphological analysis; however, the 39
He~II emitters have morphological properties spanning from compact to
more diffuse irregular objects.


Only a few He~II emitters were detected in the Spitzer-SWIRE imaging
survey (ids: 910177193, 910376001, 910192939 plus the three
AGN). However, these objects were detected in the first two IRAC
channels only, for which we have a much deeper imaging, and not in the
other two channels, which does not allow us to apply the AGN/galaxy
separation based on MIR colors (Donley~et~al.~2012). Two objects were
detected in the Herschel imaging: objects 910376001 and 20366296. The
first object has a total IR luminosity $L_{TIR}=1.8\times10^{12}
L_{\odot}$. Assuming that the FIR luminosity comes only from star
formation, this value corresponds to a star-formation rate of about
$\sim300 M_{\odot} yr^{-1}$, not far from the estimate of $\sim1000
M_{\odot} yr^{-1}$ obtained from the SED fitting. The other object
detected in SPIRE is, instead, a clear AGN, but this object was not
detected in the XMM-LSS survey, implying that it is probably a
star--forming object with a moderately weak AGN component.  Finally,
we note that none of the other He~II emitters were detected in the
XMM-LSS survey.

\subsection{Spectral properties of the individual He~II emitters}
We show the spectra of the 39 He~II emitters, for the four groups
mentioned above, in Fig.~\ref{spec_narrow} (the {\it secure}
emitters with narrow He~II emission), Fig.~\ref{spec_broad} (the
{\it secure} emitters with broad He~II emission),
Fig.~\ref{spec_agn} (the three AGN) and Fig.~\ref{spec_uncertain}
( the {\it possible} He~II emitters).

For each individual spectrum we identify at least two lines, that have
been used to determine the spectroscopic redshift (for about 90\% of
the objects we identify three or more absorption and/or emission
lines). The most common feature in the spectra is the doublet
Si~II+C~IV at 1526 \AA~ and 1549 \AA~. For some of the objects coming
from the Deep survey (labelled 20xxxxxxx) the Ly$\alpha$ region is not
covered. It can be also seen that for 17 He~II emitters the Ly$\alpha$
line is in absorption. We also remark that C~IV is in absorption for
all objects, except for one of the AGN; for the other two, C~IV is not
in the covered spectral range. However, the C~IV line is often
asymmetric, with the red wing sharper than the blue one, suggesting in
some cases the presence of an emission component just redward of the
absorption reminiscent of a P-Cygni profile.

\begin{figure}
  \centering \includegraphics[width=8cm]{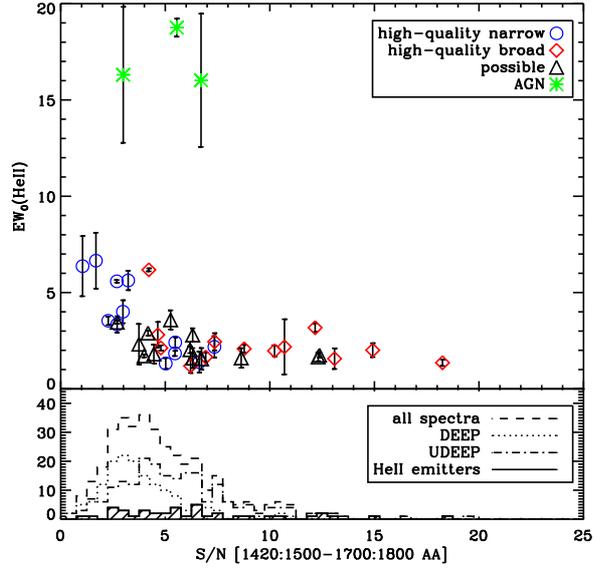}
  \caption{{\it Top panel:} Rest-frame equivalent width of the
    He~II~$\lambda$1640 line as a function of the average S/N ratio
    per resolution element of the spectra in the two regions
    1420--1500\AA~ and 1700--1800\AA~for the 39 galaxies with detected
    He~II in emission. The blue circles, red diamonds, black
    triangles, and green asterisks represent the reliable emitters
    with narrow He~II lines, the reliable emitters with broad He~II
    lines, the possible emitters, and the objects classified as AGN,
    respectively. {\it bottom panel}: histogram of the average S/N
    ratio at 1420--1500\AA~ and 1700--1800\AA~ for all 277 galaxies in
    the sample (long dashed line), for the DEEP and UDEEP spectra
    (dotted and dot-dashed, respectively), and for the 39 galaxies with
    He~II in emission (continuous filled histogram). }
  \label{sn_ew}%
\end{figure}

To investigate the possible presence of these C~IV P-Cygni--like
profiles, we report in Figs.~\ref{zoom_narrow},~\ref{zoom_broad}, and
~\ref{zoom_prob} the close--ups of the region around He~II and C~IV
for the narrow, broad, and probable He~II emitters, respectively. For
each object, we fit a single Gaussian to the He~II line, which almost
always gives a good representation of the line profile except for a
few broad emitters, for which the observed line is skewed towards red
wavelengths. On the contrary, the absorption profile of C~IV is often
asymmetric, with a blue absorption wing that is much broader than the
red one, and not well reproduced by a single Gaussian. To improve the
quality of the fit for these cases, we have to add a second Gaussian
in emission, just redward of the absorption. We then fit the C~IV
profile combining two Gaussians: one in absorption, centered at
1549~\AA, with FWHM and flux chosen to reproduce well the blue
absorption wing; and a second Gaussian in emission for which the
$\Delta~\lambda$ with respect to 1549~\AA~ and the flux in the line
are left as free parameters, while we set the FWHM of this component
to the same value measured for He~II (assuming that they are produced
by the same region). As an output, we get the ratio between the flux
of the C~IV component in emission and the flux in the He~II line.

From Fig.~\ref{zoom_narrow} it is clear that only two galaxies
classified as narrow He~II emitters (ids: 910265546; 910294959) need a
C~IV emission to improve the fit to the C~IV profile. However, for
object 910294959 the effect is quite marginal, with an EW of the
emission component of $\sim$0.2~\AA~only.  In all the other cases, the
C~IV is either undetected (ids: 910282010; 910246547; 910191609;
910362042) or is in absorption with a symmetric profile, which does
not require a second component (ids: 910167429; 910260902; 20107579;
20215115; 20198832). For the broad and probable emitters, the presence
of P-Cygni--like C~IV profiles is a much more frequent feature: for 11
of the 13 broad emitters a C~IV emission component is needed to
reproduce the C~IV line profile with C~IV/He~II ratios going from
$\sim$0.7 to $\sim$2 (Fig.~\ref{zoom_broad}). The group of probable
He~II emitters (Fig.~\ref{zoom_prob}) is a more heterogeneous
population: in 7 of the 12 objects a strong C~IV emission component is
needed, in the remaining 5 of the 12 the C~IV absorption line is quite
symmetric.

From this experiment we can conclude that the narrow and broad
emitters appear to have very different properties of the C~IV line.
While the bulk of the narrow emitters do not show any sign of P--Cygni
profile, most of the broad emitters need an emission component redward
of the C~IV line to explain the C~IV line profile.


In Fig.~\ref{sn_ew} we report the rest-frame equivalent width EW$_0$
of the He~II lines as a function of the signal-to-noise ratio per
resolution element measured in the two regions 1420 -- 1500~\AA~ and
1700--1800\AA~. These two regions are contiguous to the He~II line,
and as they do not contain any spectral features, they provide a
reference measurement of the continuum emission. Apart from the three
AGN that have typical $EW_0\sim$15 -- 20\AA~, the other He~II
emitters in the sample have $1\lesssim EW_0\lesssim7$\AA~. These
values are in general agreement with the rest-frame equivalent widths
of the He~II line measured by Scarlata~et~al.~(2009) for a Ly$\alpha$
blob at $z=2.373$ (EW$_0$(He~II)$\sim3.5$\AA) and by Erb~et~al.~(2010)
for a low-metallicity galaxy at $z\sim2.3$ EW$_0$(He~II)$\sim2.7\AA$.
Prescott,~Dey,~\&~Jannuzi (2009) found instead a nebula at z$\sim$1.6
with much more powerful He~II emission, with EW$_0$(He~II)~$\sim35$\AA.
 

Below $S/N\sim4$, the $EW_0$ and S/N are correlated: the smaller the
S/N, the higher the equivalent width. This is not surprising since
objects with $S/N<4$ have barely detected continua. It is also clear
that narrow He~II emitters have continuum S/N that is lower than that
of broad emitters.

In the bottom panel of Fig.~\ref{sn_ew} we compare the continuum S/N
ratios of the He~II emitters with the S/N of the full sample of 277
galaxies sample. Objects drawn from the UDEEP survey typically have
larger S/N ratios than those drawn from the DEEP survey, as expected
from scaling the exposure times, with the total distribution extending
from $S/N=0$ (on the continuum, for emission-line only spectra) to
$S/N\sim20$ and peaking around $S/N\sim4$. The 39 He~II objects with
He~II emission have a S/N distribution that is similar to the
distribution for the 277 parent galaxies: we found He~II emission in
objects with both faint and bright continua.


\begin{figure}
\centering
\includegraphics[width=1.0\linewidth]{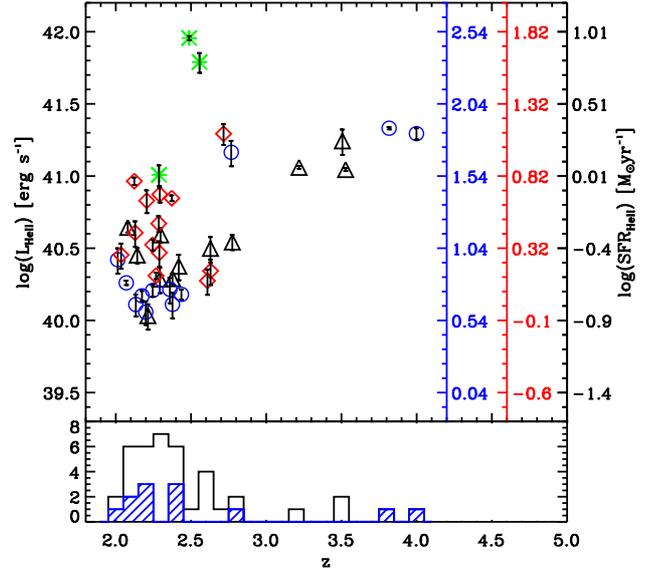}
\caption{{\it Top panel:} Luminosity of the He~II line as a function of
  the redshift for the 39 identified He~II emitters in our sample. The
  blue circles, red diamonds, black triangles, and green stars represent
  the reliable emitters with narrow He~II lines, the reliable
  emitters with broad He~II lines, the possible emitters, and the
  objects classified as AGN, respectively. The three axes on the
  right represent the star--formation rates needed to produce these He~II
  luminosities for a top-heavy IMF with Salpeter slope extending from
  $1 M_{\odot}$ to $500s M_{\odot}$ with metallicities Z=0 (black),
  Z=$10^{-7}$ (red), and Z=$10^{-5}$ (blue), using the conversion in
  Schaerer~2003 (see text). {\it Bottom panel:} Redshift distribution
  of the 39 He~II emitters (empty histogram) and of the reliable
  ones (filled histogram). }
\label{lum_z}
\end{figure}

In Fig.~\ref{lum_z} we show the luminosity of the He~II line as a
function of the redshift for the 39 He~II emitters in the sample.  Two
of the AGN are the most luminous objects in He~II with luminosities up
to $10^{42} erg~s^{-1}$, while the other emitters have
$10^{40}<L_{He~II}<10^{41.5} erg~s^{-1}$.


In the same figure, the He~II luminosities are converted to the
star--formation rates needed to produce this He~II emission, where it
is assumed that the He~II emission is due to star formation that is
happening in a gas region of primeval composition. We applied three
conversions calculated by Schaerer~(2003): a top-heavy IMF with
Salpeter slope extending from $1 M_{\odot}$ to $500 M_{\odot}$, and
three metallicities Z=0, Z$=10^{-7}$, and Z$=10^{-5}$. From these
calculations we infer that, if He~II originates from a region with
zero metallicity, the star formation rate needed to produce the
observed He~II line luminosity is only 0.1 -- 3 $M_{\odot}
yr^{-1}$. We remark also that adding just a few metals changes this
estimate considerably: for Z$=10^{-7}$ and Z$=10^{-5}$ the production
of He~II photons is much less efficient, so we need 10 to 50 times
more star formation to produce the observed He~II line flux. In any
case, no matter the metallicity, the SFR inferred from the He~II lines
under the assumption that He~II is the result of star formation only
are always smaller than the values measured for the individual
galaxies from SED fitting (see Fig.~\ref{sfr_lum}).



\begin{figure}
\centering \includegraphics[width=1.0\linewidth]{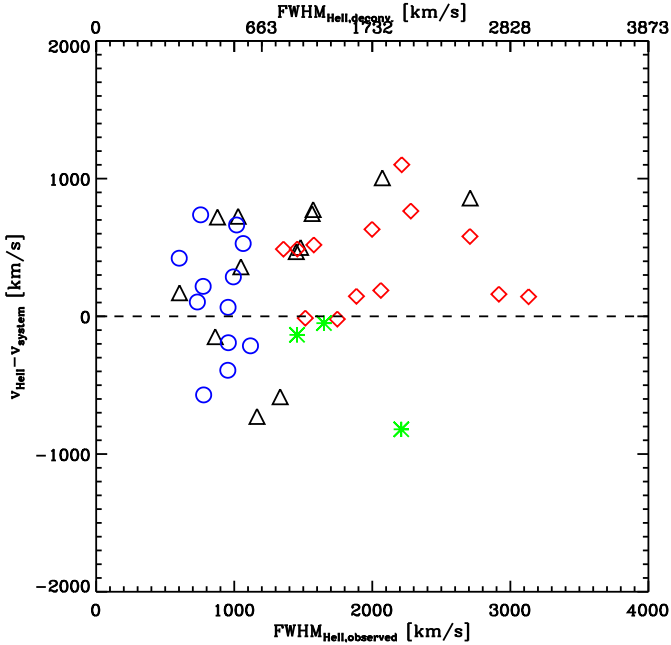}
\caption{For the 39 He~II emitters, the velocity difference between the
  centroid of the He~II line and the systemic redshift vs the
  FWHM of the He~II line in km/s. The bottom axis is linear and
  reports the observed FWHM, while the top axis is converted in
  deconvoluted FWHM. The color code of the symbols is the same as in
  Fig.~\ref{lum_z}.}
\label{dynamic}
\end{figure}

In Fig.~\ref{dynamic} we show the width of the He~II line (FWHM,
expressed in km/s, observed and intrinsic on the bottom axis and top
axis, respectively) as a function of the velocity difference between
the He~II line and the systemic redshift (measured from the position
of the absorption lines, see Sect.~\ref{sect_obs}), together with the
distribution of the two quantities. The distribution of the observed
FWHM extends from 600km/s to 3000km/s, with $\sim10$ emitters having a
FWHM formally smaller than the spectral resolution of the
instrument. As we explained in Sect.~2, this is not unexpected: the
nominal resolution of $\sim$1000 km/s has been measured on the spectra
of a lamp that uniformly illuminates the 1'' slit, while our galaxies
have spatial extent of $\sim$0.8'' (FWHM). We conclude that galaxies
with observed FWHM(He~II)$ \leq 1000 km/s$ are spectrally unresolved.

By our definition, the narrow and broad He~II emitters have observed
FWHM$<1200$ and FWHM$>1200$ km/s, respectively, corresponding to
intrinsic FWHM$\gtrless 663 km/s$. This limit has been chosen because
Wolf-Rayet stars are always associated with strong stellar winds with
speeds around or larger than 1500 km/s; He~II emitters powered by W-R
stars have expected line widths around or above that limit. The
possible He~II emitters span a broad range of line widths, but only 3
out of 12 have FWHM$>1500 km/s$.

In Fig.~\ref{dynamic} we also report the velocity difference $\Delta$v
between the He~II line and the systemic velocity, calculated measuring
the position of the He~II line in the rest-frame spectra. Because the
systemic redshift is centered on the interstellar lines, this
corresponds to the velocity difference between the He~II line and
those lines. We see that the reliable narrow He~II emitters have
$\Delta$v$_{He~II}$ values spanning from -1000 to 1000 km/s, with the
distribution well centered on $\Delta$v$_{He~II}=0$. On the other
hand, reliable broad He~II emitters always have $\Delta$v$_{He~II}>0$.
We ran a 2D KS test, and we found that the two distributions are
different at a 90\% level. This additional difference supports the
hypothesis that broad and narrow He~II emitters are two distinct
families.

\subsection{Spectral properties of stacked spectra}
In order to study the average properties of broad vs narrow vs
probable He~II emitters, we produced stacked spectra for these three
families of He~II emitters, following the same procedure used for the
composite of all the 277 He~II emitters. For these composites we also
experimented different stacking parameters, as previously described in
Sect.~3.1, and we again find that our set of parameters gives the best
and most reliable measures.

\begin{figure*}[!ht]
  \centering
\includegraphics[width=\textwidth]{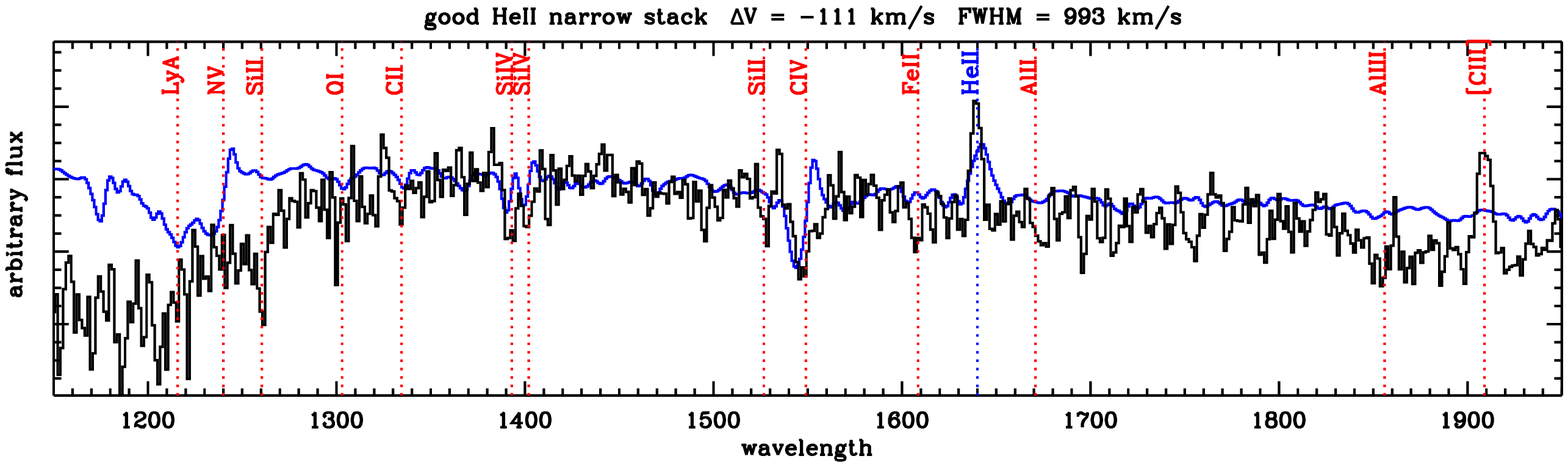}
\includegraphics[width=\textwidth]{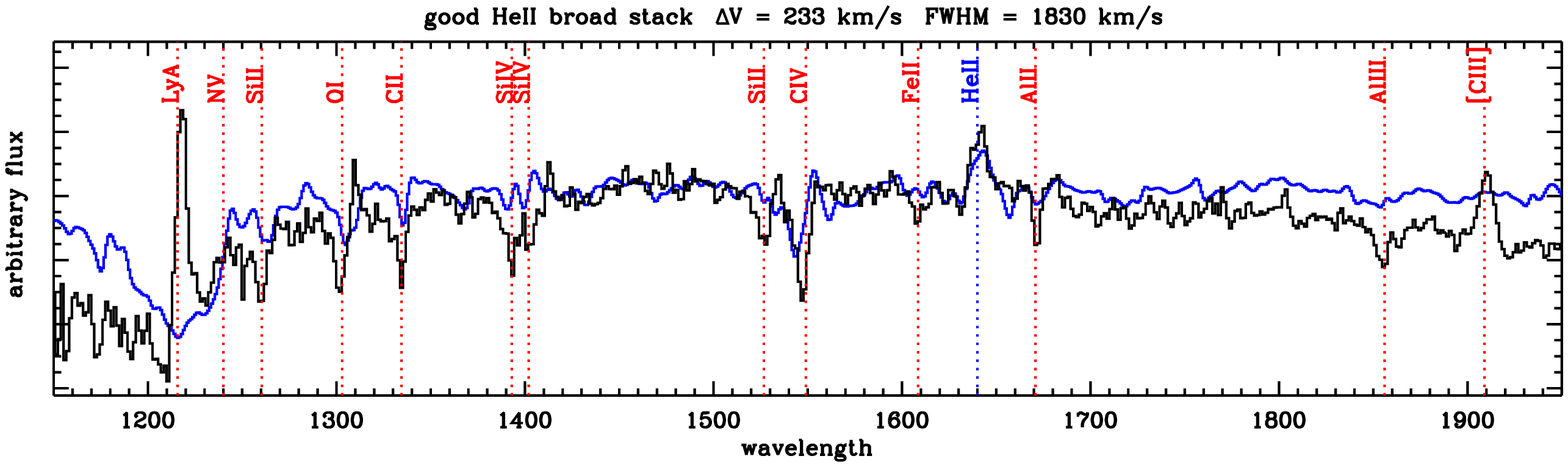}
\includegraphics[width=\textwidth]{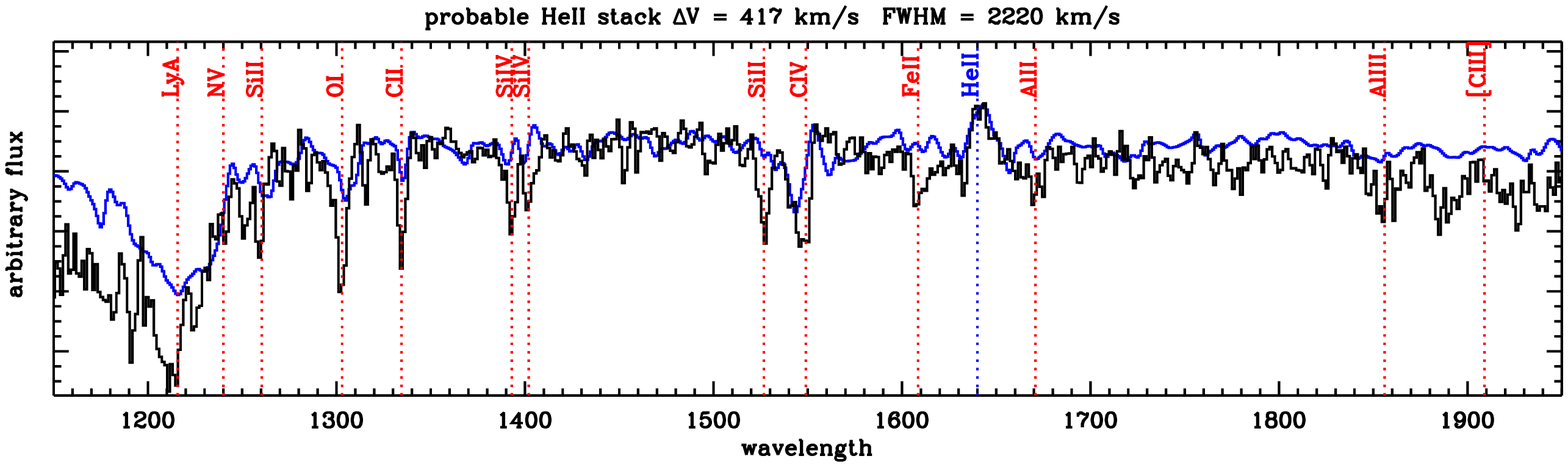}
\caption{Stack of the individual spectra for the reliable galaxies
  with narrow He~II emission (top), the reliable galaxies with
  broad He~II emission (middle), and the possible He~II emitters
  (bottom). The blue lines show models obtained combining the
  predictions for a burst of star formation by Maraston~et~al.~(2009)
  with the predictions for a Wolf-Rayet phase by
  Eldrige~\&~Stanway~(2012). As explained in the text, they are not
  an attempt of fitting the composite spectra with models, but rather
  are aimed at comparing some spectral features in the composite
  spectra with the models.}
\label{spec_stack}
\end{figure*}

The stacked spectra are shown in Fig.~\ref{spec_stack}, in comparison
with stellar population models by Maraston~et~al.~(2009) and
Eldridge~\&~Stanway~(2012). We will discuss the comparison between
data and models in Sect.~\ref{sect:discussion}. The measurements of
the rest-frame equivalent width EW$_0$ and FWHM for the relevant
interstellar absorption lines are listed in Table~\ref{tab:stacks}.

Recently, Sommariva~et~al.~(2012), developing a method already used by
Rix~et~al.~(2004), Halliday~et~al.~(2008), and Quider~et~al.~(2009),
have shown that some photospheric lines in the UV rest--frame domain
can be used to estimate the stellar metallicity and age. In
particular, using the latest version of the Starburst99 code
(Leitherer~et~al.~1999; Leitherer~et~al.~2010) they have determined
the strength of five UV features (1360-1380~\AA, 1415-1435~\AA,
1450-1470~\AA, 1496-1506~\AA, and 1935-2020~\AA) as a function of age
and stellar metallicity. The dependence on age is very weak, with the
EW of these features reaching a constant value after a few Myr
only. We applied this technique to the stacks of the narrow and broad
emitters, as well as to the stack of all of the 277 star forming
galaxies. Some of these spectral features are very weak and noisy in
our combined spectra; as a result, we obtain very different values
from different lines (from 0.01 $Z_{\odot}$ if the 1360--1380~\AA~
feature is used, to 0.2 $Z_{\odot}$ if 1496--1506~\AA~ is instead
used). Moreover, we do not see any difference between the values
obtained for the three different stacks (broad He~II emitters, narrow
He~II emitters, and the global sample).



The stack of the good--quality narrow He~II emitters clearly shows a
narrow He~II emission FWHM$_{observed}\sim1000$ km/s (hence unresolved
at our resolution) with a rest-frame equivalent width $EW\sim4\AA~$
comparable to those of individual spectra. The stack of the
good--quality broad He~II emitters also shows a broad He~II line with
FWHM$_{observed}\sim1800 km/s$ (hence 1500 km/s intrinsic) and
EW$\sim2\AA~$. Interestingly, the stack of the possible He~II emitters
also shows a clear He~II emission with EW$=3.2\AA~$ and
FWHM$_{observed}\sim2200 km/s$ (1960 km/s intrinsic).


From both a visual inspection of the composite spectra in
Fig.~\ref{spec_stack} and from the analysis of the equivalent widths
and line widths in Table~\ref{tab:stacks} it is clear that narrow and
broad He~II emitters have completely different properties.  The He~II
line width is different (FWHM$\sim$ 1000 vs 1600 km/s), and the
strength of the interstellar absorption lines also varies
significantly between the two families. One striking difference is
that the narrow He~II composite spectrum has very weak low ionization
interstellar lines (Si~II, O~I, C~II, EWs$\sim-1$\AA) compared to
the strong high ionization ones (Si~IV, C~IV, EWs$\sim$-5\AA);
compared to the narrow-He~II composite which shows normal
low-to-high ionization ratios (similar to the ratios for the stack of
all galaxies at $2<z<4.6$). Moreover, the C~IV line for the stack of
the narrow emitters has a red wing absorption that is not observed in
the other stacks.


\section{Analysis}\label{sect:analysis}
In the previous section we described the properties of a sample of 39
He~II emitters discovered among 277 $z>2$ galaxies of the VVDS Deep
and Ultra-Deep surveys. In this section we analyze and discuss these
properties with the aim of constraining the mechanism that produces
the He~II line. As discussed in the Introduction, only a few
astrophysical mechanisms can power the He~II nebular emission because
a powerful ionizing source is needed to produce photons with energies
$E>54.4~eV$ that can ionize He$^+$.

\subsection{AGN}\label{sect:agn}
The extremely blue continuum of active galactic nuclei is energetic
enough to produce photons with $e>54.4~eV$, thus ionizing $He^+$ and
powering the He~II emission feature. Similarly to the case of W-R
stars, typical spectra of AGN (both Type I and II) have other emission
lines in their spectra, such as C~IV and Si~II, that can be used as
diagnostics. Typical narrow line Type II AGN have
$\left<C~IV/He~II\right>=1.50$ (McCarthy~1993; Corbin~\&Boroson~1996;
Humphrey~et~al.~2008; Matsuoka~et~al.~2009).  Moreover, Type I AGN
(those AGN in which the active nucleus is directly exposed to the
observer) have line widths of 2,000 km/s and above.

Three He~II emitters in our sample have been classified as AGN. Object
910359136 has been discovered in the Ultra-Deep part of the survey,
and thus is covered both with the blue and the red grism. This
produces a rest-frame spectral window that covers all the spectra from
Ly$\alpha$ to C~III. The morphology of this object in the CFHTLS
images is diffuse, with no point--like components, suggesting that the
broad--line region is completely obscured from our line of sight. The
modest line widths (FWHM$_{obs}\sim1400 km/s$, corresponding to
intrinsic FWHM$_0\sim1000 km/s$) supports this interpretation. We
measure line ratios Ly$\alpha$/He~II$=11.5\pm1.6$,
C~IV/He~II$=1.97\pm0.35$, and C~III/C~IV=$0.72\pm0.25$, that are
remarkably similar to the average values for radio galaxies
$<Ly\alpha/He~II>=9.8\pm5.69$ and $<C~IV/He~II>=1.50\pm0.56$
(McCarthy~1993; Corbin~\&Boroson~1996; Humphrey~et~al.~2008).
Matsuoka~et~al.~(2009) also found C~IV/He~II$=1.34^{+0.57}_{-0.4}$ and
C~III/C~IV$=1.14^{+0.43}_{-0.31}$for narrow line regions at $2<z<2.5$
and with $41.5<L_{He~II}<42.5$. We can conclude that object 910359136
is indeed a Type II AGN.

The other two objects classified as AGN (20366296 and 20273801) are
instead drawn from the VVDS Deep survey, so they are only observed
with the red grism, and their rest-frame spectra only cover the
region between $\lambda=1600$ \AA~and $\lambda=2600$ \AA~, a spectral
domain which covers only from He~II to C~III. The He~II and C~III lines
are broad ($\sim2000 km/s$), and the sources are point--like in the
CFHTLS images; thus, it is likely that they are Type I AGN, in which
the nucleus is directly exposed to the line of sight directed to the
observer.

Are the remaining 36 He~II emitters powered by an AGN?  From a visual
inspection of Figs. \ref{zoom_narrow}, \ref{zoom_broad}, and
\ref{zoom_prob} it is evident that for the remaining He~II emitters in
our sample C~IV is either undetected (910252781, 910301515, 910191609,
910362042, 20215115, 910248329, 910285698, 910246547, and 20191386) or
in absorption (all the other cases). This implies C~IV/He~II ratios
always smaller than 0.5: the ratio is an upper limit for the objects
with no C~IV detection and it is negative for the objects with C~IV in
absorption. Since in typical AGN we expect a C~IV line always stronger
than the He~II line (McCarthy~1993; Corbin~\&Boroson~1996;
Humphrey~et~al.~2008; Matsuoka~et~al.~2009), we can exclude the AGN as
a plausible source of ionization for all 37 He~II emitters. We also
stress that it is very unlikely that a highly obscured AGN, like the
ones discovered by Civano~et~al.~(2011), is responsible for the He~II
emission in our objects. If those objects do not have in their spectra
the emission lines typical of unobscured AGN because they have been
absorbed by dust, there is no reason why the He~II line could not have
been absorbed too. Moreover, we stress that those objects for which a
C~IV emission component has been identified, responsible for the
asymmetry of the C~IV absorption profile, are unlikely to be powered
by AGN: AGN in fact usually do not show P--Cygni--like line profiles
in their spectra (e.g. Matsuoka~et~al.~2009).

Finally, it is also possible that a normal Type II AGN is buried deep
inside our objects, producing at the same time the He~II emission line
and a C~IV emission that is not visible since it would be hidden in
the C~IV absorption typical of the host galaxies. However, since AGN
do not produce low-ionization lines such as Si~II~$\lambda$1527, this
scenario would significantly reduce the C~IV absorption while leaving
Si~II unchanged, ultimately altering the C~IV/Si~II ratio. Since we do
not see objects with strange C~IV/Si~II ratios (i.e., objects with a
stronger Si~II than C~IV absorption), we can exclude this possibility.

\subsection{Wolf-Rayet stars}\label{sect:WR}
\begin{figure}
\centering \includegraphics[width=1.0\linewidth]{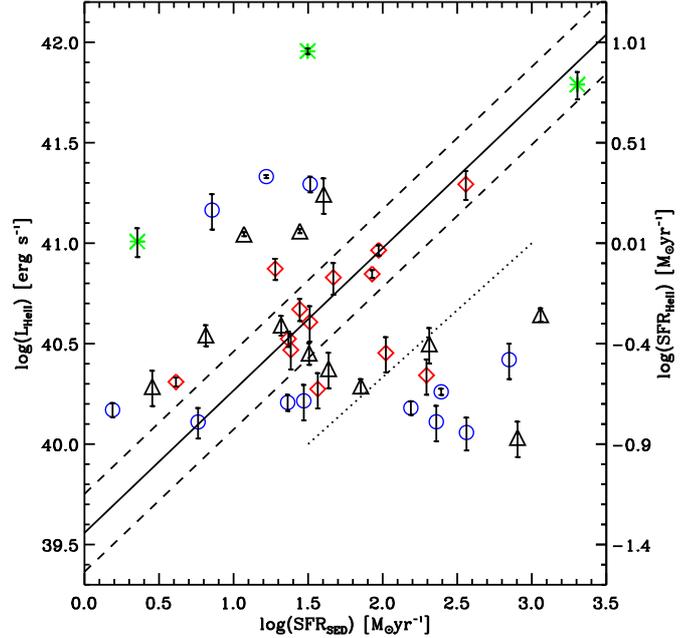}
\caption{For the 39 He~II emitters, comparison between the SFR
  measured fitting the broad-band SED of galaxies and the luminosity
  of the He~II line. The continuous and dashed lines show a linear
  best fit (with scatter) to the reliable He~II emitters with broad
  emission (red diamonds). The dotted line is the expected correlation
  between these two quantities according to the cooling radiation
  model by Yang~et~al.~(2006). The color code of the symbols is the
  same as in Fig.~\ref{lum_z}. The right axis reports the
  star--formation rate needed to produce this level of He~II
  luminosity (as for Fig.~\ref{lum_z}), assuming a top-heavy IMF with
  Salpeter slope extending from $1 M_{\odot}$ to $500s M_{\odot}$ and
  metallicity Z=0.}
\label{sfr_lum}
\end{figure}

\begin{figure}
\centering \includegraphics[width=1.0\linewidth]{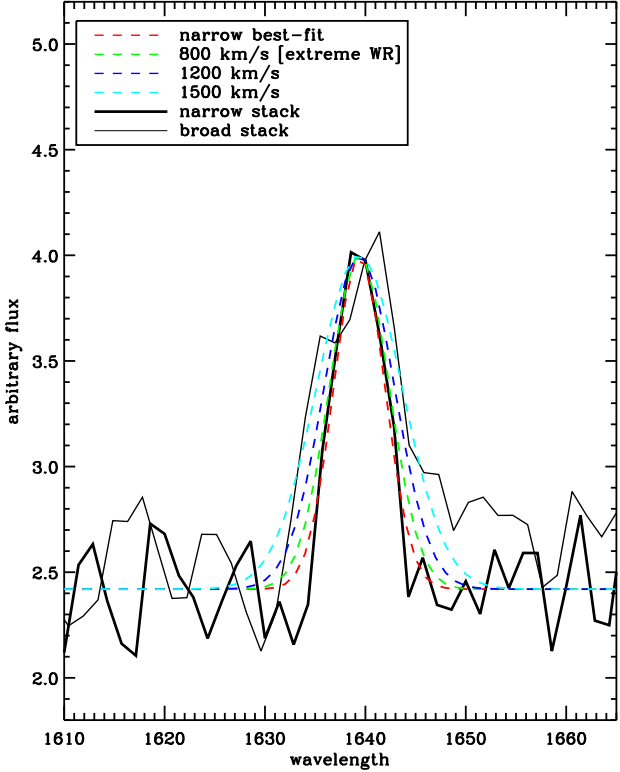}
\caption{Blow-up of the region around the He~II line for the stacks of
  the narrow emitters (thick black line) and the broad emitters (
  light black line). We report the Gaussian with an observed FWHM=1000
  km/s, the width that we measure for the He~II line in the narrow
  stack. For comparison, we also show other Gaussians with intrinsic
  FWHMs of 800 km/s (green dashed line), 1200 km/s (blue dashed line) and 1500
  km/s (cyan dashed line). These line widths, once convoluted with our 1000
  km/s resolution, produce observed FWHM of 1250, 1500, and 1800 km/s,
  respectively.}
\label{he2zoom}
\end{figure}

Another possible mechanism powering He~II emission is Wolf-Rayet
stars. Wolf-Rayet stars are evolved and massive stars ($>10
M_{\odot}$), descendents of O stars, with surface temperatures between
30,000 and 200,000 K that are losing mass very rapidly. They
typically show strong and broad emission lines produced by their
dense and fast stellar winds (Leitherer~et~al.~1995;
Leitherer,~Robert~\&~Heckman,~1996; Crowther~2007). The extremely high
surface temperatures of these stars can produce the energetic photons
needed to power the He~II emission. The He~II line width, as for the
other emission lines, is expected to be broad with a FWHM on the
order of few thousands of km/s (Schaerer~2003).

The Wolf-Rayet stars are commonly divided in WC, whose spectra are
dominated by carbon lines, and WN stars, with spectra dominated by
nitrogen lines. Both classes show strong and broad He~II~$\lambda$1640
emission, with typical line widths in excess of 1000 km/s
(Crowther~2007). In WC stars the
C~IV~$\lambda$1549/He~II~$\lambda$1640 ratio is always greater than
one (Sander,~Hamann~\&~Todt,~2012), while in WN stars this ratio is
lower than 1, with cases with no C~IV emission at all
(Hamann,~Grafener~\&~Liermann, 2006). While the abundance of WC and WN
stars is about the same in the Milky Way, WN are ten times more common
than WC stars in low metallicity objects (Massey~\&~Holmes~2002).  In
addition, Crowther~\&~Hadfield~(2006) showed that in low metallicity
environments the He~II flux and line width are reduced, with line
FWHMs as low as 850 km/s. Chandar,~Leitherer~\&~Tremonti~2004 and
Hadfield~\&~Crowther identified the most extreme W-R galaxy in the
local Universe: NGC3125 shows a strong He~II line with EW$\sim7$\AA~
and FWHM$\sim$800 km/s. The C~IV emission in W-R stars often shows a
P--Cygni profile, typical of the hot expanding envelope associated
with this stellar phase (Crowther~2009). In Sect. 3.4 we showed that
11 out of 13 reliable broad emitters have strong indications of
P-Cygni C~IV line profiles (Fig.~7). This probably indicates that, for
those broad emitters, the He~II and C~IV originate in the winds of
massive W-R stars.

In Fig.~\ref{he2zoom} we show the blow-up of the region around the
He~II line for the narrow and broad stacks, in comparison with
Gaussians with different widths, in order to evaluate the ability of
our data, which have a typical velocity resolution of 1000 km/s, of
resolving the line widths for the most extreme W-R objects. We
overplot a Gaussian with observed FWHM=1000 km/s, the width that we
measure for the He~II line in the stack of the narrow emitters, and
that again corresponds to our spectral resolution. We also overplot a
gaussian with intrinsic FWHM=800 km/s (the narrowest He~II in the
local Universe produced by W-R stars) that corresponds, after
convolving with the instrumental resolution of 1000 km/s, to an
observed FWHM=1250 km/s. It is evident that this Gaussian is wider
than the observed He~II line, and that in principle we would be able
to identify a line with such FWHM observed at our resolution. For
comparison, we also report two gaussians with observed FWHM of 1500
and 1800 km/s, corresponding to intrinsic FWHM of 1200 and 1500 km/s
at our resolution. The latter corresponds to the width that we measure
for the He~II line in the broad composite.

Including the W-R phase in stellar population synthesis models is not
an easy task; the W-R phase heavily depends on the number of very
massive stars (and thus on the IMF), on the atmosphere and wind models
of very massive stars, and as we mentioned earlier, on the metallicity
of the environment.  Shapley~et~al.~(2003), who first detected a broad
He~II emission in the composite of 1000 $z\sim3$ galaxies, interpreted
the He~II line as the signature of W-R stars, but then failed to
identify a stellar model that could at the same time reproduce the
strength of the line and the P-Cygni C~IV
emission. Brinchmann,~Pettini~\&~Charlot~(2008) combined the
state--of--the--art population synthesis models with observations of
local galaxies in the Sloan Digital Sky Survey, generating a model
that can reproduce the observational features of the S03 composite
spectrum with a half solar metallicity.

Eldridge~\&~Stanway~(2012; hereafter ES12) have recently published a
grid of synthetic spectra produced using a new stellar population
synthesis code that includes massive binary stars. These models
naturally incorporate a W-R phase, and predict strong He~II emission
about 50 Myr after the burst of star formation (the age for which the
number of W-R stars reaches a maximum). The strength of the He~II line
depends on the metallicity, and it is strongest for Z=0.004. In the
ES12 models the He~II emission is broad and is always accompanied by a
C~IV absorption+emission P-Cygni line. Using these models and invoking
a Carbon reduction with respect to other metals, ES12 can reproduce
quite nicely the C~IV line profile and the He~II emission in the S03
spectrum.

In Fig.~\ref{spec_stack} we checked whether or not we could reproduce
the properties of our stacked spectra with the ES12 models. However, these
models do not include interstellar absorption lines. In order to
build more reasonable synthetic spectra, we decided to combine the
ES12 models with the synthetic high-resolution UV spectra by
Maraston~et~al.~(2009) that conversely include interstellar lines
(but no He~II emission). We stress that we did not try to perform any
rigorous fitting procedure, but we simply compared the predictions by
a combination of these two sets of models with our composite
spectra. The only restriction we imposed is that the metallicity of
the components used to build our synthetic spectra is always subsolar
(as it is expected to be at $z\sim2.5$).

The models are superimposed to the composite spectra in
Fig.~\ref{spec_stack}. For the stack of the narrow He~II emitters,
which does not have strong low-ionization lines, we simply chose an
E12 model, with a continuous star formation of 1 $M_{\odot} yr^{-1}$
lasting 100 Myr and metallicity Z=0.004 (but an instantaneous
10--Myr--old burst and the same metallicity produces a similar
synthetic spectrum). For the broad He~II emitters, we built a more
complicated ad-hoc (and possibly unphysical) model: we combined a
150--Myr--old burst of star formation with metallicity Z=0.01 from
Maraston~et~al.~(2009), with a second burst of star formation with
metallicity Z=0.004 and age=10 Myr, from the ES12 library. Finally, for
the stack of the possible emitters, we produced a similar synthetic
spectrum to the one used for the broad composite, with only a slightly
younger second component from the ES12 library. We stress that these
are not the best possible models; rather, they are only one of the
many combinations that come close to reproducing the features of our
composite spectra.

Starting from the narrow composite, it is clear that the overall shape
of the spectrum is reproduced well by the synthetic model. The model
reproduces the absence of low-ionization lines (O~I, Si~II, and C~II),
and the strength of the Si~IV absorption feature at
$\sim1400$\AA~. However, the He~II emission in the model is
significantly broader ($\sim2000$ km/s) than in the composite (even
though the EW is similar), and the shape of the C~IV line is
completely different, with the model predicting a P-Cygni profile that
we do not observe in the stack. We remark that other models in ES12
library are not able to reproduce the narrow He~II line that we see in
our composite, nor the absence of C~IV P-Cygni profile. On the other
hand, our ad-hoc models for the broad and possible He~II emitters
reproduce quite nicely the strength, shape, and width of the He~II
line. Moreover, the shapes of the C~IV lines in the model and in the
composite spectrum are qualitatively similar, and all the other
absorption lines are quite well reproduced.

The outcome of this analysis is that we were unable to reproduce the
properties of the narrow He~II emitters with the state-of-the-art
stellar population models including W-R stars. On the other hand,
although more refined modeling might be required, we can reproduce the
main spectral features of the broad emitters with stellar models
including a specific treatment of the W-R phase.

We remark that the lack of broad He~II emitters with a negative
$\Delta$v between the He~II line and the systemic velocity (see
Fig.~\ref{dynamic}) is compatible with the W-R scenario. It is
expected that if the He~II emission originates in the fast winds of a
W-R star, the profile of the line is the typical P-Cygni profile
(Venero,~Cidale,~\&~Finguelet~2002). It is also worth noting that a
correlation between the star--formation rate of the galaxy and the
number of W-R stars (and thus the He~II luminosity) is expected for
the W-R scenario, with the only caution that the W-R phase might be
very short, and would not match the global star--formation rate of the
galaxy.  In Fig.~\ref{sfr_lum} we compare the star--formation rate
measured from fitting the SED of galaxies (see details in
Sect.~\ref{sect:photometry}) and the luminosity of the He~II line.
Interestingly, we observe a correlation between the luminosity of the
He~II line and the star--formation rate for the galaxies in the
high--quality broad emitters category (see Fig.~\ref{sfr_lum}), in
broad agreement with predictions by models.

\subsection{Gravitational cooling radiation from gas accretion}\label{sect:grav_cool}

Gravitational cooling radiation from gas accreting onto dark matter
haloes is another mechanism that can produce Ly$\alpha$+He~II emission
lines, as predicted by many theoretical papers. Keres~et~al.~(2005)
showed that about half of the gas infalling onto dark matter potential
is heated to temperatures $T\sim10^4$ -- $10^5$ K. At these
temperatures, the gas is expected to cool via line emission, in
particular Ly$\alpha$ and He~II (Yang~et~al.~2006). The Ly$\alpha$
line is in principle expected to be the strongest line, but its
strength depends on the radiative transfer of Ly$\alpha$ photons and
on the geometry and dynamics of the system; as a result, the
Ly$\alpha$ line is easily destroyed. On the other hand, He~II emission
should always be observed, and its luminosity should correlate with
the star--formation rate of the galaxy the gas is falling into.
Intuitively, the more gas falls onto the galaxy, the higher the He~II
emission and thus the higher the SFR. The correlation is tighter
assuming that the infalling gas is optically thin, and the predicted
scatter is larger if the gas is optically thick (Yang~et~al.~2006).

Another prediction of this model is that the He~II line is expected
to have widths not larger than 300 km/s, and the He~II emission is
expected to be less extended than that of the Ly$\alpha$ line
(Yang~et~al.~2006; Haiman,~Spaans~\&~Quataert~2000). However, these
models assume primordial composition for the infalling gas; the
presence of metals can change the cooling efficiency and thus affect
the strength of the He~II line.

With the spectral resolution of our observations
$FWHM_{intrinsic}\sim$660km/s we are only able to say that the widths
of the narrow He~II emitters are broadly consistent with the line
width of $\sim$300km/s predicted by the models (Yang et
al. 2006). Moreover, galaxies with a resolved He~II line
(i.e., galaxies with FWHM$_{He~II}>1200km/s$) are not compatible with
the theoretical predictions for the cooling radiation
scenario. Unfortunately, the spatial resolution of our spectra
(FWHM$_{seeing}=0.8$'' at best) does not allow us to check if the
He~II and Ly$\alpha$ emission (where present) have different spatial
extents. Interestingly, we find that the C~IV line profile of the
narrow He~II emitters in the composite spectrum shows a red absorption
wing. In the context of accretion, this could be interpreted as the
signature of infalling gas.

However, we saw in Fig.~\ref{sfr_lum} that, for the narrow He~II
emitters, we do not observe any correlation between the He~II
luminosity and the overall star--formation rate of the galaxies, in
disagreement with the predictions by Yang~et~al.~(2006).



Another test that we can perform is to compare the number density of
He~II emitters above a given He~II Luminosity with the predictions of
the models: since the gas accretion is ubiquitous in cosmological
simulations, models predict number density as high as $0.1$ He~II
emitters per cubic Mpc at $z\sim$2 -- 3 with He~II Luminosity
$L_{He~II}>10^{40} erg/s$ (Yang~et~al.~2006). We calculated the number
density of He~II in our sample (excluding the broad emitters, that are
probably powered by W-R stars) applying the classical $V/V_{max}$
method (Avni~\&~Bahcall~1980), and weighting each galaxy by the
photometric sampling rate (basically the ratio between the number of
galaxies of a given magnitude that have been spectroscopically
observed). The total volume sampled by the Ultra-Deep and Deep surveys
between $z=2$ and $z=3.5$ is about 10$^{6.3}$ and $10^7$ Mpc$^3$,
respectively. The number density that we measure for our sample is
$\sim 10^{-4} gal/Mpc^3$, thus 2 -- 2.5 orders of magnitude smaller
than the value expected by the simulations.

We conclude from this indirect argument that it seems
unlikely that our 39 He~II emitters are powered by cooling
radiation. However, we note that to make accretion models compatible
with our observations, some dramatic reduction in the cooling
radiation output, by about 2 orders of magnitude, would be needed in
the simulations.
 
\subsection{Pop III star formation}\label{sect:PopIII}
Star formation in metal-free or very low metallicity regions is
another mechanism that is able to produce the extremely energetic
photons needed to ionize $He^+$ and thus power the He~II emission.
Many authors suggested that PopIII star forming regions can be found
looking for dual Ly$\alpha$+He~II emitters (Tumlinson~et~al.~2001;
Schaerer~2003; Raiter~et~al.~2010). These models predict extremely
large equivalent widths for both Ly$\alpha$ and
He~II. Schaerer~(2003), for example, expects EW(Ly$\alpha$)=600 --
1500\AA~ and EW(He~II)=20 -- 100\AA~for a burst of star formation
(according to the metallicity, from Z=0 to Z=10$^{-5}$, and the
IMF). However, such large EWs are sustained only for 2 Myr at the
most, and the equilibrium values reached in the case of a constant
star formation are much smaller: for a top-heavy IMF and extremely low
metallicities ($Z<10^{-5}$), the EW for the Ly$\alpha$ and He~II lines
are $\sim$300 -- 400\AA~ and $\sim5-15$\AA~, respectively
(Schaerer~2003). We remark that the equivalent width of the He~II line
for this model strongly depends on the metallicity, and drops to
virtually zero at $Z>10^{-5}$. The typical ratio between the strength
of the He~II and Ly$\alpha$ line that the models predict also depends
on the metallicity as well as the IMF: for a top-heavy IMF and zero
metallicity this ratio is around He~II/Ly$\alpha\sim0.1$
(Schaerer~2003; Raiter~et~al.~2010). Unfortunately, the Ly$\alpha$ is
resonant, and the Ly$\alpha$ photons can be easily destroyed, as noted
above, so the predictions for the Ly$\alpha$ luminosity and EW are
highly uncertain.

The distribution of EW(He~II) for all He~II emitters in our sample
(except AGN) as shown in Fig.~\ref{sn_ew} is EW$\sim$1--8~\AA~,
values that are similar to those expected for an event of continuous
PopIII star formation (Schaerer~2003; Raiter~et~al.~2010). We showed
in Fig.~\ref{lum_z} that the SFR needed to power the
He~II fluxes that we observe in our galaxies are not so extreme: as
long as the IMF is top-heavy (as seems reasonable for PopIII star
formation) and the metallicity is close to zero we find values of 0.1
-- 10 $M_{\odot}yr^{-1}$.

In Fig.~\ref{sfr_lum} we showed that there is no correlation between
the SFR measured for the individual galaxies and the He~II line
luminosity. Moreover, from Fig.~\ref{sfr_lum} we also showed that,
assuming again that the He~II emission is powered by PopIII star
formation, the SFR in PopIII stars needed to generate the observed
He~II luminosities is always smaller (by $\sim2$ orders of magnitude)
than the SFR measured from fitting the SED. These two findings
together imply that, if the nebular He~II emission is really powered
by PopIII star formation, it is happening in a small fraction of the
volume of the galaxy containing some pristine gas and not in the
galaxy as a whole. This would imply that the observed integrated
spectra are a mix of two components 1) normal star-forming
moderate-metallicity stellar populations comprising a majority (90\%)
of the SF activity, and 2) PopIII stellar populations producing the
ionized He, comprising a small fraction of the global star
formation. This implies that the EW of the He~II line is somewhat
diluted in the integrated spectrum. If the host spectrum had a
continuum 100x brighter than the PopIII component at 1640\AA, the EW
of the He~II line in the total spectrum would be 100x smaller than the
intrinsic PopIII value.


From Fig.~\ref{dynamic} we also saw that the He~II emitters in our
sample show a broad range of He~II line widths: 11 He~II emitters have
intrinsic FWHM$_{He~II}<660 km/s$ and have velocity differences
between the He~II line and the system well distributed around
zero. These He~II emitters are not compatible with the W-R scenario in
which the He~II line is expected to be broadened by stellar winds, and
are more compatible with the PopIII scenario. Conversely, the He~II
emitters with broad He~II line are instead inconsistent with the
PopIII scenario, which would produce narrow He~II emission, and are
instead more likely to be powered by W-R stars (as explained in the
previous section).

Another prediction of the PopIII model that we can test is the
star--formation rate density in PopIII stars at $z\sim2.5$, the median
redshift of our sample. Tornatore,~Ferrara~\&~Schneider~(2007) predict
that, although the metallicity in the center of overdensities is
rapidly enriched by star formation feedback, zero metallicity regions
can survive in the low--density regions around the large overdensities
even down to $z\sim2$.  In particular, at $z\sim2.5$
Tornatore,~Ferrara~\&~Schneider~(2007) predict a PopIII SFRD of
3$\times10^{-7} M_{\odot}yr^{-1}Mpc^{-3}$. If we make the conservative
assumption that only the 11 reliable He~II emitters with narrow He~II
emission in our sample (excluding the possible He~II emitters or a
fraction thereof) are powered by PopIII star formation, and we use the
conversion between He~II Luminosity $L_{He~II}$ and PopIII SFR for a
top-heavy IMF and zero metallicity calculated by Schaerer~(2003), we
get a SFRD$\sim10^{-6} M_{\odot}yr^{-1}Mpc^{-3}$, only a factor three
above the prediction by Tornatore,~Ferrara~\&~Schneider~(2007). This
value is in broad agreement with the one derived by
Prescott,~Dey~\&~Jannuzi~(2009), who found SFRD$\sim3.3\sim10^{-7}
M_{\odot}yr^{-1}Mpc^{-3}$, based on the discovery of a
Ly$\alpha$+He~II nebula at $z=1.67$. We conclude from this analysis
that the SFRD derived from the observed He~II assuming it is produced
by PopIII stars is in broad agreement with predictions from the model
of Tornatore,~Ferrara~\&~Schneider~(2007), but that, if this is indeed
the mechanism in place, models would need to be iterated on to match
the observed SFRD.


\subsection{Peculiar stellar populations}\label{sect:pec}

In the local Universe several HII regions showing nebular He~II
emission have been identified (Schaerer,~Contini~\&~Pindao~1999;
Izotov~\&~Thuan~1998; Guseva~\&~Izotov~2000; Kehrig~et~al.~2011). For
many of these regions the spectra show the features typical of W-R
stars, which are thought to be the source of the strong ionizing continuum
(see Sect.~\ref{sect:WR}), but some are not clearly associated with
W-R stars, nor with O-B stars (Kehrig~et~al.~2011).
  
A nebular He~II component is also observed in the spectra of galaxies
(de Mello~et~al.~1998; Izotov~et~al.~2006; Shirazi~\&~Brinchmann~2012;
Shapley~et~al.~2003; Erb~et~al.~2010). Once again, for many of these
objects, especially at low metallicity, the nebular He~II emission is
not accompanied by any W-R signatures, and thus the W-R scenario can
not fully explain the origin of this nebular He~II emission.

The origin of the nebular He~II emission in these objects is
puzzling. Kudritzki~(2002) showed that not only W-R, but also certain
O stars are hot enough to produce nebular He~II
emission. Brinchmann,~Kunth~\&~Durret~(2008) claimed that at low
metallicity the main source of nebular He~II emission appears to be O
stars, arguing for less dense stellar winds that can be optically thin
and thus be penetrated by the ionizing photons. However, the ability
of O stars to produce nebular He~II emission is still debated, as some
local nebulae do show narrow He~II emission but no they do not seem to
contain O stars neither W-R stars (Kehrig~et~al.~2011).

Another possibility is that at low metallicity massive stars rotate
fast enough to evolve homogeneously (i.e., Meynet~\&~Mader~2007), thus
producing higher surface temperatures. However, observational
evidence for this scenario has not been reported so far.

Post-AGB stars are also claimed to be able to produce nebular He~II
emission (Binette~et~al.~1994). In this scenario, these authors
demonstrate that 100 million years after a strong burst of star
formation, the dominating source of ionizing photons becomes post-AGB
stars, and that their combined radiation field is sufficient to ionize
He~II. However, even assuming that our galaxies are as massive as
10$^{11} M_{\odot}$, and that all the stellar mass has been formed
during an initial burst of star formation, the He~II luminosities
produced by post-AGB stars would be 2--3 orders of magnitude smaller
than those observed for our sample.

If the physical process responsible for nebular He~II in the local
universe, whatever it turns out to be, is similar to the process at
work at z$\sim$3, we note that there must have been a strong increase
of the frequency of this phenomenon, or its timescale, as we observe
more than 3\% of the global galaxy population at z$\sim$3 with narrow
He~II, while this number is much lower at z$\sim$0.

\subsection{Other mechanisms}
Shocks by supernova driven winds have been proposed by many authors
(Taniguchi~\&~Shioya~2000; Mori,~Umemura,~\&Ferrara~2004) as the
possible ionization source powering the so called Ly$\alpha$ blobs
(Steidel~et~al.~2000). It is then reasonable to consider whether this
mechanism can be the source of ionization producing the He~II emission
as well. However, fast shock models always predict C~IV in emission
(Dopita~\&~Sutherland~1996; Allen~et~al.~2008), with C~IV/He~II ratios
much higher than those measured for our sample ($0.1<C~IV/He~II<4$,
depending on the shock velocity). We can therefore exclude that the
bulk of our He~II emitters are powered by supernova winds.

\section{Discussion}\label{sect:discussion}
In the previous sections we reviewed the mechanisms that are able to
produce the very energetic photons ($e>54.4$ eV) needed to ionize
He$^+$, and that can power the He~II emission, and we compared what
the models of these mechanisms predict with the properties of our He~II
emitters.

As we discussed in Sect.~\ref{sect:agn}, apart from the three certain
AGN, it is unlikely that the remaining 37 He~II emitters are powered
by an AGN, as C~IV is always undetected or in absorption
(Figs.~\ref{spec_narrow},~\ref{spec_broad} and \ref{spec_uncertain}),
while for typical narrow line regions it is expected to be in emission and always
stronger than He~II (McCarthy~1993; Humphrey~et~al.~2008). With a
similar argument, we can also exclude that the source of ionizing
photons are the fast winds produced by supernovae. Similar conclusions
(although less robust) were drawn by Prescott,~Dey~\&~Jannuzi~(2009)
and Scarlata~et~al.~(2009), who identified Ly$\alpha$+He~II emitters
at $1.5<z<2.5$.

Shapley~et~al.~(2003) presented the stack of 1000 spectra of galaxies
at $z\sim3$, in which they detected a broad He~II emission
(FWHM$\sim1500$km/s). Because of the broad He~II line, and a clear
signature of fast winds (i.e., the C~IV P-Cygni profile in their
composite), they interpreted the He~II emission as evidence of a
rich W-R population in their sample. More recent studies have confirmed
this claim (Brinchmann,~Pettini~\&~Charlot~2008;
Eldridge~\&~Stanway~2012).  We saw in Fig.~\ref{spec_all} that the
composite of our 277 galaxies at $2<z<4.6$ is quite similar to the S03
composite, so we could be tempted to draw similar conclusions about
the source of the He~II emission.

We saw however that our He~II emitters sample is quite heterogeneous,
with as many objects with narrow He~II emission as with broad. The two
families have very different properties: the narrow He~II emitters
have weaker low-ionization lines (O~I, C~II, and Si~II) than the broad
ones; the narrow stack shows a red absorption wing in the C~IV line
that is not observed at all in the composite of the broad emitters.

From a direct comparison with modern models of stellar population
synthesis that include the W-R phase (Fig.~\ref{spec_stack}), we
conclude that these models can reproduce, at least qualitatively, the
spectral features of the composite of the broad He~II emitters, mixing
components of different ages and metallicities. Conversely, these
models can not reproduce the main features of the narrow stack.
Although with the appropriate combination of metallicity and age we
can reproduce the EW of the He~II line, these models always predict a
broad He~II line and a C~IV P-Cygni line.

Another piece of information comes from the comparison of the SFR of
the galaxies (measured from the SED fitting) and the He~II
luminosity. The broad He~II emitters do show a positive correlation,
as would be expected for the W-R scenario, assuming that with higher
the star formation, there would be more W-R stars. No correlation at
all is observed for the narrow emitters.

Putting all this evidence together, we can say that the properties
of the broad He~II emitters can be, at least qualitatively, explained
by the W-R scenario. The narrow emitters, instead, are a different
family for which the W-R scenario does not work, and thus an alternative
explanation is required.

We have identified three main processes which could be the cause of
the He~II emission in the spectra with narrow He~II emission:
gravitational cooling radiation, PopIII star formation, and the
presence of a peculiar stellar population. As indicated in
Sect.~\ref{sect:grav_cool} the He~II luminosities measured from our
galaxies are compatible with the ones predicted by the model by
Yang~et~al.~(2006); however, we do not see any correlation between the
total SFR of the galaxies and the strength of the He~II line predicted
by this model. Moreover, the number density of He~II emitters that we
derive is at least two orders of magnitude smaller than the prediction
by Yang~et~al.~(2006). To make models compatible with our observations
would therefore require either that {\it all} galaxies at these
redshifts be He~II emitters with EW$\sim$4A, which is excluded by our
observations, or that models overpredict the number density of He~II
emitters by more than two orders of magnitude.

As discussed in Sect.~\ref{sect:PopIII}, all observables for the
narrow He~II emitters are compatible with the predictions of a PopIII
scenario (and this hypothesis could also be valid for some of the
possible He~II emitter population), if we assume that the
star--formation event is happening in pockets of pristine gas that
either survived at the periphery of the galaxies or has just recently
accreted, hence concerning small volumes and not the full volume of
the galaxies. It is quite striking that the SFRD in PopIII stars that
we derive from these emitters is in broad agreement with the model by
Tornatore,~Ferrara~\&~Schneider~(2007) at $z\sim3$, the factor of 3
difference between the two indicating that some further adjustments to
the model may be necessary.

We also noted in Sect.~\ref{sect:pec} that various classes of objects
with nebular He~II emission and no W-R signatures have been identified
in the local Universe. Although no convincing explanation for the
origin of the He~II emission has been found so far, these objects have
indeed similar properties to our narrow He~II emitters, and we can not
exclude that they are all powered by the same mechanism.  The fact
that these objects are quite rare in the local Universe
(Kehrig~et~al.~2011; Shirazi~\&~Brinchmann~2012) and more frequent in
our sample at high redshift (3--5\% depending on the possible He~II
class, see below) may simply reflect the evolution of the
star--formation rate density of the Universe (that is at its peak
around the median redshift of our galaxies) and the evolution of the
average metallicity of the Universe.

We were only able to compare our results with a small number of
previous studies. Scarlata~et~al.~(2009) found He~II emission
connected to an extended Ly$\alpha$ blob at z$\sim2.2$. Similarly to
the bulk of our emitters, they did not detect C~IV nor N~V in
emission, as expected when the nebular emission is due to AGN
activity. They concluded that the most plausible scenario for the
He~II emission is the cooling radiation, that can at the same time
explain the width of the He~II line and the lack of other emission
lines. Prescott,~Dey~\&~Jannuzi~(2009) also found a large
Ly$\alpha$+He~II nebula at $z\sim1.67$. However, they found a
rest-frame EW for the He~II line of $\sim$35\AA, much larger than the
typical values of our sample; moreover, they detect C~IV in emission
in their spectrum (even though weaker than He~II). These two findings
suggest that their object is different from the bulk of our
galaxies. In the end, they can not constrain the mechanism powering
the He~II emission in their object. Erb~et~al.~(2010) thoroughly
analyze the UV-optical rest-frame spectrum of a galaxy at z=2.3, in
which they detect the He~II line, with a rest-frame equivalent width
EW$_0$=2.7\AA. Thanks to their high--resolution spectroscopy
($\sigma_V\sim$200 km/s), they can study in detail the shape of the
He~II line, finding that 75\% of the He~II flux comes from a broad
component (FWHM$\sim$1000 km/s) and 25\% from a narrow, unresolved
component. While they conclude that the broad emission probably has a
stellar origin (W-R stars), the narrow emission is indicative of a
hard ionizing source, possibly a very low metallicity star forming
region.


To summarize, we have identified a rich population of star forming
galaxies with He~II emission: some show a broad He~II emission, and
can easily be explained by standard W-R populations; some have narrow
He~II emission, and they could be the high redshift counterparts of
the local objects displaying nebular He~II emission. Various
mechanisms, more or less plausible, have been proposed to explain the
He~II emission in these objects (see Sect.~\ref{sect:pec}), and in
this paper we show that PopIII star formation is another mechanism
that can reproduce the properties of these objects. With the current
stellar population models (especially for the most massive stars and
low metallicities) and with the current observations we are not able
to further constrain the real mechanism.




\section{Conclusions}

In this paper we have reported the discovery of 39 He~II~$\lambda$1640
emitters, discovered in a sample of 277 VVDS galaxies with
$2<z<4.6$. Our main findings are the following:
   \begin{enumerate}
     \item We showed that He~II emission is a common feature in the
       spectra of $2<z<4.6$ galaxies: 39/277 galaxies (14\%) have
       He~II in emission in their spectra, with EW$_0=$1 -- 7\AA~. We
       also showed that not all of the He~II emitters have Ly$\alpha$
       in emission. On the contrary, we report for the first time the
       discovery of 17 He~II emitters with Ly$\alpha$ line in
       absorption.
     \item The rest-frame EW for the He~II line in our spectra is
       moderate; apart from three AGN, for which we find
       EW$_0\sim$15 -- 20, we find EW$_0\sim$1 -- 7, similar to the
       values expected for PopIII stars, for a continuous episode of
       star formation in a zero-metallicity region, assuming a
       top-heavy IMF, but also consistent with the EW expected for gas
       accretion.
     \item Apart from the three AGN, we never detect C~IV in emission in
       the spectra of the He~II emitters. In about 30\% of the cases
       C~IV is undetected, and in all the other cases it is in
       absorption.  We interpret this as evidence that He~II
       emission in most of our galaxies is not powered by AGN or
       supernovae winds.
     \item If we isolate the He~II emitters with broad He~II line
       (FWHM$_{obs}>1200 km/s$), we do observe a correlation between
       the SFR of the whole galaxy (measured by fitting the UV part
       of the SED) and the He~II line luminosity. The composite
       spectrum of these objects is, at least qualitatively, reproduced
       by stellar population synthesis models including W-R stars.
       We conclude that for these objects the He~II emission is most likely
       powered by W-R stars.
     \item The radiation produced by gravitational cooling does not
       seem to play a role in producing the He~II emission in our
       sample. For the narrow emitters, we see no correlation between
       the global SFR of each galaxy determined from SED fitting and
       the luminosity of the He~II lines, as would be expected for
       this scenario.  Moreover, we measured a number density of He~II
       emitters with $L_{He~II}>10^{40}erg s^{-1}$ of $\sim10^{-4}
       Mpc^{-3}$, while the predictions for the cooling radiation
       scenario are 2 -- 2.5 orders of magnitude higher. This implies
       that either these cooling radiation simulations overestimate
       the number density of He~II emitters, or that some mechanism
       somehow lowers the efficiency of the cooling radiation, which
       results in fainter He~II fluxes.
     \item If we assume that the He~II emission originates from a
       PopIII star forming region and we convert the He~II luminosity
       into SFR using a simple recipe by Schaerer~(2003), we find that
       a SFR of 0.1 -- 10 $M_{\odot}yr^{-1}$ is enough to sustain the
       observed He flux. The SFR derived from this PopIII hypothesis
       is always smaller (by 2 orders of magnitudes on average) than
       the global SFR measured from the SED fitting. This implies that
       the He~II emission could originate from a relatively small
       volume of pristine gas which could be distributed around the
       galaxies, uncontaminated by metal mixing processes, and in
       which PopIII stars are currently forming. On the other hand,
       considering that PopIII star formation is powering the narrow
       He~II emitters, as a conservative lower limit, we find a SFRD
       in PopIII stars of $10^{-6} M_{\odot} yr^{-1} Mpc^{-3}$, only 3
       times higher than the value predicted at $z\sim2.5$ by
       Tornatore,~Ferrara~\&~Schneider~(2007).
       \item The narrow He~II population has similar properties to a
         rare class of local objects (both individual nebulae and
         galaxies) which display nebular He~II emission. A clear
         mechanism for the origin of the nebular He~II line in these
         objects has yet to be identified, and we can not exclude
         that this mechanism is the low redshift tail of the PopIII
         star formation observed in the z$\sim$3 galaxies.
   \end{enumerate}

 
   \begin{acknowledgements}
We are very grateful to Jarle Brinchmann, Daniel Schaerer, Matthew
Hayes, and Jean-Michel Deharveng for useful discussions. We also thank
the anonymous referee for the constructive comments. This work is
supported by funding from the ERC advanced grant
ERC-2010-AdG-268107-EARLY.
   \end{acknowledgements}

\end{document}